\documentclass[prd, amsfonts, twocolumn, nofootinbib, showpacs]{revtex4-1}
\usepackage{graphicx, epsfig}
\usepackage{color}
\usepackage{amsmath}
\usepackage{mathtools}
\usepackage{physics}
\usepackage{amssymb}
\usepackage{slashed}
\usepackage{bbold}
\usepackage{multirow}
\usepackage{ulem}
\usepackage{lipsum}

\newcommand{\be}{\begin{equation}}
\newcommand{\ee}{\end{equation}}
\newcommand{\bea}{\begin{eqnarray}}
\newcommand{\eea}{\end{eqnarray}}

\newcommand{\gapp}{\mathrel{\raise.3ex\hbox{$>$}\mkern-14mu \lower0.6ex\hbox{$\sim$}}}
\newcommand{\lapp}{\mathrel{\raise.3ex\hbox{$<$}\mkern-14mu \lower0.6ex\hbox{$\sim$}}}
\newcommand{\LSM}{L$\Sigma$M}

\newcommand{\GMLfull}{Gell-Mann-L\'evy~}
\newcommand{\GML}{GML~}

\newcommand{\HVEV}{\langle H\rangle}

\newcommand{\half}{\frac{1}{2}}
\newcommand{\mpisq}{m_\pi^2}

\def\bbox{{\,\lower0.9pt\vbox{\hrule \hbox{\vrule height 0.2 cm
\hskip 0.2 cm \vrule  height 0.2 cm}\hrule}\,}}

\begin{document} 


\title{
Global $SU(2)_L\otimes$BRST symmetry and its LSS theorem:
Ward-Takahashi identities 
\\governing 
Green's functions, on-shell T-Matrix elements, and the effective potential, 
in the scalar-sector
of certain spontaneously broken non-Abelian gauge theories
}


\date{\today}

\author{\"{O}zen\c{c} G\"{u}ng\"{o}r$^1$, Bryan W. Lynn$^{1,2}$ and Glenn D. Starkman$^{1}$}
\affiliation{$^1$ ISO/CERCA/Department of Physics, Case Western Reserve University, 
Cleveland, OH 44106-7079}
\affiliation{$^2$ Department of Physics and Astronomy, University College London, London WC1E 6BT, UK}
\email{oxg34@case.edu, bryan.lynn@cern.ch,  gds6@case.edu} 

\begin{abstract}

{\bf This work is dedicated to the memory of Raymond Stora (1930-2015).} $SU(2)_L$ is the  simplest spontaneous symmetry breaking (SSB) non-Abelian gauge theory. Its simplest bosonic representation is a complex scalar doublet $\phi =\frac{1}{\sqrt{2}}\begin{bmatrix}H+i\pi_3\\-\pi_2 +i\pi_1\end{bmatrix}\equiv \frac{1}{\sqrt{2}}\tilde{H}e^{2i\tilde{t}\cdot\tilde{\vec{\pi}}/\expval{H}}\begin{bmatrix}1\\ 0\end{bmatrix}$ and a vector gauge boson $\vec{W}^\mu$. In Landau gauge, $\partial_\mu \vec{W}^\mu =0$, $\vec{\tilde{\pi}}$ are massless derivatively coupled Nambu-Goldstone bosons (NGB). A global shift symmetry $\vec{\tilde{\pi}} \rightarrow \vec{\tilde{\pi}} + \expval{H}\vec{\theta} + \vec{\tilde{\pi}}\times \vec{\theta} + \mathcal{O}(\theta^2)$ enforces $m^{2}_{\tilde{\pi}}=0$. We observe that on-shell T-matrix elements of physical states ${\vec W}^\mu$,$\phi$ (but not ghosts $\vec{\omega}$ and anti-ghosts $\bar{\vec{\eta}}$) are independent of anomaly-free global $SU(2)_{L}$ transformations, a remnant of local anomaly-free $SU(2)_L$ gauge symmetry, and that the associated global current is exactly conserved for amplitudes of physical states. We identify two towers of ``1-soft-pion" SSB global Ward-Takahashi Identities (WTI), which govern the $\phi$-sector, and represent a new global symmetry which we call $SU(2)_L\otimes$BRST, a symmetry not of the Lagrangian but of the physical states. The first tower gives relations among 1-$\phi$-I (1 scalar particle irreducible but one $W^{\mu}, B^{\mu}$ reducible) off-shell Green's functions, therefore severely constraining the all-loop-orders $\phi$-sector SSB effective Lagrangian. The second tower governs on-shell T-matrix elements, replaces the Adler self-consistency relations with those for gauge theories, further constrains the effective potential, and guarantees IR finiteness in the scalar sector despite zero NGB mass. These on-shell WTI include a Lee-Stora-Symanzik (LSS) theorem, which enforces the condition $m_{\pi}^{2}=0$ (far stronger than $m_{\tilde{\pi}}^{2}=0$) on the $\vec{\pi}$ and causes all relevant-operator contributions to the effective Lagrangian to vanish exactly. The global $SU(2)_L$ transformations and the nilpotent BRST transformations commute in $R_\xi$ gauges, $[s,\delta_{SU(2)_{L}}]=0$. With the on-shell T-matrix constraints, the physics therefore has more symmetry than does its BRST invariant Lagrangian, i.e. global $SU(2)_L\otimes$BRST symmetry. We also show that the statements made above hold for a $SU(2)_L\otimes U(1)_Y$ gauge theory i.e the electroweak sector of Standard Model bosons.

\end{abstract}

\maketitle

\section{Introduction}
\label{Introduction}

What are the symmetries driving spontaneously broken $SU(2)_L$ gauge theory physics \cite{LSS-3Proof}?
Although the symmetries of the $SU(2)_L$ Lagrangian are well known \cite{Ramond2004}, local gauge-invariance  is lost, broken by gauge-fixing terms, and replaced with global BRST invariance \cite{BecchiRouetStora,Tyutin1975,Tyutin1976}. 

In their seminal work, Elisabeth Kraus and Klaus Sibold \cite{KrausSiboldAHM} 
showed important new practicalities of the renormalizability and unitarity (to all-loop-orders) of the spontaneous symmetry breaking (SSB) $U(1)$ Abelian Higgs model (AHM). They did this by deriving rigid invariance from BRST invariance.
The SSB case is tricky because the globally BRST-invariant Lagrangian is not $U(1)$ symmetric. But they identified a set of  ``deformed" (i.e. with no remnant of the original $U(1)$ group symmetry) rigid/global AHM transformations which, after inclusion of well-defined $U(1)$-breaking by quantum loops (e.g. in scalar wavefunction renormalization beyond the classical AHM), are compatible with BRST symmetry. 

Kraus and Sibold then constructed deformed Ward-Takahashi identities (WTI) 
for quantum $U(1)$ Green's functions, showing them
(with appropriate normalization conditions) 
to obey all-loop-orders renormalizability and unitarity.
Because their renormalization relies only on  deformed  WTI, 
Kraus and Sibold's results are independent of regularization scheme, 
for any acceptable scheme (i.e. if one exists).
They did not construct WTI for on-shell T-Matrix elements.

Nevertheless, Slavnov-Taylor identities \cite{JCTaylor1976} prove that the
on-shell S-Matrix elements of AHM ``physical states" 
$A^\mu$,$\phi$
(i.e. spin $S=0$ scalars $h,\pi$
and $S=1$ gauge bosons $A_\mu$,
but not fermionic ghosts $\omega$ or anti-ghosts ${\bar \eta}$)
are independent, in the AHM, 
of the usual undeformed anomaly-free $U(1)$ local/gauge transformations, 
even though these break the AHM Lagrangian's BRST symmetry. 
We observed \cite{LSS-3Proof}  that they are therefore also 
independent of  anomaly-free undeformed $U(1)$ global/rigid transformations, 
resulting in ``new" global/rigid currents 
and appropriate un-deformed $U(1)$ Ward-Takahashi Identities. 
In this paper, we extend that thinking to non-Abelian gauge theories.

We here distinguish carefully between off-shell Green's function WTI, 
which constrain the (un-observable) effective  
Lagrangian and action, 
and on-shell T-Matrix WTI,
which further severely constrain observable physics.
We show here that, in SSB $SU(2)_L$,
a tower of Ward-Takahashi Identities (WTI) relates all relevant-operator
contributions to $SU(2)_L$ physical-scalar-sector observables to one another. 
An on-shell T-Matrix WTI, i.e. the equivalent of an Adler self-consistency relation but for the $SU(2)_L$ gauge theory,
then causes all such contributions  to vanish.
It does so through its insistence that the scalar mass-squared  vanish exactly 
\begin{equation}
\label{LSSTheorem}
\mpisq=0
\end{equation}
in spontaneously broken ($\HVEV\neq0$) theories. We term this the Lee-Stora-Symanzik (LSS)\footnote{
Note from OG, BWL and GDS: Raymond Stora  would never name anything after himself. But we judge that, given the stature of B.W. Lee, R. Stora and K. Symanzik (now all deceased) in the history of SSB physics, the community would refer to that result as the ``LSS theorem" anyway. 
}
theorem 
after the three physicists who recognized its central role 
in the renormalization of global Linear Sigma Models, 
and the one who was central to our understanding of its role 
in the renormalization
of gauge theories.\footnote{
	As first noted by Kibble \cite{Kibble1967}, 
	in Landau gauge a relation naively similar in appearance to (\ref{LSSTheorem}), 
	$m_{\tilde\pi}^2=0$,
	enforces the masslessness of a Nambu Goldstone Boson (NGB) $\tilde{\vec \pi}$,
	i.e. is a Goldstone Theorem \cite{Nambu1959,Goldstone1961,Goldstone1962}
	for the SSB $SU(2)_L$ gauge theory.  
	This is regardless of the fact that the NGBs are not a physical degree of freedom.
	However, as we describe in greater detail below (cf. equation (\ref{Kibble}),
	${\tilde{\vec \pi}}$ is the angular degree of freedom 
	in the Kibble representation of the complex scalar doublet, 
	while $\vec \pi$ is the pseudoscalar degree of freedom in the linear representation.
	In global Linear Sigma Models $(L\Sigma M)$, 
	the masslessness of the NGB and the LSS condition (\ref{LSSTheorem}) are  equivalent. 
	Indeed, B. Lee \cite{Lee1970}, K. Symanzik \cite{Symanzik1970a,Symanzik1970b},
	A. Vassiliev \cite{Vassiliev1970} and classic texts \cite{ItzyksonZuber} advocate 
	that the spontaneously broken (``Goldstone") mode of a global $L\Sigma M$ 
	is to be understood as the zero-explicit-breaking limit (i.e. $\mpisq \to 0$) 
	of the explicit-breaking Partially Conserved Axial-vector Current (PCAC) term,
	$L_{PCAC}=\HVEV\mpisq H$, 
	included in the Gell-Mann and L${\acute e}$vy $L\Sigma M$ \cite{GellMannLevy1960}.
	The existence and masslessness of the purely derivatively coupled NGB
	is a result of and requires 
	the vanishing of the explicit-symmetry-breaking pseudo-scalars' mass-squared.
	\newline\indent
}
In addition to constraining the parameters of the theory, 
LSS permits us to employ pion-pole-dominance to compute the WTI.

The crucial advance over \cite{LSS-2}, 
which considered the  global $SU(3)_C\times SU(2)_L\times U(1)_Y$ Linear Sigma Model (L$\Sigma$M),
is a proof that the WTI remain in place in a SSB gauge theory,
with the LSS theorem playing 
the same protective role as did the Goldstone Theorem in the global theory \cite{LSS-2}.

Our new rigid $SU(2)_L$ WTIs 
govern the scalar-sector of the $SU(2)_L$ gauge theory.
They are therefore independent of regularization-scheme (assuming one exists).
Although not a gauge-independent procedure, 
it may help the reader to imagine that loop integrals 
are cut off at a short-distance  finite Euclidean UV scale, $\Lambda$, 
never taking the $\Lambda^2 \to \infty$ limit. 
Although that cut-off can be imagined to be near the Planck scale $\Lambda\simeq M_{Pl}$, 
quantum gravitational loops are not included.

The WTIs derived in this work give relations among one-scalar-particle-irreducible (1-$\phi$-I) off-shell Green's functions, and (separately) among connected on-shell T-matrix elements. 
Each 1-$\phi$-I Green's function is a sum over an infinite number of one-particle-irreducible (1-P-I) graphs. 
1-$\phi$-I Green's functions are the appropriate ones to consider for calculation of the scalar-sector effective potential. 
Other authors (see \cite{KrausSiboldAHM,Grassi1999}) have derived WTIs for 1-P-I Green's funtions for SSB gauge theories, but our identities, formulated with respect to a different set of graphs, are fundamentally different from those.

The structure of this paper is as follows:

Section \ref{AbelianHiggsModelSymmetry} introduces $SU(2)_L\otimes$BRST symmetry for 
a general 't Hooft $R_\xi$ gauge, and explains why physical results obey that new symmetry.

Section \ref{AbelianHiggsModel} 
concerns the correct renormalization of spontaneously broken $SU(2)_L$ in Landau gauge 
We treat $SU(2)_L$ in isolation, 
as a stand-alone flat-space weak-scale quantum field theory,
not embedded or integrated into any higher-scale ``Beyond-$SU(2)_L$"  physics.  

Section \ref{StoraSymmetry} extends $\cal G\otimes$BRST to gauge theories with certain simple Lie algebraic structure group 
$\cal G$.

Section \ref{StandardModel} analyses an example a non-simple Lie algebraic structure group: $SU(2)_L\otimes U(1)_Y\otimes$BRST for the standard electroweak model of gauge bosons, complex scalar doublet, ghosts and anti-ghosts.

Section \ref{nuDSM} extends our results to a CP-conserving version of the Standard Model with only the 3rd generation of quarks and leptons.

Section \ref{Stora} discusses the exacting mathematical rigor that
would have fully satisfied Raymond Stora.

Section \ref{Conclusions} draws conclusions.

Appendix \ref{GlnuDSMWTIs} reduces the derivation of  $SU(2)_L$ gauge-theoretic WTIs in Landau gauge to that of the $SU(2)_{L-R}$ Schwinger $L\Sigma M$.

Appendix \ref{DerivationWTIAHM} gives a complete and pedagogical derivation
of the Landau gauge $SU(2)_L$ WTIs governing the $\phi$-sector of $SU(2)_L$. 
Our renormalized WTIs  include all contributions from virtual transverse gauge bosons; 
$\phi$-scalars; anti-ghosts and ghosts;
${\vec W}^\mu;h,{\vec \pi};{\bar {\vec \eta}},{\vec \omega}$ respectively.

Appendix \ref{LightParticleCurrents} derives the $SU(2)_L$ current ${\vec J}^\mu_L$ in terms of ${\vec W}^\mu;h,{\vec \pi};{\bar {\vec \eta}},{\vec \omega}$, together with its divergence and commutators with scalar $\phi$.

Appendix \ref{EwSMLightParticleCurrents} derives the $SU(2)_L$ sub-current ${\vec J}^\mu_{L;2\otimes 1}$ in the $SU(2)\times U(1)_Y$ gauge theory, together with its divergence and commutators with scalar $\phi$.

Appendix \ref{EwSMTotalMasterEqWTI} gives a complete and pedagogical derivation
of the Landau gauge $SU(2)_{L-R}$ WTIs governing the $\phi$-sector of $SU(2)_L\times U(1)_Y$. 
Our renormalized WTIs  include all contributions from virtual transverse gauge bosons; 
$\phi$-scalars; anti-ghosts and ghosts;
${\vec W}^\mu, B^\mu;h,{\vec \pi};{\bar {\vec \eta}},{\bar { \eta}}_B;{\vec \omega},{ \omega}_B$ respectively.

\section{$SU(2)\otimes$BRST symmetry in 't Hooft $R_{\xi}$ gauges}
\label{AbelianHiggsModelSymmetry}

The BRST-invariant \cite{BecchiRouetStora,Tyutin1975,Tyutin1976} Lagrangian 
of the $SU(2)_L$ gauge theory 
may be written, 
in a general 't Hooft $R_\xi$ gauge, 
in terms of  
a transverse vector ${\vec W}_\mu$, 
a complex scalar doublet $\phi$, 
ghosts $\vec \omega$, 
and anti-ghosts $\bar {\vec \eta}$:
\begin{eqnarray}
	\label{LagrangianAHMRxiGauge}
	L_{SU(2)_L}^{R_\xi}&=&L_{SU(2)_L}^{GaugeInvariant} \\
	&+&L_{SU(2)_L}^{GaugeFix; R_\xi}
	+L_{SU(2)_L}^{Ghost;R_\xi} \nonumber
\eea
where

\begin{eqnarray}
\label{Lphi}
L_{SU(2)_L}^{GaugeInvariant}&=&-\half Tr \left(W_{\mu \nu}W^{\mu \nu}\right)+\vert D_{\mu}\phi \vert ^2 - V \nonumber
\end{eqnarray}
with
\begin{eqnarray}
W_{\mu \nu}&=&\partial_{\mu}W_{\nu}-\partial_{\nu}W_{\mu} +ig_2\left[W_{\mu},W_{\nu}\right] \nonumber \\
D_{\mu}\phi &=& \left[\partial_{\mu} + ig_2W_{\mu}\right]\phi \nonumber \\
V &=& \mu^2_\phi (\phi^{\dagger}\phi) + \lambda^2_\phi (\phi^{\dagger}\phi)^2 \nonumber \\
W_{\mu}&=& {\vec t} \cdot {\vec W}_{\mu} 
\end{eqnarray}
with gauge coupling constant $g_2$, isospin and Pauli matrices ${\vec t}\equiv\half\vec \sigma$.

In this paper, we consider only the case of a single complex scalar doublet $\phi$.
\be
\label{LinearScalarRep}
\phi =  \frac{1}{\sqrt{2}} \Big[ \begin{array}{c}H+i\pi_3\\ -\pi_2 + i\pi_1\end{array}\Big]\equiv \frac{1}{\sqrt 2}{\tilde H}e^{2i{\vec t}\cdot {\tilde {\vec \pi}}/\HVEV}\Big[ \begin{array}{c}1\\ 0\end{array}\Big]
\ee
after SSB, and $H=h+\HVEV,{\tilde H} ={\tilde h} +\HVEV$.

In G. 't Hooft's $R_\xi$ gauges, gauge fixing and DeWitt-Fadeev-Popov ghost terms 
\cite{DeWitt1967,Fadeev1967}
are written in terms of a Nakanishi-Lautrup field $\vec b$
\cite{Nakanashi1966,Lautrup1967}, 
the SSB vector mass $M_W$ and the gauge-fixing function ${\vec F}_W$
\bea
\label{SU2GaugeFixing}
M_W&=&\frac{1}{2}g_2\HVEV \nonumber \\
{\vec F}_W &=& \partial_\mu {\vec W}^\mu +\xi M_W {\vec \pi}
\eea
\bea
\label{tHooftGaugeFixing}
&&L_{SU_2}^{GaugeFix;R_\xi}+ L_{SU_2}^{Ghost;R_\xi} \\
&& \quad= {\vec b} \cdot\Big(\partial_\mu {\vec W}^\mu +\xi M_W {\vec \pi} + \frac{1}{2}\xi {\vec b}\Big) \nonumber \\
&& \quad- {\bar {\vec \eta}} \cdot \Big[\partial^2 {\vec \omega}+g_2 \partial_\mu \Big( {\vec W}^\mu \times {\vec \omega} \Big) + \xi \frac{M_W^2}{\HVEV} \Big(  H {\vec \omega} +{\vec \pi} \times  {\vec \omega} \Big) \Big] \nonumber \\
&& \quad=s\Big[{\bar {\vec \eta}} \cdot \Big(\partial_\mu {\vec W}^\mu +\xi M_W {\vec \pi} + \half \xi {\vec b} \Big)\Big]. \nonumber
\end{eqnarray}
Under global BRST transformations \cite{BecchiRouetStora,Tyutin1975,Tyutin1976,Nakanashi1966,Lautrup1967,Weinberg1995} $s$
\begin{eqnarray}
\label{BRSTTransformations}
s{\vec W}_{\mu}&=&\partial_\mu {\vec \omega}+g_2{\vec W}_{\mu} \times {\vec \omega} \nonumber \\
sH&=&-\half g_2{\vec \pi} \cdot {\vec \omega}\nonumber \\
s{\vec \pi}&=&\half g_2\Big( H {\vec \omega} +{\vec \pi}\times {\vec \omega} \Big)  \nonumber \\
s{\vec \omega}&=&-\half g_2 {\vec \omega} \times {\vec \omega}\nonumber \\ 
s{\bar {\vec \eta}}&=&{\vec b}; \nonumber \\
s{\vec b}&=&0
\end{eqnarray}
so that the Lagrangian (\ref{LagrangianAHMRxiGauge}) is BRST invariant
\bea
\label{BRSTTransformationLagrangian}
s L_{SU(2)_L}^{R_\xi} &=& 0 
\eea

Now define the properties of the various fields under the usual anomaly-free un-deformed rigid/global $SU(2)_L$.  $\delta_{SU(2)_L}$ transforms fields by constant $\vec \Omega$
\begin{eqnarray}
\label{U(1)Transformations}
\delta_{SU(2)_L}{\vec W}_{\mu}&=&g_2{\vec W}_{\mu} \times {\vec \Omega} \nonumber \\
\delta_{SU(2)_L}H&=&-\half g_2{\vec \pi} \cdot {\vec \Omega}\nonumber \\
\delta_{SU(2)_L}{\vec \pi}&=&\half g_2\Big(H {\vec \Omega} +{\vec \pi}\times {\vec \Omega} \Big)  \nonumber \\
\delta_{SU(2)_L}{\vec \omega}&=&g_2 {\vec \omega} \times {\vec \Omega}\nonumber \\ \delta_{SU(2)_L}{\bar {\vec \eta}}&=&0 \nonumber \\
\delta_{SU(2)_L}{\vec b}&=&0
\end{eqnarray}

The transformation sets (\ref{BRSTTransformations}) and (\ref{U(1)Transformations}) commute
\bea
\label{SU2BRSTFieldCommutators}
\Big[ \delta_{SU(2)_L}, s \Big] {\vec W}^\mu&=& 0;\quad \Big[ \delta_{SU(2)_L}, s \Big] {\vec \omega}=0; \nonumber \\
\Big[ \delta_{SU(2)_L}, s \Big] H&=& 0;\quad \Big[ \delta_{SU(2)_L}, s \Big] {\bar {\vec \eta}}=0; \nonumber \\
\Big[ \delta_{SU(2)_L}, s \Big] {\vec \pi}&=& 0;\quad \Big[ \delta_{SU(2)_L}, s \Big] {\vec b}=0; 
\eea

and although $R_\xi$-gauge Lagrangian (\ref{LagrangianAHMRxiGauge}) is not invariant under $SU(2)_L$ transformations
\bea
\label{U(1)TransformationLagrangian}
&&\delta_{SU(2)_L} L_{SU(2)_L}^{R_\xi} \nonumber \\
&& \qquad =s \Big( \delta_{SU(2)_L}\Big[{\bar {\vec \eta}}\cdot \Big({\vec F}_W +\half \xi {\vec b}  \Big)\Big]\Big) \nonumber \\
&& \qquad =s \Big({\bar {\vec \eta}}\cdot\Big[ g_2 \partial_\mu{\vec W}^\mu \times \Omega
+ \xi \frac{M_W^2}{\HVEV}\big(H{\vec \Omega} +{\vec \pi}\times \Omega \big) \Big]\Big) \nonumber \\
&& \qquad \neq 0
\eea

with (\ref{BRSTTransformationLagrangian},\ref{SU2BRSTFieldCommutators}) and the nilpotent property $s^2 =0$,
\bea
\label{U(1)BRSTCommutatorAHM}
\Big[ \delta_{SU(2)_L}, s \Big] L_{SU(2)_L}^{R_\xi} &=& 0,
\eea 
and the two separate global symmetries can therefore co-exist in $SU(2)_L$ physics.

In an operational sense, we summarize (\ref{SU2BRSTFieldCommutators},\ref{U(1)BRSTCommutatorAHM}) with the short-hand
\bea
\label{SU2BRSTShortHand}
\Big[ \delta_{SU(2)_L}, s \Big] &=& 0\,.
\eea 

The reader will have noticed that, although the anti-ghosts ${\bar {\vec \eta}}$ transform in the usual way, as a triplet  $s{\bar {\vec \eta}}={\vec b}$ under BRST in (\ref{BRSTTransformations}), we have chosen them to be singlets under global $SU(2)_L$ transformations $\delta_{SU(2)_L}{\bar {\vec \eta}}=0$ in (\ref{U(1)Transformations}). This does not affect the proof of renormalizabililty and unitarity with Slavnov-Taylor identities \cite{Storaprivate}: the subject of a future paper, but outside the scope of this paper.
This freedom to choose ${\bar {\vec \eta}}$ singlets under $SU(2)_L$:
\begin{itemize}
\item Renders the ghost Lagrangian without definite $\delta_{SU(2)_L}$ properties.
\item Is the genesis of the commutation properties (\ref{SU2BRSTFieldCommutators}) and (\ref{U(1)BRSTCommutatorAHM}).
\item Allows us to build a classical current ${\vec J}_L^\mu$, conserved up to gauge-fixing terms
\end{itemize}

We will show in this paper that, due to (\ref{BRSTTransformations},\ref{U(1)Transformations},\ref{SU2BRSTShortHand}), $SU(2)_L$ physics simultaneously obeys both the usual BRST symmetry, and a global $SU(2)_L$ symmetry which controls Green's functions and on-shell T-Matrix elements.
We reason as follows:
\begin{itemize}
\item All aspects of the SSB $SU(2)_L$ physics obey BRST symmetry. 

\item We work in Landau gauge
\bea
\label{LandauLorenzGauges}
L_{SU(2)_L}^{Landau}&=&L_{SU(2)_L}^{GaugeInvariant} \\
&-&\lim_{\xi\to 0}\frac{1}{2\xi}\big(\partial_\mu {\vec W}^\mu +\xi M_W {\vec \pi}\big) ^2 \nonumber \\
&-& {\bar {\vec \eta}} \cdot \Big[ \partial^2  {\vec \omega} + g_{2} \partial_\mu \Big( {\vec W}^\mu\times{\vec \omega}\Big) \Big]\nonumber
\eea 

\item Physical states and time-ordered amplitudes of the exact renormalized scalar $\phi =  \frac{1}{\sqrt{2}} \Big[ \begin{array}{c}H+i\pi_3\\ -\pi_2 + i\pi_1\end{array}\Big]$ and vector ${\vec W}_\mu$ obey G. 't Hooft's gauge condition
\cite{tHooft1971} 
\begin{eqnarray}
\label{GaugeConditionsPrimePrime}
&&\big< 0\vert T\Big[ \Big( \partial_{\mu}{\vec W}^{\mu}(z)\Big) \nonumber \\
&&\quad \times h(x_1)...h(x_N)\pi_{t_1}(y_1)...\pi_{t_M}(y_M)\Big]\vert 0\big>_{\rm connected} \nonumber \\
&&\quad =0 \,.
\end{eqnarray}
in Landau gauge.
Here we have N external renormalized scalars $h=H-\HVEV$ (coordinates x), 
and M external ($CP=-1$) renormalized pseudo-scalars ${\vec \pi}$ (coordinates y, isospin $t$). 

\item We  prove in Appendix \ref{DerivationWTIAHM} for $SU(2)_L$ 
that, in Landau gauge, scalar-sector connected amputated Green's functions and on-shell T-Matrix elements
obey 
the $SU(2)_L$ symmetry.
That is true even though (\ref{U(1)TransformationLagrangian}) shows that the BRST-invariant $SU(2)_L$ Lagrangian  is not invariant under that $SU(2)_L$ symmetry.

\end{itemize}

\section{$SU(2)_L$ in Landau gauge}
\label{AbelianHiggsModel}

\subsection{$SU(2)_L$ in Landau gauge}
\label{DefineAHM}

We form the Lagrangian
\begin{eqnarray}
	\label{LagrangianAHM}
	L_{SU(2)_L}^{Landau}&=&L_{SU(2)_L}^{GaugeInvariant} \\
	&+&L_{SU(2)_L}^{GaugeFix;Landau}  
	+L_{SU(2)_L}^{Ghost;Landau} \nonumber
\eea
by taking the $\xi \to 0$ limit of (\ref{LagrangianAHMRxiGauge}).

This paper distinguishes carefully between 
the local BRST-invariant $SU(2)_L$ Lagrangian (\ref{LagrangianAHM}), and its 3 physical modes \cite{Lee1970,Symanzik1970a,Symanzik1970b,Vassiliev1970,ItzyksonZuber}: symmetric Wigner mode, classically scale-invariant  point, and physical Goldstone mode.

1) Symmetric Wigner mode $\HVEV=0, M_W^2=0,\mpisq =m_{BEH}^2 = \mu^2_\phi \neq 0$:

This is $SU(2)_L$ QED with massless photons and massive charged scalars. 
Thankfully, Nature is not in Wigner mode! 
Further analysis and renormalization of the Wigner mode 
lies outside the scope of this paper. 

2) Classically scale-invariant point $\HVEV=0, M_W^2=0,\mpisq =m_{BEH}^2 = 0$:

Analysis of the scale-invariant point is also outside the scope of this paper.

3) Spontaneously broken Goldstone mode
$\HVEV\neq0, M_W^2 \neq0,\mpisq =0, m_{BEH}^2 \neq 0$:

 The famous Higgs model, 
 with its Nambu-Goldstone boson (NGB) ``eaten" by the Brout-Englert-Higgs mechanism, WTI governed by the LSS theorem \cite{LSS-3Proof}, 
is  the SSB ``Goldstone mode" 
of the BRST-invariant local Higgs Lagrangian.

We work in Landau gauge and the linear representation for many reasons: 
\begin{itemize}

\item $\vec \pi$ and ${\vec W}^\mu$ do not mix.

\item The theory is manifestly renormalizable in the Dyson sense since massive vector propagators fall off as $k^{-2}$ as $k \rightarrow \infty$

\item Only in the SSB Goldstone mode of the BRST-invariant Lagrangian (\ref{LagrangianAHM}), 
and only after first renomalizing in the linear $\phi$ representation, 
does the renormalized Kibble $\phi$ unitary representation
\be
\label{Kibble}
\phi =  \frac{1}{\sqrt{2}} \Big[ \begin{array}{c}H+i\pi_3\\ -\pi_2 + i\pi_1\end{array}\Big]\equiv \frac{1}{\sqrt 2}{\tilde H}e^{2i{\vec t}\cdot {\tilde {\vec \pi}}/\HVEV}\Big[ \begin{array}{c}1\\ 0\end{array}\Big]
\ee
make sense. 
Here ${H}=\HVEV + { h}; {\tilde H} =\HVEV + {\tilde h}$.

\item We will prove to  all-loop-orders  the $SU(2)_L$  Lee-Stora-Symanzik theorem (\ref{TMatrixGoldstoneTheoremPrime}),
an $SU(2)_L$ gauge theory analogue of an old theorem for global $L\Sigma M$ \cite{Lee1970,LSS-2}, 
 which forces the $\vec \pi$ mass-squared $\mpisq=0$.

\item We use ``pion-pole dominance" (i.e. $\mpisq =0)$ arguments to derive $SU(2)_L$ SSB WTIs
 (\ref{AdlerSelfConsistencyPrime},\ref{AdlerSelfConsistency},\ref{InternalTMatrix}).

\item 
We prove  with $SU(2)_L$ Green's function  WTIs
that, in SSB Goldstone mode, $\vec {\tilde \pi}$ in (\ref{Kibble}) is a Nambu-Goldstone boson (NGB), and that the resultant SSB gauge theory 
has a ``shift symmetry" ${\tilde {\vec \pi}} \to {\tilde {\vec \pi}} +\HVEV {\vec \theta}+{\tilde {\vec \pi}} \times {\vec \theta}+ {\cal O}(\theta^2)$ for constant $\vec \theta$.
\end{itemize}

Analysis is done in terms of the exact renormalized interacting fields, 
which asymptotically become the in/out states, i.e. free fields for physical S-Matrix elements.

An important issue is the classification and disposal of relevant operators, 
in this case the ${\vec \pi}$, $h$ 
inverse propagators (together with  tadpoles).
 Define the exact renormalized pseudo-scalar propagator 
in terms of massless $\vec \pi$, the K$\ddot a$ll$\acute e$n-Lehmann \cite{Bjorken1965,Lee1970} spectral density $\rho^{\pi}_{SU(2)_L}$, and  wavefunction renormalization $Z_{SU(2)_L}^\phi$. In Landau gauge: 
\begin{eqnarray}
\label{pNGBPropagator}
&&\Delta^{\pi}_{SU(2)_L}\delta^{t_1t_2} = -i(2\pi)^2\langle 0\vert T\left[ \pi^{t_1}(y)\pi^{t_2}(0)\right]\vert 0\rangle\vert^{Fourier}_{Transform} \nonumber \\
&&\Delta^{\pi}_{SU(2)_L}(q^2)  = \frac{1}{q^2
+ i\epsilon} + \int dm^2 \frac{\rho^{\pi}_{SU(2)_L}(m^2)}{q^2-m^2 + i\epsilon}  \\
&&\Big[Z^{\phi}_{SU(2)_L}\Big]^{-1} = 1+ \int dm^2 \rho^{\pi}_{SU(2)_L}(m^2)\,. \nonumber
\end{eqnarray} 

Define also the BEH scalar propagator in terms of a BEH scalar pole and the (subtracted) spectral density $\rho_{BEH}$, and the same wavefunction renormalization. We assume $h$ decays weakly, and resembles a resonance:
\begin{eqnarray}
\label{BEHPropagator}
&&\Delta^{BEH}_{SU(2)_L}(q^2) = -i (2\pi)^2\langle 0\vert T\left[ h(x) h(0)\right]\vert 0\rangle\vert^{Fourier}_{Transform} \nonumber \\
&& \quad \quad =\frac{1}{q^2-m_{BEH;Pole}^2 + i\epsilon}+ \int dm^2 \frac{\rho^{BEH}_{SU(2)_L}(m^2)}{q^2-m^2 + i\epsilon} \nonumber \\
&&\Big[ Z^{\phi}_{SU(2)_L}\Big]^{-1} = 1+ \int dm^2 \rho^{BEH}_{SU(2)_L}(m^2)  \nonumber \\
&& \int dm^2 \rho^{\pi}_{SU(2)_L}(m^2) = \int dm^2 \rho^{BEH}_{SU(2)_L}(m^2) 
\end{eqnarray}

The spectral density parts of the propagators are
\begin{eqnarray}
\label{SpectralDensityPropagators}
 \Delta^{\pi ;Spectral}_{SU(2)_L}(q^2) &\equiv& \int dm^2 \frac{\rho^{\pi}_{SU(2)_L}(m^2)}{q^2-m^2 + i\epsilon}  \nonumber \\ 
\Delta^{BEH; Spectral}_{SU(2)_L}(q^2) &\equiv& \int dm^2 \frac{\rho^{BEH}_{SU(2)_L}(m^2)}{q^2-m^2 + i\epsilon} 
\end{eqnarray}
With these choices for renormalization conditions, $\expval{H}_{bare} = \sqrt{Z_{\phi}}\expval{H}_{renormalized}$, in agreement with \cite{BWLeeJustin1972-1,BWLeeJustin1972-2,BWLeeJustin1972-3,Sperling2013}. Dimensional analysis of the wavefunction renormalizations (\ref{pNGBPropagator},\ref{BEHPropagator}), 
shows that the 
finite Euclidean cut-off contributes only irrelevant terms $\sim \frac{1}{\Lambda^2}$.

\subsection{Rigid/global $SU(2)_L$ WTI and conserved current (not charge)
for physical states in Landau gauge}
\label{GlobalCurrents}

In their seminal work,  E. Kraus and K. Sibold 
\cite{KrausSiboldAHM} 
identified, in the Abelian Higgs model (AHM), 
anomaly-free deformed rigid/global transformations.
They are called ``deformed" because they have no remnant 
of the original anomaly-free $U(1)_Y$ symmetry due to $U(1)_Y$-breaking quantum loops in wavefunction renormalization. 
The SSB  case is tricky because gauge-fixing terms explicitly break both local and global $U(1)_Y$ symmetry in the BRST-invariant Lagrangian.
Still, Kraus and Sibold constructed deformed global/rigid Ward-Takahashi Identities (WTI) for 1-P-I Green's functions
allowing them to demonstrate
(with appropriate normalization conditions) 
proof of all-loop-orders renormalizability and unitarity for the SSB Abelian Higgs model. 
Because their renormalization relies only on  deformed  WTI, 
Kraus and Sibold's results are independent of regularization scheme, 
for any acceptable scheme (i.e. if one exists).\footnote{
	E. Kraus and K. Sibold  also constructed, 
	in terms of deformed WTI, 
	all-loop-orders renormalized QED, QCD, 
	and the electroweak Standard Model 
	\cite{KrausSiboldSM1996,KrausSM1997} 
	to be independent of regularization  scheme. 
	From this grew the powerful technology of ``Algebraic Renormalization", 
	used by them, W. Hollik and others \cite{Hollik2002b}, 
	to renormalize SUSY QED, SUSY QCD, and the MSSM.
}

Nevertheless, Slavnov-Taylor identities \cite{JCTaylor1976} in the  Abelian Higgs Model
prove that the
on-shell S-Matrix elements of ``physical particles'' 
(i.e. spin $S=0$ scalars $h,\pi$, and $S=1$ transverse gauge bosons $A_\mu$, 
but not fermionic ghosts $\omega$ or anti-ghosts ${\bar \eta}$),
are independent of the usual undeformed anomaly-free 
$U(1)_Y$ local/gauge transformations, 
even though these break the Lagrangian's BRST symmetry. Ref. \cite{LSS-3Proof} exploited this fact to derive two towers of  WTIs and an LSS theorem, which represent a new global/rigid $U(1)_Y\otimes$BRST symmetry in the Abelian Higgs Model, and severely constrain its effective potential.

Slavnov-Taylor identities prove the same for SSB $SU(2)_L$ on-shell S-Matrix elements, and for those of any SSB Lie algebraic structure group, $\cal G$ \cite{JCTaylor1976}. We observe here that SSB $SU(2)_L$ T-Matrix elements are therefore 
also independent of anomaly-free undeformed
$SU(2)_L$ global/rigid transformations, 
resulting in a ``new" global/rigid current, 
two towers of appropriate un-deformed $SU(2)_L$ Ward-Takahashi identities, and a new $SU(2)_L\otimes$BRST symmetry.

We are interested in rigid-symmetric relations among 
1-$(h,{\vec \pi})$-Irreducible (1-$\phi$-I) (this set of Green's functions include an infinite number of 1-P-I Green's functions) connected amputated Green's functions $\Gamma_{N,M}$, 
and among 1-$(h,{\vec \pi})$-Reducible (1-$\phi$-R) 
connected amputated transition-matrix (T-Matrix) elements $T_{N,M}$,   
with external $\phi$ scalars. 
Because these are 1-${\vec W}_\mu$-R in $SU(2)_L$ (i.e. reducible by cutting a ${\vec W}_\mu$), it is convenient to use the powerful old tools (e.g. canonical quantization) 
from Vintage Quantum Field Theory (Vintage-QFT), a name coined by Ergin Sezgin. The distinction between 1-$\phi$-I Green's functions and 1-P-I Green's functions is important. The 1-$\phi$-I (but potentially reducible with respect to other fields) Green's functions is the correct set of Green's functions to use for the scalar-sector effective potential, and the WTIs obtained for those are fundamentally different from the Slavnov-Taylor identities for 1-P-I Green's functions obtained elsewhere in the literature ({\it e.g.}, \cite{KrausSiboldAHM,Grassi1999}).

We focus on the rigid/global $SU(2)_L$ current constructed with (\ref{U(1)Transformations}) in Appendix \ref{LightParticleCurrents}. In Landau gauge, these are
\begin{eqnarray}
\label{AHMCurrentPrime}
{\vec J}^{\mu}_{L} &=& {\vec J}^{\mu}_{L;Schwinger} + {\vec {\cal J}}^{\mu}_{L} \nonumber \\
{\vec J}^{\mu}_{L;Schwinger} &=& {\vec J}^{\mu}_{L+R;Schwinger} + {\vec { J}}^{\mu}_{L-R;Schwinger}  \nonumber \\
{\vec J}^{\mu}_{L+R;Schwinger} &=& \half {\vec \pi}\times \partial^\mu {\vec \pi} \\
{\vec J}^{\mu}_{L-R;Schwinger} &=& 
\half \Big( {\vec \pi}\partial^\mu H -H \partial^\mu {\vec \pi} \Big) \nonumber \\
{\vec {\cal J}}^{\mu}_{L} &=&-\frac{1}{4} g_2 {\vec W}^{\mu}\left[ H^2+{\vec \pi}^2\right] 
+  {\vec W}^{\mu \nu} \times {\vec W}_{\nu} \nonumber \\
&-& \lim_{\xi \to 0} \frac{1}{\xi}\left[{\vec W}^{\mu} \times \partial_\beta {\vec W}^\beta \right] 
- \partial^\mu {\bar {\vec \eta}}\times {\vec \omega} \nonumber 
\end{eqnarray} 

Rigid/global transformations of the fields arise, as usual, 
from the equal-time commutators (\ref{EqTimeCommAHM}):
\begin{eqnarray}
\label{TransformationsAHMFields}
 \delta_{SU(2)_L} H(t,{\vec y})&=&-ig_2 \Omega^{t_1} \int d^3 z\left[ {J}^{t_1;0}_{L}(t,{\vec z}),H(t,{\vec y})\right]  \nonumber \\
&=& - \half g_2 \Omega^{t_1}\int d^3 z \pi^{t_1}(t,{\vec z})\delta^3({\vec z}-{\vec y}) \nonumber \\
&=& - \half g_2 \pi^{t_1} (t,{\vec y}) \Omega^{t_1}\\
 \delta_{SU(2)_L} \pi^{t}(t,{\vec y})&=&-i g_2 \Omega^{t_2}\int d^3 z\left[ {J}^{t_2;0}_{L}(t,{\vec z}),\pi^{t}(t,{\vec y})\right] \nonumber \\
&=& \half g_2  \int d^3 z \delta^3({\vec z}-{\vec y}) \nonumber \\
&\times&\Big[ H(t,{\vec z})\Omega^{t} +\epsilon^{tt_1t_2} \pi^{t_1}(t,{\vec z})\Omega^{t_2}\Big]\nonumber \\
&=&  \half g_2\Big[ H(t,{\vec y})\Omega^{t} +\epsilon^{tt_1t_2} \pi^{t_1}(t,{\vec y})\Omega^{t_2}\Big] \nonumber \\
\end{eqnarray} 
so ${\vec J}^\mu_{L}(t,{\vec z})$ serves as a ``proper" local current for commutator purposes. 

In contrast, we show below that, in  Landau gauge, $SU(2)_L$ 
has no associated proper global charge 
because $\frac{d}{dt}{\vec Q}(t)\neq0$. 
(See Eqn. (\ref{ChargeAHMTimeOrderedProductsNotConserved}) below.)

The classical equations of motion reveal a crucial fact: due to gauge-fixing terms in the BRST-invariant Lagrangian (\ref{LagrangianAHM}), the 
classical current 
(\ref{AHMCurrentPrime}) is 
not conserved. In Landau gauge  
\begin{eqnarray}
\label{DivergenceAHMCurrentPrime}
\partial_{\mu} {\vec J}^{\mu}_{L}&=&\half M_W\left[{\vec \pi} \times \partial_\mu {\vec W}^\mu +H \partial_\mu {\vec W}^\mu\right] \,.
\end{eqnarray}

The global $SU(2)_L$ current (\ref{AHMCurrentPrime}) 
is, however, conserved by the physical states, 
and therefore still qualifies as a ``real" current. 
Strict quantum constraints are imposed
that force the relativistically-covariant theory of gauge bosons 
to propagate only its true number of quantum spin $S=1$ degrees of freedom. 
These constraints are, in the modern literature, 
implemented by use of spin $S=0$ fermionic DeWitt-Fadeev-Popov ghosts
$({\bar {\vec \eta}},{\vec \omega})$.  
The physical states and their time-ordered products, 
but not the BRST-invariant Lagrangian  (\ref{LagrangianAHM}),  
then obey G. 't Hooft's \cite{tHooft1971} Landau gauge  gauge-fixing condition (\ref{GaugeConditionsPrimePrime}).

Eq. (\ref{GaugeConditionsPrimePrime}) restores conservation 
of the rigid/global $SU(2)_L$ current 
for $\phi$-sector connected time-ordered products
\begin{eqnarray}
\label{PhysicalAHMCurrentConservation}
&&\Big< 0\vert T\Big[ \Big( \partial_{\mu}{\vec J}^{\mu}_{L}(z) \Big) \\
&&\quad \quad \times h(x_1)...h(x_N) \pi^{t_1}(y_1)...\pi^{t_M}(y_M)\Big]\vert 0\Big>_{\rm connected} =0 \,.	\nonumber
\end{eqnarray}
It is in this ``physical" connected-time-ordered product sense 
that the rigid global $SU(2)_L$ ``physical current" is conserved: 
the physical states, but not the BRST-invariant Lagrangian  (\ref{LagrangianAHM}), 
obey the physical-current conservation equation 
(\ref{PhysicalAHMCurrentConservation}). 
It is this physical conserved current that generates our $SU(2)_L\otimes$BRST WTI. 

Appendices \ref{GlnuDSMWTIs}, \ref{DerivationWTIAHM} and \ref{LightParticleCurrents} derive two towers of quantum $SU(2)_L$ WTIs that 
exhaust the information content of (\ref{PhysicalAHMCurrentConservation}),
severely constrain the dynamics (i.e. the connected time-ordered products) 
of the $\phi$-sector physical states of SSB $SU(2)_L$, and realize the new $SU(2)_L \otimes$BRST symmetry of Section \ref{AbelianHiggsModelSymmetry}.

We might have hoped to also build a conserved charge
\begin{eqnarray}
\label{ChargeAHM}
 {\vec Q}_{L}(t)&\equiv&\int d^3 z {\vec J}^0_{L}(t,{\vec z}) 
\end{eqnarray} 
in the usual way, but (\ref{DivergenceAHMCurrentPrime}) reveals that, because of gauge-fixing terms, classically
\begin{eqnarray}
\label{ChargeAHMGaugeFixing}
 \frac{d}{dt}{\vec Q}_{L}(t)\neq\int d^3 z  {\partial}^i  {\vec J}^i_{L}(t,{\vec z}) \quad \quad
\end{eqnarray} 
and so ${\vec Q}_{L}(t)$ is not a classical conserved current.

Hope springing eternal, we might have hoped for a more restricted form of charge
conservation, namely the vanishing of physical connected time-ordered products
of the time-derivative of ${\vec Q}_{L}(t)$.  
Consider 
\begin{eqnarray}
\label{ChargeAHMTimeOrderedProducts}
&&\Big< 0\vert T\Big[ \Big(  \frac{d}{dt}{\vec Q}_{L}(t) \Big) \\
&&\quad \quad \times h(x_1)...h(x_N) \pi^{t_1}(y_1)...\pi^{t_M}(y_M)\Big]\vert 0\Big>_{\rm Connected} \nonumber \\
&&\quad =\int d^3 z \Big< 0\vert T\Big[ \Big(  {\partial}^i  {\vec J}^i_{L}(t,{\vec z})\Big) \nonumber \\
&&\quad \quad \times h(x_1)...h(x_N) \pi^{t_1}(y_1)...\pi^{t_M}(y_M)\Big]\vert 0\Big>_{\rm Connected} \nonumber \\
&& \quad =\int_{\mathrm 2-surface\to\infty}  d^2z \quad {\widehat {z}}^{\mathrm 2-surface;i} \Big< 0\vert T\Big[ \Big(  {\vec J}^i_{L}(t,{\vec z})\Big) \nonumber \\
&&\quad \quad \times h(x_1)...h(x_N) \pi^{t_1}(y_1)...\pi^{t_M}(y_M)\Big]\vert 0\Big>_{\rm Connected} \nonumber 
\end{eqnarray}
where we have used Stokes theorem, and $ {\widehat {z}_\mu}^{\mathrm 2-surface}$ is a unit vector normal to the $2$-surface. The time-ordered product constrains the $2$-surface to lie on or inside the light-cone.

At a given point on the surface of a large enough 3-volume $\int d^3z$ (i.e. the volume of all space), which lies on-or-inside the light cone, all fields on the $z^{\mathrm 2-surface}$: are asymptotic in-states and out-states; are properly quantized as free fields; with each field species orthogonal to the others.

Consider first the ${\vec J}^i_{L-R}$ contribution (from (\ref{AHMCurrentPrime})) 
to  the surface integral (\ref{ChargeAHMTimeOrderedProducts})
\begin{eqnarray}
\label{SurfaceIntegralPrime}
&& \int_{\mathrm 2-surface\to\infty} \!\!\!\!\!\!\!\!\!\!\!\!\!\!\!\!\!\!\!\!\!\!\!\!
	d^2z ~ {\widehat {z}}^{\mathrm 2-surface;i}  
	\Big< 0\vert T\Big[ 
{\vec J}^i_{L-R;Schwinger}(z) \nonumber \\
&&\quad \quad \times h(x_1)...h(x_N) \pi^{t_1}(y_1)...\pi^{t_M}(y_M)\Big]
	\vert 0\Big>_{\rm Connected} \\
&& =\int_{\mathrm 2-surface\to\infty}  \!\!\!\!\!\!\!\!\!\!\!\!\!\!\!\!\!\!\!\!\!\!\!\!\!\!
	d^2z ~ {\widehat {z}}^{\mathrm 2-surface;i} \Big< 0\vert T\Big[ 
	  \half\Big({\vec \pi} {\partial}^i h-h{ \partial}^i {\vec \pi} 
	- \HVEV \partial^i{\vec \pi}\Big)(z)  \nonumber \\
	 \nonumber \\
&&\quad \quad \times h(x_1)...h(x_N) \pi^{t_1}(y_1)...\pi^{t_M}(y_M)\Big]\vert 0\Big>_{\rm Connected} \nonumber 
\end{eqnarray}
The $({\vec \pi} {\partial}^i h-h{ \partial}^i {\vec \pi})$ contribution 
to the surface integral (\ref{SurfaceIntegralPrime}) vanishes 
because $h$ is massive in spontaneously broken $SU(2)_L$, $m_{BEH}^2\neq0$.
Propagators connecting $h$ from points on $z^{\mathrm 2-surface}\to \infty$ 
to the localized interaction points $(x_1...x_N;y_1...y_M)$ 
must stay inside the light-cone, but die off exponentially
and are incapable of carrying information to the arbitrarily distant 2-surface.

It is crucially important that this argument fails for the remaining term, 
$-\half\HVEV \partial^i{\vec \pi}(z)$,
in (\ref{SurfaceIntegralPrime}).
$\vec \pi$ is massless in Landau gauge in SSB $SU(2)_L$, $\mpisq=0$.
The $\vec \pi$ are therefore capable of carrying  
long-ranged pseudo-scalar forces out to the  very ends of the light-cone $(z^{LightCone}\to \infty)$ (though not inside it).
Since it is independent of $\HVEV$,
the ${\vec J}^i_{L+R}$ (from (\ref{AHMCurrentPrime})) 
contribution to the surface integral (\ref{ChargeAHMTimeOrderedProducts})
cannot cancel the ${\vec J}^i_{L-R}$ contribution.
This shows that, at least in Landau gauge,
the spontaneously broken $SU(2)_L$  charge (\ref{ChargeAHM}) 
is not conserved, even in the sense of connected time-ordered products, 
\begin{eqnarray}
\label{ChargeAHMTimeOrderedProductsNotConserved}
&&\Big< 0\vert T\Big[ \Big(  \frac{d}{dt}{\vec Q}_{L}(t) \Big)  \\
&&\quad \quad \times h(x_1)...h(x_N) \pi^{t_1}(y_1)...\pi^{t_M}(y_M)\Big]\vert 0\Big>_{\rm Connected}  \nonumber \\
&&\quad \quad \neq 0\,, \nonumber
\end{eqnarray}
dashing all further hope. 

The classic 
proof of the Goldstone theorem \cite{Goldstone1961,Goldstone1962,Kibble1967} 
requires a conserved charge $\frac{d}{dt}{Q}=0$, 
so that proof fails for spontaneously broken gauge theories.
This is a very famous result 
\cite{Higgs1964,Englert1964,Guralnik1964,Kibble1967}, 
and allows spontaneously broken $SU(2)_L$ to generate 
a mass-gap $M_W^2$ for the vector ${\vec W}^\mu$ 
and avoid massless particles in its observable physical spectrum. 
This is true, even in $R_\xi(\xi=0)$ 
Landau gauge, where there is a Goldstone theorem \cite{Guralnik1964,Kibble1967},
so $\tilde {\vec \pi}$ are derivatively coupled (hence massless) NGB, 
and where there is an LSS theorem \cite{LSS-3Proof}, 
so $\vec \pi$ is massless. 

Massless $\vec \pi$ (not $\tilde{\vec \pi}$) is the basis  
of our pion-pole-dominance-based $SU(2)_L$ WTIs, derived in Appendices \ref{GlnuDSMWTIs},  \ref{DerivationWTIAHM} and \ref{LightParticleCurrents}, which give: 
relations among 1-$\phi$-I connected amputated 
$\phi$-sector Greens functions $\Gamma_{N,M}$ (\ref{GreensWTIPrime}, \ref{GreensFWTI}); 
1-soft-pion theorems  (\ref{AdlerSelfConsistencyPrime}, 
	\ref{AdlerSelfConsistency}, \ref{InternalTMatrix}); 
infra-red finiteness for $\mpisq =0$ (\ref{AdlerSelfConsistencyPrime}, \ref{AdlerSelfConsistency});
an LSS theorem (\ref{TMatrixGoldstoneTheoremPrime}); 
vanishing 1-$\phi$-R connected amputated
on-shell $\phi$-sector T-Matrix elements $T_{N,M}$ 
(\ref{AdlerSelfConsistencyPrime},  \ref{InternalTMatrix}); which all 
realize the full $SU(2)_L \otimes$BRST symmetry of Section \ref{AbelianHiggsModelSymmetry}.

\subsection{Construction of the scalar-sector effective Lagrangian 
from those $SU(2)_L$ WTIs that  govern connected amputated 1-$\phi$-I Greens functions}
\label{GreensFunctionsAHM}

In Appendix \ref{DerivationWTIAHM} we derive $SU(2)_L$ ``pion-pole-dominance"  1-$\phi$-R
connected amputated T-Matrix  WTI \eqref{InternalTMatrix} for $CP$-conserving SSB $SU(2)_L$. 
Their solution is a tower of recursive $SU(2)_L$ WTI (\ref{GreensFWTI}) 
that govern 1-$\phi$-I  $\phi$-sector connected amputated Greens functions $\Gamma_{N,M}$.
For $ \pi$ with $CP=-1$, the result
\begin{eqnarray}
	\label{GreensWTIPrime}
	&&\HVEV\Gamma_{N,M+1}^{t_1...t_Mt}(p_1 ...p_N;q_1...q_M0) \nonumber  \nonumber \\
	&&\quad \quad =\sum ^M_{m=1} \delta^{tt_m}\Gamma_{N+1,M-1}^{t_1...{\widehat {t_m}}...t_M} 
(p_1...p_Nq_m;q_1...{\widehat {q_m}}...q_M) \nonumber \\
&&\quad \quad -\sum ^N_{n=1}\Gamma_{N-1,M+1}^{t_1...t_Mt}(p_1 ...{\widehat {p_n}}...p_N;q_1...q_Mp_n)
\end{eqnarray} 
is valid for $N,M \ge0$. 
On the left-hand-side of (\ref{GreensWTIPrime}) there are 
N renormalized $h$ external legs (coordinates x, momenta p), 
M renormalized ($CP=-1$) ${\vec \pi}$ external legs (coordinates y, momenta q, isospin t), 
and 1 renormalized soft external ${\vec \pi}(k_\mu=0)$  (coordinates z, momenta k). On the right-hand side,
``hatted" fields with momenta $({\widehat {p_n}},{\widehat {q_m}})$ are omitted.

The rigid $SU(2)_L$ WTI 1-soft-pion theorems (\ref{GreensWTIPrime})
relate a 1-$\phi$-I Green's function with $(N+M+1)$ external fields 
(including a zero-momentum ${\vec \pi}$) 
to two 1-$\phi$-I  Green's functions with $(N+M)$ external fields.%
\footnote{
	The rigid  $SU(2)_L$ WTI (\ref{GreensWTIPrime})  for the $SU(2)_L$ gauge theory 
	are a generalization of the classic work of B.W. Lee \cite{Lee1970},  
	who constructed two all-loop-orders renormalized towers of WTIs 
	for the global $SU(2)_L\times SU(2)_R$ Gell-Mann L${\acute e}$vy (GML) model \cite{GellMannLevy1960} 
	with Partially Conserved Axial-vector Currents (PCAC).
	We replace GML's strongly-interacting Linear Sigma Model (\LSM) 
	with a weakly-interacting BEH \LSM, with no explicit PCAC breaking. 
	Replace $\sigma \to H$, ${\vec\pi}\to{\vec \pi}$,$m_{\sigma}\to m_{BEH}$ and $f_{\pi} \to \HVEV$, 
	and add local gauge group $SU(2)_L$. 
	This generates a set of global $SU(2)_L$ WTI 
	governing relations among weak-interaction 1-$\phi$-R T-Matrix elements $T_{N,M}$. 
	A solution-set of those $SU(2)_L$ WTI 
	then govern relations among $SU(2)_L$ 1-$\phi$-I Green's functions $\Gamma_{N,M}$. 
	\hfil\break
	\indent As observed by Lee for \GML with PCAC, 
	one of those on-shell T-Matrix WTI is equivalent to the Goldstone theorem. 
	This equivalence relies on the ability to incorporate a PCAC term into the global theory,
	and then retrieve the spontaneously broken theory in the 
	appropriate zero-explicit-breaking limit, namely $\mpisq\to0$.
	In the gauge theory, although  explicit-breaking terms are allowed by power-counting,
	they violate the BRST symmetry and spoil unitarity \cite{Storaprivate}.
	Yet, the T-matrix WTI persists and forces $\mpisq=0$ in Landau gauge, 
	which is now  the new LSS theorem.
	The Goldstone theorem also persists in Landau gauge, and forces $m_{\tilde\pi}^2=0$.
	\hfil\break
	\indent Appendix \ref{DerivationWTIAHM} includes, in Table 1,  translation between the WTI proofs in this paper (a gauge theory) and in  B.W. Lee (a global theory). 
}
The Green's functions $\Gamma_{N,M}^{t_1...t_M}(p_1...p_N;q_1...q_M)$ 
are not themselves gauge-independent. 
Furthermore, although 1-$\phi$-I, they are 1-${\vec W}^\mu$-Reducible (1-${\vec W}^\mu$-R) by cutting a transverse ${\vec W}_\mu$ gauge boson line.

We can now form the $\phi$-sector effective momentum space Lagrangian in Landau gauge:
\begin{eqnarray}
\label{FormSchwingerPotential}
&& L^{Eff;}_{\phi;SU(2)_L} =  \Gamma_{1,0}(0;)h +\frac{1}{2!} \Gamma_{2,0}(p,-p;)h^2 \nonumber \\
&&  \quad \quad + \frac{1}{2!} \Gamma_{0,2}^{t_1t_2}(;q,-q)\pi_{t_1}\pi_{t_2} +\frac{1}{3!} \Gamma_{3,0}(000;)h^3  \nonumber \\ 
&& \quad \quad + \frac{1}{2!} \Gamma_{1,2}^{t_1t_2}(0;00) h \pi_{t_1}\pi_{t_2} +\frac{1}{4!} \Gamma_{4,0}(0000;)h^4  \nonumber \\
&&  \quad \quad + \frac{1}{2!2!} \Gamma_{2,2}^{t_1t_2}(00;00) h^2 \pi_{t_1}\pi_{t_2}  \\
&& \quad \quad +  \frac{1}{4!}\Gamma_{0,4}^{t_1t_2t_3t_4}(;0000)\pi_{t_1}\pi_{t_2} \pi_{t_3}\pi_{t_4} + {\cal O}_{SU(2)_L}^{Ignore}  \nonumber 
\end{eqnarray} 
All perturbative quantum loop corrections, to all-loop-orders 
and including all UVQD, log-divergent and finite  contributions, 
are included in this $\phi$-sector effective Lagrangian:
renormalized
1-$\phi$-I Green's functions $\Gamma^{t_1...t_M}_{N,M}(p_1...p_N;q_1...q_M)$; 
wavefunction renormalizations;  
renormalized $\phi$-scalar propagators (\ref{pNGBPropagator},\ref{BEHPropagator}); 
the Brout-Englert-Higgs (BEH) VEV $\HVEV$ (\ref{HVEV}); 
all gauge boson and ghost propagators. 
This includes the full all-loop-orders renormalization of the $SU(2)_L$ $\phi$-sector,
originating in quantum loops containing transverse virtual gauge bosons, 
$\phi$-scalars, anti-ghosts and ghosts (i.e.
${\vec W}^\mu;h,{\vec \pi};{\bar {\vec \eta}},{\vec \omega}$ respectively). 
Because they arise entirely from global $SU(2)_L$ WTI, 
our results are independent of regularization-scheme \cite{KrausSiboldAHM}.

We wish to focus in this paper on finite relevant operators, 
as well as quadratic and logarithmically divergent operators,
arising from $SU(2)_L$ loops.
Therefore, in  (\ref{FormSchwingerPotential}), 
we have separated off three classes of
operators that  are finite and beyond the scope of this paper:
\begin{itemize}
\item Finite  ${\cal O}_{SU(2)_L}^{1/\Lambda^2;Irrelevant}$ vanish as $m_{Weak}^2/ \Lambda^2 \to 0$;
\item ${\cal O}_{SU(2)_L}^{Dim>4;Light}$ are finite-dimension $Dim>4$ operators, 
where only the light degrees of freedom 
(${\vec W}^\mu;h,{\vec \pi};{\bar {\vec \eta}},{\vec \omega}$) 
contribute to   all-loop-orders renormalization;  
\item ${\cal O}_{SU(2)_L}^{Dim\leq4;NonAnalytic}$ are finite-dimension $Dim\leq4$ operators 
that are non-analytic in momenta or in a renormalization scale $\mu^2$  
(e.g. finite renormalization-group logarithms). 
\end{itemize}
All such operators will be ignored.
\begin{eqnarray}
\label{ignoredoperators}
{\cal O}_{SU(2)_L}^{Ignore}&=&{\cal O}_{SU(2)_L}^{1/\Lambda^2;Irrelevant} + {\cal O}_{SU(2)_L}^{Dim>4;Light} \nonumber \\
&+&{\cal O}_{SU(2)_L}^{Dim\leq4;NonAnalytic}
\end{eqnarray}

For clarity, we separate the isospin indices
\begin{eqnarray}
\label{IsospinIndices}
	\Gamma_{0,2}^{t_1t_2}(;q,-q) 
	&\equiv& \delta^{t_1t_2} \Gamma_{0,2}(;q,-q)
	\,, \nonumber \\
	\Gamma_{1,2}^{t_1t_2}(-q;q0) &\equiv& \delta^{t_1t_2}\Gamma_{1,2}(-q;q,0) \,,\nonumber  \\
	\Gamma^{t_1t_2}_{2,2}(00;00)&\equiv&\delta^{t_1t_2}\Gamma_{2,2}(0,0;0,0)\,, \\
	 \Gamma_{0,4}^{t_1t_2t_3t_4}(;0000)  &\equiv& \Gamma_{0,4}(;0,0,0,0) \nonumber \\
	&\times& \left[ \delta^{t_1t_2}\delta^{t_3t_4} + \delta^{t_1t_3}\delta^{t_2t_4} + \delta^{t_1t_4}\delta^{t_2t_3} \right]\,. \nonumber 
\end{eqnarray}
and write the  1-$\phi$-I ${\vec \pi}$  and $h$ 
(isospin-index-suppressed) inverse propagators as:
\begin{eqnarray}
\label{InversePropagators}
\Gamma_{0,2}(;q,-q) &\equiv& \left[ \Delta_{\pi}(q^2) \right]^{-1} \nonumber \\
\Gamma_{2,0}(p,-p;) &\equiv& \left[ \Delta_{BEH}(p^2) \right]^{-1} \,.
\end{eqnarray}
The quadratic and quartic coupling constants 
are defined in terms of 2-point and  4-point 1-$\phi$-I connected amputated GF
\begin{eqnarray}
\label{SchwingerWignerData}
 \Gamma_{0,4}(;0000)  &\equiv& -2 \lambda_{\phi}^2 \nonumber \\
 \Gamma_{0,2}(;00)  &\equiv& -2 \mpisq
\end{eqnarray} 

The Ward-Takahashi IDs (\ref{GreensWTIPrime}) for Greens functions 
severely constrain the effective Lagrangian (\ref{FormSchwingerPotential}).
By using the WTI for $N+M \leq 4$, 
the  all-loop-orders renormalized $\phi$-sector momentum-space effective Lagrangian 
(\ref{FormSchwingerPotential}) - 
constrained only by those $SU(2)_L$ WTI governing Greens functions (\ref{GreensWTIPrime}) - 
may be written
\begin{eqnarray}
\label{LEffectiveSM}
&&L^{Eff;all-modes}_{SU(2)_L;Landau} = L^{Kinetic;Eff;all-modes}_{SU(2)_L;Landau} \nonumber \\
&& \quad \quad \qquad \qquad -V^{Eff;all-modes}_{SU(2)_L;Landau} + {\cal O}_{Ignore}^{SU(2)_L} \nonumber \\
&&L^{Kinetic;Eff;all-modes}_{SU(2)_L;Landau}  \\
&&\quad \quad \qquad \qquad =\half \Big(\Gamma_{0,2}(p,-p;) - \Gamma_{0,2}(00;) \Big) h^2 \nonumber \\
&&\quad \quad \qquad \qquad +\half \Big( \Gamma_{0,2}(;q,-q)- \Gamma_{0,2}(;00)  \Big) {\vec \pi}^2 \nonumber \\
&&V^{Eff;all-modes}_{SU(2)_L;Landau}= \mpisq \Big[ \frac{h^2 + {\vec \pi}^2}{2} +\HVEV h \Big] \nonumber \\
&&\quad \quad \qquad \qquad + 
\lambda_{\phi}^2 \Big[ \frac{h^2 + {\vec \pi}^2}{2} +\HVEV h \Big]^2 \,,\nonumber
\end{eqnarray} 
with
finite non-trivial wavefunction renormalizations 
\begin{eqnarray}
\label{WavefunctionRenormMom}
\Gamma_{0,2}(;q,-q)-\Gamma_{0,2}(;00) &\sim& q^2 \,. \nonumber
\end{eqnarray}

The $\phi$-sector effective Lagrangian (\ref{LEffectiveSM}) 
has insufficient boundary conditions 
to distinguish among the three modes 
\cite{Lee1970,Symanzik1970a,Symanzik1970b,Vassiliev1970}  
of the BRST-invariant Lagrangian $L_{SU(2)_L}$ in (\ref{LagrangianAHM}).\footnote{
	The inclusive Gell-Mann L${\acute e}$vy \cite{GellMannLevy1960} effective potential derived \cite{Lynnetal2012} from B.W. Lee's WTI \cite{Lee1970}, reduces to the three different effective potentials of the global $SU(2)_L \times SU(2)_R$ Schwinger model \cite{Schwinger1957}: 
	Schwinger Wigner mode $(\HVEV =0,\mpisq =m_{BEH}^2 \neq 0)$; 
	Schwinger Scale-Invariant point $(\HVEV =0,\mpisq = m_{BEH}^2 =0)$; 
	or Schwinger Goldstone mode $(\HVEV \neq 0,\mpisq = 0;m_{BEH}^2\neq 0)$.
	}
For example, the effective potential 
$V^{Eff;all-modes}_{SU(2)_L;Landau}$ 
becomes in the appropriate limits:
\begin{eqnarray}
\label{WignerSIGoldstonePotentials}
V^{Eff;Wigner}_{SU(2)_L;Landau}&=& \mpisq \Big[ \frac{h^2 + {\vec \pi}^2}{2} \Big] + \lambda_{\phi}^2 \Big[ \frac{h^2 + {\vec \pi}^2}{2}  \Big]^2\,, \nonumber \\
V^{Eff;ScaleInvariant}_{SU(2)_L;Landau}&=& \lambda_{\phi}^2 \Big[ \frac{h^2 + {\vec \pi}^2}{2}  \Big]^2 \,, \nonumber \\
V^{Eff;Goldstone}_{SU(2)_L;Landau}&=& \lambda_{\phi}^2 \Big[ \frac{h^2 + {\vec \pi}^2}{2} +\HVEV h \Big]^2\,.
\end{eqnarray}

Eqn. (\ref{LEffectiveSM}) has exhausted the constraints 
on the allowed terms in the $\phi$-sector effective Lagrangian
due to those $SU(2)_L$ WTIs 
governing 1-$\phi$-I $\phi$-sector Green's functions $\Gamma_{N,M}$ 
(\ref{GreensWTIPrime}, \ref{GreensFWTI}). 
In order to provide boundary conditions 
that distinguish among the effective potentials in (\ref{WignerSIGoldstonePotentials}), 
we must turn to the $SU(2)_L$ WTIs 
that govern $\phi$-sector 1-$\phi$-R on-shell T-Matrix elements $T_{N,M}$.

\subsection{The Lee-Stora-Symanzik (LSS) Theorem:
IR finiteness
and automatic tadpole renormalization}
\label{SectionLSSTheorem}

\begin{quote}
{\it ``Whether you like it or not, 
you have to include in the Lagrangian all possible terms consistent with locality and power counting, 
unless otherwise constrained by Ward identities."}
Kurt Symanzik, in a private letter to Raymond Stora \cite{SymanzikPC} 
\end{quote}

Evaluating the effective potential in (\ref{LEffectiveSM}) with $\HVEV \neq 0$, 
and then in the Kibble representation
\begin{eqnarray}
\label{PreLSSGoldstoneModePotential}
&&V^{Eff;pre-LSS-Goldstone}_{SU(2)_L;\phi;Landau}\\
&&= \mpisq \Big[ \frac{h^2 + {\vec \pi}^2}{2} +\HVEV h\Big] 
+\lambda_{\phi}^2 \Big[ \frac{h^2 + {\vec \pi}^2}{2}  +\HVEV h\Big]^2 \nonumber\\
&&=\mpisq \Big[ \phi^{\dagger} \phi -\half \HVEV ^2\Big]+\lambda_{\phi}^2 \Big[ \phi^{\dagger} \phi -\half \HVEV ^2\Big]^2 \nonumber \\
&&=\frac{\mpisq}{2} \Big[ {\tilde H}^2 -\HVEV ^2\Big]
+\frac{\lambda_{\phi}^2}{4} \Big[ {\tilde H}^2 -\HVEV ^2\Big]^2 \nonumber 
\end{eqnarray}

As expected, the NGBs ${\tilde {\vec \pi}}$ have disappeared from the effective potential, 
have purely derivative couplings through their kinetic term, 
and obey the shift symmetry 
\bea
\label{SU2ShiftSymmetry}
{\tilde {\vec \pi}}&\to& {\tilde {\vec \pi}}+\HVEV {\vec \theta} +{\tilde {\vec \pi}}\times {\vec \theta} +{\cal O}({\vec \theta}^2) 
\eea
for constant $\vec \theta$. 
In other words, the Goldstone theorem is already properly enforced. 

Eqn. (\ref{PreLSSGoldstoneModePotential})
 appears at first sight to embrace a disaster: 
the term linear in $\phi^{\dagger} \phi -\half \HVEV ^2$ (a remnant of Wigner mode in (\ref{WignerSIGoldstonePotentials})) persists, 
destroying the symmetry of the famous ``Mexican hat",  
and the $SU(2)_L$ gauge theory is not actually in Goldstone mode! 

In strict obedience to K. Symanzik's edict, 
we now further constrain the allowed terms in the $\phi$-sector effective Lagrangian 
with  those $SU(2)_L$ Ward-Takahashi identities 
that govern 1-$\phi$-R on-shell T-Matrix elements $T_{N,M}$.
In Appendix \ref{DerivationWTIAHM}, we extend Adler's self-consistency condition  
(originally written for the global 
$SU(2)_L\times SU(2)_R$ \GMLfull Linear Sigma Model 
	with PCAC \cite{Adler1965,AdlerDashen1968}), 
but now derived for the $CP$-conserving $SU(2)_L$ gauge theory 
	in Landau gauge (\ref{AdlerSelfConsistency})
\begin{eqnarray}
\label{AdlerSelfConsistencyPrime} 
&&\HVEV T^{tt_1...t_M}_{N,M+1}(p_1...p_N;0q_1...q_M)\nonumber \\
&& \quad \quad \times (2\pi)^4\delta^4 \Big(\sum_{n=1}^N p_n +\sum_{m=1}^M q_m \Big) \Big\vert^{p_1^2 =p_2^2...=p_N^2=m_{BEH}^2}_{q_1^2 =q_2^2...=q_M^2=0}  \nonumber \\
&& \quad \quad =0 
\end{eqnarray}
The T-matrix vanishes as one of the pion momenta goes to zero 
(i.e. these are 1-soft-pion theorems), 
provided all other physical scalar particles are on mass-shell. 
Eqn. (\ref{AdlerSelfConsistencyPrime}) also 
\begin{quote}
{\it ``asserts the absence of infrared (IR) divergences 
in the scalar-sector (of $SU(2)_L$) Goldstone mode  (in Landau gauge). 
Although individual Feynman diagrams are IR divergent, 
those IR divergent parts cancel exactly in each order of perturbation theory. 
Furthermore, the Goldstone mode amplitude must vanish in the soft-pion limit.
B.W. Lee \cite{Lee1970}".}
\end{quote}
It is crucial to note that 
the external states in $T_{N,M}$ are $N$ $h$'s and $M$ $\pi$'s,
not $\tilde\pi$'s. 
We are working in the soft-$\pi$, not the soft-$\tilde\pi$ limit.

The $N=0,M=1$ case of (\ref{AdlerSelfConsistencyPrime}) 
is the LSS theorem: 
\begin{eqnarray}
\label{TMatrixGoldstoneTheoremPrime}
\HVEV T^{t_1t_2}_{0,2}(;00)&=&\delta^{t_1t_2}\HVEV T_{0,2}(;00) =0 
\end{eqnarray}
This looks like the Goldstone Theorem,
but, since it involves $\vec \pi$ not $\tilde {\vec \pi}$ it is quite distinct.\footnote{
	B.W. Lee \cite{Lee1970} proves two towers of WTI 
	for the global $SU(2)_L\times SU(2)_R$ \GMLfull model 
		(GML) \cite{GellMannLevy1960}
	 in the presence of the Partially Conserved Axial-vector Current (PCAC) hypothesis. 
	 PCAC conserves the vector current $\partial_\mu {\vec J}_{L+R}^{\mu;GML} =0$, 
	 but explicitly breaks the axial-vector current, 
	$\partial_\mu {\vec J}_{L-R}^{\mu;GML} = \gamma_{PCAC}^{GML} {\vec \pi} $.
	Lee identifies the all-loop-orders GML WTI 
	\begin{eqnarray}
	\label{LeePCACGoldstoneTheorem}
	\gamma_{PCAC}^{GML} = -\HVEV \Gamma_{0,2}^{GML}(;00)
	\end{eqnarray}
	as the ``Goldstone theorem in the presence of PCAC."  
	\hfil\break
	\indent The PCAC analogy for Landau-gauge $SU(2)_L$ would have been
	\begin{eqnarray}
	\label{StandardModelPCAC}
	\partial_\mu {\vec J}_{L}^{\mu;SU(2)_L} &=& \gamma_{PCAC}^{SU(2)_L} {\vec \pi} 
		+\HVEV\times ({\rm a~gauge-fixing~term})  \nonumber \\
	\gamma^{SU(2)_L}_{PCAC} &=& -\HVEV T_{0,2}^{SU(2)_L}(;00)  \,,
	\end{eqnarray}
	but $SU(2)_L$ is a  local/gauge theory.
	This requires that $\gamma^{SU(2)_L}_{PCAC}\equiv 0$ exactly.
	SSB current conservation can be broken only softly 
	by gauge-fixing terms as in (\ref{DivergenceAHMCurrentPrime}), 
	in order to preserve renormalizability and unitarity \cite{JCTaylor1976}. 
	The Landau-gauge $SU(2)_L$ LSS theorem
	therefore reads
	\begin{eqnarray}
	\label{StandardModelNoPCAC}
	-\HVEV T_{0,2}^{SU(2)_L}(;00)=\gamma^{SU(2)_L}_{PCAC} \equiv 0
	\end{eqnarray}
	as in (\ref{TMatrixGoldstoneTheoremPrime}). 
	Crucially, with $\HVEV \neq 0$ in the SSB Goldstone mode of $SU(2)_L$ 
	(and of $SU(2)_L\times U(1)_Y$, and of the $\nu_D SM^{CP}$ in section \ref{nuDSM}), 
	\begin{equation}
	\label{SSBStandardModelNoPCAC}
	 0 = T_{0,2}^{SU(2)_L}(;00)
	=\big[\Delta_\pi^{}(0)\big]^{-1}
	=-\mpisq\,.
	\end{equation}
	This condition, that the mass-squared of the pseudoscalars $\vec \pi$ are exactly zero, 
	is distinct from and more powerful than the more familiar condition $m_{\tilde \pi}^2=0$, 
	i.e. the masslessness of the NGBs $ {\tilde {\vec \pi}}$. 
	\hfil\break\indent
	We see that on-shell T-Matrix WTI 
		(\ref{AdlerSelfConsistencyPrime},\ref{AdlerSelfConsistency})  
	add information to that contained 
	in Green's function WTI (\ref{GreensWTIPrime},\ref{GreensFWTI}). 
	Beyond IR finiteness \cite{Lee1970}, 
	they provide crucial constraints on the gauge theory by insisting: 
	that $\gamma^{SU(2)_L}_{PCAC}\equiv0$ exactly, 
		as in (\ref{StandardModelNoPCAC},\ref{SSBStandardModelNoPCAC}); 
	that the $SU(2)_L$ current is softly broken or conserved,
		as insisting(\ref{DivergenceAHMCurrentPrime},\ref{GaugeConditionsPrimePrime},
			\ref{PhysicalAHMCurrentConservation}); 
	and that unitarity and renormalizability of the $SU(2)_L$ gauge theory 
	are preserved \cite{JCTaylor1976}.
} 
Since the 2-point $T_{0,2}$ is already 1-$\phi$-I\footnote{
	{
	An SSB 1-$\phi$-R T-Matrix element $T_{N,M}$ consists
	of a sum of many possible diagrams, $T^{i}_{N,M}$, 
	where $i$ indexes all the possibilities.
	We can represent each such diagram 
	as a set of  1-$\phi$-I  vertices $\Gamma_{n,m}$ (which we term beads) 
	attached by $\phi$ propagators,
	in such a way as to leave
	$N$ external $h$ lines and $M$ external $\pi$ lines.
 
	Consider in particular $T_{0,2}(;q,-q)$.
	For any diagram $T^{i}_{0,2}(;q,-q)$ contributing to $T_{0,2}(;q,-q)$,
	there is a unique  ``string'' of $\phi$ propagators
	that threads from end to end through the diagram.
	Each bead on this string has  2 $\phi$-legs, with equal and opposite 4-momenta $q$ and $-q$.
	Since $\Gamma_{0,0}=\Gamma_{0,1}=\Gamma_{1,0}=0$, one cannot have additional
	$\phi$ legs connecting off this main $\phi$ line to another ``side bead''  
	unless they connect in groups of two or more.  But in this case, the main bead
	and the secondary bead cannot be separated by cutting one $\phi$ line,
	and so are part of  the same bead.
	Since $CP=(+1,-1)$ for $(h,\pi)$, and is conserved in this paper, 
	the 1-$h$-Reducible contribution vanishes,
	and so the beads must be connected only by $\pi$s, and each bead is just a
	$\Gamma_{0,2}(;q,-q)$.

	Thus the diagram corresponding to $T^i_{0,2}(;q,-q)$ 
	would appear to consist of $i+1$ copies of  $\Gamma_{0,2}(;q,-q)$
	irreducible vertices connected by $\pi$ propagators $\Delta_\pi(q^2)$,
	and so 
	$T^i_{0,2}(;q,-q) = \Gamma_{0,2}(;q,-q) \left[\Gamma_{0,2}(;q,-q)\Delta_\pi(q^2)\right]^i$.
	$T_{0,2}(;q,-q)$ would then consist of the sum over all such strings.

	However, $\Gamma_{0,2}(;q,-q)\Delta_\pi(q^2)=1$, and so, in fact,
	one should not separately count each $T^i_{0,2}(;q,-q)$, 
	but rather 
	\be
	T_{0,2}(;q,-q)=\Gamma_{0,2}(;q,-q) = \left[\Delta_\pi(q^2)\right]^{-1}\,.
	\ee
	}
}, we may write the  LSS theorem 
as a further constraint on the 1-$\phi$-I Greens function (and thus the $\vec\pi$ mass)
\begin{eqnarray}
\label{GFGoldstoneTheoremPrime}
\HVEV \Gamma_{0,2}\left(;00\right) &=&\HVEV \big[ \Delta_\pi (0) \big]^{-1} 
	=\HVEV\mpisq=0\,.
\end{eqnarray}

To the rescue of the effective potential (\ref{PreLSSGoldstoneModePotential}), 
the LSS theorem (\ref{TMatrixGoldstoneTheoremPrime}) 
(and particularly in the form \ref{GFGoldstoneTheoremPrime}),
not the  shift symmetry (\ref{SU2ShiftSymmetry}) and Goldstone theorem,
forces the $SU(2)_L$ gauge theory fully into its true Goldstone mode, with
\begin{eqnarray}
\label{GoldstoneModePotential}
V^{Eff;LSS-Goldstone}_{AHM;\phi;Lorenz}&=& \frac{\lambda_{\phi}^2}{4} \Big[ {\tilde H}^2-\HVEV^2\Big]^2 \nonumber \\
&=&\lambda_{\phi}^2 \Big[ \phi^{\dagger} \phi -\half \HVEV ^2\Big]^2\\
&=& \lambda_{\phi}^2 \Big[ {\frac{h^2+{\pi}^2}{2}} +\HVEV h\Big]^2\,.\nonumber
\end{eqnarray}

A central result of this paper is to recognize that,
in order to force (\ref{PreLSSGoldstoneModePotential}) 
	to become (\ref{GoldstoneModePotential}),
requires the LSS theorem, a new on-shell T-Matrix  symmetry
that is not a symmetry of the BRST-invariant $SU(2)_L$ Lagrangian.
$SU(2)_L$  physics, but not its Lagrangian, 
has the $SU(2)_L\otimes$BRST symmetry of Section \ref{AbelianHiggsModelSymmetry},
a conserved current (\ref{AHMCurrentPrime},\ref{PhysicalAHMCurrentConservation}), 
un-deformed WTIs governing connected amputated Green's functions (\ref{GreensWTIPrime}), 
and un-deformed  WTIs 
governing connected amputated on-shell T-Matrix elements (\ref{AdlerSelfConsistencyPrime}).

A crucial effect of the LSS theorem,  
together with the  $N=0, M=1$ $SU(2)_L$ 
Ward-Takahashi Green's-function identity (\ref{GreensWTIPrime}), 
is to automatically eliminate tadpoles in (\ref{FormSchwingerPotential})
\begin{eqnarray}
\label{ZeroTadpoles}
\Gamma_{1,0}(0;) &=& \HVEV \Gamma_{0,2}(;00) =0\,,
\end{eqnarray}
so that separate tadpole renormalization is unnecessary. This is in contrast to \cite{KrausSiboldAHM} and \cite{Grassi1999} where the authors explicity renormalize the tadpole contributions to 0. However, the LSS theorem tells us that this is unnecessary, when the full symmetries of the theory (in this case the ones concerning on-shell T-matrix elements) are imposed the tadpole contributions vanish exactly.

\subsection{Further constraints on the $\phi$-sector effective Lagrangian:
$m_{BEH}^2 = 2 \lambda _\phi ^2 \HVEV ^2$}
\label{TMatrixAHM}
The $\phi$-sector Goldstone-mode coordinate-space Landau-gauge effective Lagrangian
of spontaneously broken $SU(2)_L$,
constrained by the Lee-Stora-Symanzik theorem, can now be written
\begin{eqnarray}
\label{GoldstoneLagrangian}
L^{Eff;Goldstone}_{SU(2)_L;\phi;Landau} &=& \vert D_{\mu}\phi \vert ^2 -\lambda_{\phi}^2  \Big[ {\frac{h^2+{\vec \pi}^2}{2}} +\HVEV h\Big] ^2 \nonumber \\
&+&{\cal O}_{Ignore}^{SU(2)_L}\,.
\end{eqnarray} 
The LSS theorem (\ref{TMatrixGoldstoneTheoremPrime}) has caused all relevant operators in spontaneously broken $SU(2)_L$ to vanish!

This effective Lagrangian (\ref{GoldstoneLagrangian}): 
\begin{itemize}
\item is derived from the local BRST-invariant $SU(2)_L$ Lagrangian $L_{SU(2)_L}$ (\ref{LagrangianAHM});
\item includes 
	all divergent ${\cal O}(\Lambda^2),{\cal O}(\ln \Lambda^2)$ and finite terms 
	that arise to all perturbative loop-orders in the full $SU(2)_L$ gauge theory, 
	due to virtual transverse gauge bosons, $\phi$ scalars, anti-ghosts and ghosts
	(${\vec W}^\mu;h,{\vec \pi};{\bar {\vec \eta}},{\vec \omega}$ respectively);
\item obeys the LSS theorem  
	(\ref{TMatrixGoldstoneTheoremPrime},\ref{GFGoldstoneTheoremPrime}) 
	and all other $SU(2)_L$ Ward-Takahashi Green's function and T-Matrix identities; 
\item obeys the Goldstone theorem in the Landau gauge, 
	having massless derivatively coupled ${\tilde {\vec \pi}}$ NGBs;
\item has its vacuum automatically forced to be at $H=\HVEV$ ($\expval{\vec{\pi}}=0$)
	by $SU(2)_L \otimes BRST$ symmetry, 
	and obeys stationarity \cite{ItzyksonZuber} of that true minimum;
\item preserves the theory's renormalizability and unitarity, 
	which require that wavefunction  renormalization, 
	$\HVEV_{Bare}=\big[ Z^{\phi}_{AHM}\big]^{1/2}\HVEV$ 
	\cite{LSS-2,Bjorken1965,ItzyksonZuber}, 
	and forbid UVQD, relevant, or any other dimension-2-operator corrections to $\HVEV$.
\end{itemize}

The  Green's function Ward-Takahashi ID (\ref{GreensWTIPrime}) for $N=1,M=1$, 
constrained by the LSS theorem (\ref{GFGoldstoneTheoremPrime}), 
relates the BEH mass  to the coefficient of the $h{\vec\pi}^2$ vertex 
\begin{eqnarray}
\label{WTIHiggsMass}
\Gamma_{2,0}(00;)=\HVEV\Gamma_{1,2}(0;00) \,.
\end{eqnarray} 
Therefore, the BEH mass-squared
\begin{eqnarray}
\label{BEHMassAHMS}
m_{BEH}^2 = 2\lambda_{\phi}^2\HVEV^2
\end{eqnarray} 
arises entirely from SSB, 
as does (together with its $SU(2)_L$ decays) 
the resonance pole-mass-squared
\begin{eqnarray}
\label{BEHPoleMassPrime}
m^2_{BEH;Pole}  &=& 2\lambda_\phi^2 \HVEV^2\Big[ 1- 2\lambda_\phi^2 \HVEV^2 
	\int dm^2 \frac{ \rho^{BEH}_{SU(2)_L}(m^2)}{m^2 - i\epsilon} \Big]^{-1} \nonumber \\
	&+& {\cal O}^{Ignore}_{SU(2)_L;\phi}.
\end{eqnarray}

\section{${\cal G}\otimes$BRST symmetry in simple groups in linear gauges}
\label{StoraSymmetry}

Our results are ubiquitous to gauge theories which spontaneously break a Lie algebraic structure group $\cal G$
at a low scale, say for definiteness $m_{Weak}$. In this Section we prove 
${\cal G}\otimes$BRST symmetry and motivate 
its 2 associated towers of 1-soft-pion WTI --
one for Green's functions 
and another for on-shell T-Matrix elements, 
	including the LSS theorem for $\cal G$ at $m_{Weak}$.
These result, as above, 
in severe WTI constraints on the $m_{Weak}$-scalars' effective potential.
We illustrate this in Section \ref{StandardModel} with the specific example of the
Standard Model, in which there 
spontaneous symmetry breaking of 
$SU(2)_L \otimes U(1)_Y \otimes$BRST to $U(1)_{QED} \otimes$BRST.

In linear gauges, the Lagrangian, 
incorporating gauge-fixing and DeWitt-Fadeev-Popov ghost terms 
\cite{DeWitt1967,Fadeev1967}
is written in terms of 
gauge, matter, ghost, and anti-ghost fields 
(${\cal A}_\mu,\Psi,\omega,{\bar \eta}$ respectively),
and a Nakanishi-Lautrup field \cite{Nakanashi1966,Lautrup1967} $b$. 
\bea
\label{LinearBRSTLagrangian}
L_{\cal G}&=&L^{GaugeInvariant}_{\cal G}({\cal A}_\mu,\Psi) 
+ s\Big( {\bar \eta}\Big[ {\cal F} +\half \xi b\Big]\Big).
\eea
There is a gauge coupling $g$ for the simple gauge group $\cal G$, 
and the gauge-fixing function ${\cal F}$ depends on parameters $(\xi,{\cal M})$.
\bea
\label{LinearGauge}
{\cal F}&=& f\partial_\mu {\cal A}^\mu + {\cal M}\Psi
\eea

The gauge fields $\mathcal{A}_\mu$, the ghost field $\omega$, and the parameter $f$ 
belong to the adjoint representation; 
the matter fields $\Psi$ and the parameter $\mathcal{M}$ 
belong to the fundamental representation; 
and the anti-ghost field $\bar{\eta}$ and the Nakanishi-Lautrup auxiliary field $b$ 
belong to the singlet representation of $\mathcal{G}$.
The structure group $\cal G$ 
\bea
\label{StructureConstants}
\Big[ t_\alpha , t_\beta \Big] = iC_{\alpha \beta}^{\gamma}t_\gamma
\eea
is simple,
with structure constants $C_{\alpha \beta}^{\gamma}$.

Global BRST transformations \cite{BecchiRouetStora,Tyutin1975,Tyutin1976,Nakanashi1966,Lautrup1967,Weinberg1995} $s$ of the fields and parameters are given by
\begin{eqnarray}
	\label{LinearBRSTTransformations}
	&s{\cal A}_{\mu}=D_\mu {\omega}; \quad
	&s\Psi=t(\omega)\Psi; \nonumber\\
	&s{\omega}=-\half \Big[{\omega} , {\omega}\Big];\quad
	&s{\bar {\eta}}={b}; \nonumber \\
	&s{b}=0;& \\
	&s{\cal M}=\Lambda; \quad
	&s{f}=\lambda; \nonumber\\
	&s\Lambda=0;\quad
	&s\lambda=0\,.\nonumber
\end{eqnarray}
These ensure  that the Lagrangian (\ref{LinearBRSTLagrangian}) is BRST invariant
\bea
\label{LinearBRSTTransformationLagrangian}
s L_{\cal G}&=& 0 \,.
\eea

Meanwhile, we define the properties of the fields and parameters
under the usual anomaly-free un-deformed rigid/global structure group $\cal G$. 
$\delta_{\cal G}$ transforms fields by constant $\Omega$
\begin{eqnarray}
\label{LinearU(1)Transformations}
\delta_{\cal G}{\cal A}_{\mu}&=&\big[ {\cal A},\Omega \big]; \quad \delta_{\cal G}\Psi=-t\big( \Omega \big)\Psi;\nonumber \\
\delta_{\cal G}{\omega}&=&\big[ \omega,\Omega \big]; \quad \delta_{\cal G}{\bar \eta}=0; \nonumber \\
\delta_{\cal G}{b}&=&0 \nonumber \\
\delta_{\cal G}{\cal M}&=&{\cal M}t\big( \Omega \big); \quad \delta_{\cal G}{f}=f \text{adj}(\Omega); \nonumber \\
\delta_{\cal{G}}\Lambda&=&t(\Omega)\Lambda; \quad \delta_{\cal{G}}\lambda = \lambda \text{adj}(\Omega).
\end{eqnarray}

The  transformation sets (\ref{LinearBRSTTransformations}) and (\ref{LinearU(1)Transformations}) commute
\bea
\label{LinearSU2BRSTFieldCommutators}
\Big[ \delta_{\cal G}, s \Big] {\cal A}^\mu&=& 0;
\quad \Big[ \delta_{\cal G}, s \Big] {\Psi}=0; \nonumber \\
\Big[ \delta_{\cal G}, s \Big] \omega&=& 0;
\quad \Big[ \delta_{\cal G}, s \Big] {\bar {\eta}}=0; \nonumber \\
\Big[ \delta_{\cal G}, s \Big] {b}&=& 0; \nonumber \\
\Big[ \delta_{\cal G}, s \Big] {\cal M}&=& 0;
\quad \Big[ \delta_{\cal G}, s \Big] {f}=0; \nonumber \\
\Big[ \delta_{\cal G}, s \Big]\Lambda &=&0; \quad \Big[ \delta_{\cal G}, s \Big]\lambda =0.
\eea
The  Lagrangian (\ref{LinearBRSTLagrangian})
is not invariant under $\cal G$ transformations:
\bea
	\label{LinearU(1)TransformationLagrangian}
	\delta_{\cal G} L_{\cal G} \nonumber &=&s 
	\Big( {\bar {\eta}} \delta_{\cal G}\Big[ {\cal F} +\half \xi {b}  \Big]\Big) 
	\neq 0
\eea
Still, with (\ref{LinearBRSTTransformationLagrangian},\ref{LinearSU2BRSTFieldCommutators}) and the nilpotent property $s^2 =0$,
\bea
\label{LinearU(1)BRSTCommutatorAHM}
\Big[ \delta_{\cal G}, s \Big] L_{\cal G} &=& 0,
\eea 
and the two separate global symmetries can therefore co-exist in $\cal G$ physics.

The $\cal G$ gauge theory thus simultaneously obeys both the usual BRST symmetry
and a global $\cal G$ symmetry 
	that controls Green's functions and on-shell T-Matrix elements. 
Global $\cal G$ symmetry results, as above for $SU(2)_L$ in 1-soft-$\pi$ theorems:
\begin{itemize}
\item A tower of rigid SSB $\cal G$ WTIs governing relations among Green's functions.
\item A tower of rigid SSB $\cal G$  WTI which force on-shell 1-soft-pion T-Matrix elements to vanish, including a Lee-Stora-Symanzik theorem. 
\end{itemize}
This is despite the fact that the $\cal G$ Lagrangian 
is not invariant under structure group $\cal G$ transformations!

We choose the gauge-fixing function $\mathcal{F}$ 
to be the most general function that is a Lorenz scalar,
and is linear in the fields and its derivatives. 
The results discussed here only hold for such linear gauge-fixing functions. 
The translation to $R_{\xi}$ gauges for the $SU(2)_L$ theory is:
\begin{eqnarray}
\label{GtoSU2}
	\mathcal{A}^\mu &=& W^{\mu}; 
	\quad \Psi = (\pi^1, \pi^2, \pi^3);  \\
	f&=&\mathbb{1}; 
	\quad \mathcal{M} = \xi \frac{1}{2}g_2 \expval{H}; \nonumber 
\end{eqnarray}
and 
\begin{eqnarray}
	\delta_{\mathcal{G}}\frac{1}{2}g_{2}\expval{H} &=& 0; 
	\quad \delta_{\mathcal{G}}\mathbb{1} = 0;  \\
	s \frac{1}{2}g_{2}\expval{H} &=&0; \quad s\mathbb{1} = 0; \nonumber
\end{eqnarray}
where $W^{\mu} =\vec{t} \cdot \vec{W}^{\mu}$. 
With these choices, 
$\left[ \delta_{\cal{G}},s\right]=0$ still holds for $\mathcal{G} = SU(2)_L$.

\section{Spontaneous symmetry breaking of the Landau-gauge standard $SU(2)_L\otimes U(1)_Y\otimes BRST$ electroweak model: gauge bosons, complex scalar doublet, ghosts and anti-ghosts}
\label{StandardModel}

The Lagrangian $L_{2\otimes 1}$ for the spontaneously broken 
$SU(2)_L\otimes U(1)_Y\otimes BRST$ electroweak model in $R_\xi$ gauge
 is given in \eqref{LTotalEwSM}.
 In incorporates: 
 	$SU(2)$ and $U(1)$ gauge-boson fields ${\vec W}_{\mu}$ and $B_\mu$;
 	a Higgs field $H$, 
 	pseudoscalar fields ${\vec \pi}$;
	$SU(2)$ and $U(1)$ ghosts $\vec\omega$ and $\omega_B$, 
	and anti-ghosts ${\bar {\vec \eta}}$ and ${\bar \eta}_B$; and
	$SU(2)$ and $U(1)$  Nakanashi-Lautrup fields $\vec b$ and $b_B$.

\subsection{Global $SU(2)\times U(1)_Y$ and BRST transformations}
\label{TransformationsSM}
The global BRST transformations 
	\cite{BecchiRouetStora,Tyutin1975,Tyutin1976,Nakanashi1966,Lautrup1967,Weinberg1995}, 
	$s_{2\otimes 1}$, for the fields in $L_{2\otimes 1}$ are:
\begin{eqnarray}
\label{EwSMBRSTTransformations}
	s_{2\otimes 1}{\vec W}_{\mu}&=&\partial_\mu {\vec \omega}+g_2{\vec W}_{\mu} \times {\vec \omega} \nonumber \\
	s_{2\otimes 1}H&=&-\half g_2{\vec \pi} \cdot {\vec \omega}
	-{\tilde e}\pi_3\omega_B\nonumber \\
	s_{2\otimes 1}{\vec \pi}&=&
		\half g_2\Big( H {\vec \omega} +{\vec \pi}\times {\vec \omega} \Big)+  {\tilde e}\Big(-\pi_2,\pi_1,H\Big)\omega_B \nonumber \\
	s_{2\otimes 1}{\vec \omega}&=&-\half g_2 {\vec \omega} \times {\vec \omega} \\ 
	s_{2\otimes 1}{\bar {\vec \eta}}&=&{\vec b} \nonumber \\
	s_{2\otimes 1}{\vec b}&=&0 \nonumber \\
	s_{2\otimes 1}{ B}_{\mu}&=&\partial_\mu {\omega}_B \nonumber \\
	s_{2\otimes 1}{ \omega}_B&=&0\nonumber \\ 
	s_{2\otimes 1}{\bar {\eta}}_B&=&{b}_B \nonumber \\
	s_{2\otimes 1}{b}_B&=&0\,.\nonumber
\end{eqnarray}
These transformationa are nilpotent,  $s_{2\otimes 1}s_{2\otimes 1}=0$.

Anomaly-free un-deformed rigid/global $\delta_{SU(2)_L\times U(1)_Y}$ 
transforms these fields by constant $\vec \Omega$ and $\Omega_B$, according to:
\begin{eqnarray}
\label{EwSMU(1)Transformations}
\delta_{2\otimes 1}{\vec W}_{\mu}&=&\partial_\mu {\vec \omega}+g_2{\vec W}_{\mu} \times {\vec \omega} \nonumber \\
\delta_{2\otimes 1}H&=&-\half g_2{\vec \pi} \cdot {\vec \Omega}
-{\tilde e}\pi_3\Omega_B \\
\delta_{2\otimes 1}{\vec \pi}&=&\half g_2\Big( H {\vec \Omega} +{\vec \pi}\times {\vec \Omega} \Big)+  {\tilde e}\Big(-\pi_2,\pi_1,H\Big)\Omega_B \nonumber \\
\delta_{2\otimes 1}{\vec \omega}&=&g_2 {\vec \omega} \times {\vec \Omega}\nonumber 
\end{eqnarray}
While all other fields -- 
$B_\mu$, $\omega_B$, ${\bar {\vec \eta}}$, ${\bar {\eta}}_B$, ${\vec b}$, $b_B$,  --
are singlets. 

The transformation sets 
	(\ref{EwSMBRSTTransformations}) and (\ref{EwSMU(1)Transformations}) commute,
i.e. $\Big[ \delta_{2\otimes 1}, s_{2\otimes 1} \Big]=0$ 
when acting on any of fields -- 
${\vec W}^\mu$, ${\vec \omega}$, $H$,  ${\bar {\vec \eta}}$, ${\vec \pi}$, ${\vec b}$,
$\omega_B$, ${\bar \eta}_B$ or $b_B$.
Thus
\bea
\Big[ \delta_{2\otimes 1}, s_{2\otimes 1} \Big] L_{2\otimes 1}&=& 0.
\eea

Although the anti-ghosts ${\bar {\vec \eta}}$ transform in the usual way, 
as a triplet  $s{\bar {\vec \eta}}={\vec b}$ under BRST in (\ref{EwSMBRSTTransformations}),
we have chosen them to be singlets 
	under global $SU(2)_L$ transformations $\delta_{SU(2)_L}{\bar {\vec \eta}}=0$. 
This does not affect the proof of renormalizabililty and unitarity 
with Slavnov-Taylor identities \cite{Storaprivate}. 
(The subject of a future paper, but outside the scope of this one.)

This freedom to choose ${\bar {\vec \eta}}$ and ${\bar \eta}_B$ 
as singlets under $SU(2)_L\times U(1)_Y$
renders the ghost Lagrangian without definite $\delta_{SU(2)_L\times U(1)_Y}$ properties.
However, it allows one to build 
a classical $SU(2)_L\times U(1)_Y$ current ${\vec J}_{2\otimes 1}^\mu$
and 
a classical $SU(2)_L$ sub-current ${\vec J}_{L;2\otimes 1}^\mu$, 
both conserved up to gauge-fixing terms. 
Together with $CP$ conservation, 
this will be the basis of our global $SU(2)_{L-R}$ Ward-Takahashi identities 
for the $SU(2)_L\times U(1)_Y$  gauge theory.

\subsection{Classical global $SU(2)_L$ sub-current}
\label{LightParticleCurrentsSMPrime}

In Appendix \ref{EwSMLightParticleCurrents}, we constructed the $SU(2)_L\times U(1)_Y$ isospin sub-current ${\vec J}^\mu_{L;2\otimes 1}$, together with its divergence, and commutators with $\phi$ in an arbitrary $R^{2\otimes 1}_\xi$ gauge.

Because $SU(2)_L$ is an applicable sub-group of $SU(2)_L\times U(1)_Y$,
$\Big[ \delta_{SU(2)_L}, s_{2\otimes 1} \Big]=0$ when acting on any of the fields --
${\vec W}^\mu$, ${\vec \omega}$, $H$,  ${\bar {\vec \eta}}$, ${\vec \pi}$, ${\vec b}$,
$\omega_B$, ${\bar \eta}_B$ or $b_B$;
thus
\bea
\label{SU2LBRSTcommuteonL2x1}
\Big[ \delta_{SU(2)_L}, s_{2\otimes 1} \Big] L_{2\otimes 1}&=& 0.
\eea
There is therefore a conserved (up to gauge-fixing terms)  
$SU(2)_L\times U(1)_Y$ isospin sub-current ${\vec J}^\mu_{L;2\otimes 1}$. 
In Landau gauge the sub-current and its divergence are
\begin{eqnarray}
\label{EwSMIsospinCurrent}
{\vec J}^{\mu}_{L;2\otimes 1} &=& {\vec J}^{\mu}_{L;Schwinger} + {\vec {\cal J}}^{\mu}_{L;2\otimes 1}\,, \nonumber \\
{\vec J}^{\mu}_{Schwinger} &=& \half {\vec \pi}\times \partial^\mu {\vec \pi} + \half\Big( {\vec \pi}\partial^\mu H -H \partial^\mu {\vec \pi}\Big)\,, \nonumber\\
{\vec {\cal J}}^{\mu}_{L;2\otimes 1} &=&  {\vec W}^{\mu \nu} \times {\vec W}_{\nu} \nonumber \\
&+& \half g_1 B^{\mu}\Big(\pi_1\pi_3-\pi_2 H, \quad \pi_2\pi_3+\pi_1 H,  \nonumber \\
&& \qquad \qquad \half(H^2 +\pi_3^2-\pi_1^2-\pi_2^2)\Big) \nonumber \\
&-& \frac{1}{4} g_2 {\vec W}^{\mu}\left[ H^2+{\vec \pi}^2\right]   \\
&-& \lim_{\xi \to 0} \frac{1}{\xi}\left[{\vec W}^{\mu} \times {\vec F}_W(0) \right] \nonumber \\
&-& \partial^\mu {\bar {\vec \eta}}\times {\vec \omega} \,, \nonumber \\
\partial_{\mu} {\vec J}^{\mu}_{L;2\otimes 1}&=& \half M_B F_B(0)  \left( -\pi_2,\pi_1,H\right) \nonumber \\
&+&\half M_W\left[H {\vec F}_W(0)+{\vec \pi} \times {\vec F}_W(0) \right]\,,  \nonumber
\end{eqnarray}
with 
\begin{eqnarray}
{\vec F}_W(0) &=& \partial_{\beta}{\vec W}^{\beta}\,, \quad 
{F}_B(0) = \partial_{\beta}{B}^{\beta} \nonumber \\
M_W&=&\half g_2 \HVEV; \qquad M_B = {\tilde e}\HVEV  \\
  {\tilde e} &=& \half Y_\phi g_1 = -\half g_1; \qquad M_W^2+M_B^2=M_Z^2. \nonumber 
\end{eqnarray} 
These sub-currents obey commutation relations:
\begin{eqnarray}
 \delta(z_0-y_0)&&\left[ {\vec J}_{L;2\otimes 1}^{0}(z)- {\vec J}_{L;Schwinger}^{0}(z),H(y)\right]=0\,,  \nonumber \\
 \delta(z_0-y_0)&&\left[ {\vec J}_{L;2\otimes 1}^{0}(z)- {\vec J}_{L;Schwinger}^{0}(z),{\vec \pi}(y)\right]=0 \,,  \nonumber \\
\delta(z^0-y^0)&&\left[ {\vec J}_{Schwinger}^{0}(z),H(y)\right] \\
&=&-\half i\delta^4(z-y) {\vec \pi}(z)\,,  \nonumber \\
\delta(z^0-y^0_1)&&\left[ { J}_{Schwinger}^{0;t}(z),{ \pi^{t_1}}(y_1)\right] \nonumber \\
&=&\half i \delta^4(z-y_1)
\times \Big( \epsilon^{tt_1t_2} \pi^{t_2} (z)+ \delta^{tt_1}H(z) \Big)\,. \nonumber 
\end{eqnarray}

The expression in general $R_{\xi}$ gauges can be found in Appendix \ref{EwSMLightParticleCurrents}. 

\subsection{$SU(2)_L\times U(1)_Y$ in Landau gauge}
\label{DefineAHM}

The $SU(2)_L\times U(1)_Y$ Lagrangian in Landau gauge can be obtained
by setting $\xi=0$ in \eqref{LTotalEwSM}. 
We work in the linear representation of the scalar field in the Goldstone mode of the theory for reasons explained in Section III.A, 
and again define the exact propagators of the $\vec{\pi}$ and $h$ fields 
using the K$\ddot a$ll$\acute e$n-Lehmann representation.

Analysis is again done in terms of the exact renormalized interacting fields, 
which asymptotically become the in/out states, 
i.e. free fields for physical S-Matrix elements.

\begin{eqnarray}
\label{EwSMpNGBPropagator}
&&\Delta^{\pi}_{2\otimes 1}\delta^{t_1t_2} = -i(2\pi)^2\langle 0\vert T\left[ \pi^{t_1}(y)\pi^{t_2}(0)\right]\vert 0\rangle\vert^{Fourier}_{Transform} \nonumber \\
&&\Delta^{\pi}_{2\otimes 1}(q^2)  = \frac{1}{q^2
+ i\epsilon} + \int dm^2 \frac{\rho^{\pi}_{2\otimes 1}(m^2)}{q^2-m^2 + i\epsilon}  \\
&&\Big[Z^{\phi}_{2\otimes 1}\Big]^{-1} = 1+ \int dm^2 \rho^{\pi}_{2\otimes 1}(m^2)\,. \nonumber
\end{eqnarray} 

Define also the BEH scalar propagator in terms of a BEH scalar pole 
and the (subtracted) spectral density $\rho^{BEH}_{2\otimes 1}$, 
and the same wavefunction renormalization. 
We assume $h$ decays weakly, and resembles a resonance:
\begin{eqnarray}
\label{EwSMBEHPropagator}
&&\Delta^{BEH}_{2\otimes 1}(q^2) = -i (2\pi)^2\langle 0\vert T\left[ h(x) h(0)\right]\vert 0\rangle\vert^{Fourier}_{Transform} \nonumber \\
&& \quad \quad =\frac{1}{q^2-m_{BEH;Pole}^2 + i\epsilon}+ \int dm^2 \frac{\rho^{BEH}_{2\otimes 1}(m^2)}{q^2-m^2 + i\epsilon} \nonumber \\
&&\Big[ Z^{\phi}_{2\otimes 1}\Big]^{-1} = 1+ \int dm^2 \rho^{BEH}_{2\otimes 1}(m^2)  \nonumber \\
&& \int dm^2 \rho^{\pi}_{2\otimes 1}(m^2) = \int dm^2 \rho^{BEH}_{2\otimes 1}(m^2) 
\end{eqnarray}
The spectral density parts of the propagators are
\begin{eqnarray}
\label{EwSMSpectralDensityPropagators}
 \Delta^{\pi ;Spectral}_{2\otimes 1}(q^2) &\equiv& \int dm^2 \frac{\rho^{\pi}_{2\otimes 1}(m^2)}{q^2-m^2 + i\epsilon}  \nonumber \\ 
\Delta^{BEH; Spectral}_{2\otimes 1}(q^2) &\equiv& \int dm^2 \frac{\rho^{BEH}_{2\otimes 1}(m^2)}{q^2-m^2 + i\epsilon} \nonumber
\end{eqnarray}
Dimensional analysis of the wavefunction renormalizations (\ref{EwSMpNGBPropagator},\ref{EwSMBEHPropagator}), 
shows that the 
finite Euclidean cut-off contributes only irrelevant terms $\sim \frac{1}{\Lambda^2}$.

We are interested in rigid-symmetric relations among 
1-$\phi$-I connected amputated Green's functions $\Gamma_{N,M}$, 
and among 1-$\phi$-R 
connected amputated transition-matrix (T-Matrix) elements $T_{N,M}$,   
with external $\phi$ scalars and we seperate them as 
\begin{equation}
T_{N,M} = \Gamma_{N,M} + 1-\phi -R.
\end{equation} 
Because these are 1-(${\vec W}_\mu ,B_\mu$)-R in $SU(2)_L\times U(1)_L$ 
(i.e. reducible by cutting a ${\vec W}_\mu$ or $B_\mu$ line), 
it is even more convenient to use the powerful old tools
from Vintage Quantum Field Theory.  

We focus on the rigid/global $SU(2)_L$ current 
constructed in Appendix \ref{EwSMLightParticleCurrents} 
and displayed in (\ref{EwSMIsospinCurrent}) in Landau gauge.

Rigid/global transformations of the fields arise, as usual, 
from the equal-time commutators:
\begin{eqnarray}
\label{EwSMTransformationsAHMFields}
 \delta_{SU(2)_L} H(t,{\vec y})&=&-ig_2 \Omega^{t_1} \int d^3 z\left[ {J}^{0;t_1}_{L;2\otimes 1}(t,{\vec z}),H(t,{\vec y})\right]  \nonumber \\
&=& - \half g_2 \pi^{t_1} (t,{\vec y}) \Omega^{t_1}\\
 \delta_{SU(2)_L} \pi^{t}(t,{\vec y})&=&-i g_2 \Omega^{t_2}\int d^3 z\left[ {J}^{0;t_2}_{L;2\otimes 1}(t,{\vec z}),\pi^{t}(t,{\vec y})\right] \nonumber \\
&=&  \half g_2\Big[ H(t,{\vec y})\Omega^{t} +\epsilon^{tt_1t_2} \pi^{t_1}(t,{\vec y})\Omega^{t_2}\Big] \nonumber
\end{eqnarray} 
so ${\vec J}^\mu_{L;2\otimes 1}(t,{\vec z})$ serves as a ``proper" local current for commutator purposes. 

The classical equations of motion reveal a crucial fact: due to gauge-fixing terms in the BRST-invariant Lagrangian (\ref{TotalEwSMLandauGauge}), the 
classical current 
(\ref{EwSMIsospinCurrent}) is 
not conserved. In Landau gauge  
\begin{eqnarray}
\label{EwSMDivergenceAHMCurrentPrime}
\partial_{\mu} {\vec J}^{\mu}_{L}&=&\half M_W\left[{\vec \pi} \times \partial_\mu{\vec W}^\mu +H \partial_\mu{\vec W}^\mu \right] \nonumber  \\
&+&\half M_B  \partial_\mu{ B}^\mu \Big(-\pi_2 , \pi_1, H \Big) 
\end{eqnarray}

The global $SU(2)_L$ current (\ref{EwSMIsospinCurrent}) 
is, however, conserved by the physical states, 
and therefore still qualifies as a ``real" current. 
Strict quantum constraints are imposed
that force the relativistically-covariant theory of gauge bosons 
to propagate only its true number of quantum spin $S=1$ degrees of freedom. 
These constraints are, in the modern literature, 
implemented by use of spin $S=0$ fermionic DeWitt-Fadeev-Popov ghosts
$({\bar {\vec \eta}},{\vec \omega})$.  
The physical states and their time-ordered products, 
but not the BRST-invariant Lagrangian  (\ref{TotalEwSMLandauGauge}),  
then obey 
G. 't Hooft's gauge-fixing \cite{tHooft1971}  conditions
\begin{eqnarray}
\label{EwSMGaugeConditionsPrime}
&&\big< 0\vert T\Big[ \Big( \partial_{\mu}{\vec W}^{\mu}(z)\Big) \nonumber \\
&&\quad \times h(x_1)...h(x_N)\pi_{t_1}(y_1)...\pi_{t_M}(y_M)\Big]\vert 0\big>_{\rm connected}  =0  \nonumber \\
&&\big< 0\vert T\Big[ \Big( \partial_{\mu}{B}^{\mu}(z)\Big)  \\
&&\quad \times h(x_1)...h(x_N)\pi_{t_1}(y_1)...\pi_{t_M}(y_M)\Big]\vert 0\big>_{\rm connected}  =0 \,.\nonumber
\end{eqnarray}
Here we have N external renormalized scalars $h$ (coordinates x, momenta p), and M external ($CP=-1$) renormalized pseudo-scalars ${ \pi}$ (coordinates y, momenta q, isospin t).

Eqs. (\ref{EwSMGaugeConditionsPrime}) restore conservation 
of the rigid/global $SU(2)_L$ current 
for $\phi$-sector connected time-ordered products
\begin{eqnarray}
\label{EwSMPhysicalAHMCurrentConservation}
&&\Big< 0\vert T\Big[ \Big( \partial_{\mu}{\vec J}^{\mu}_{L;2\otimes 1}(z) \Big) \\
&&\quad \quad \times h(x_1)...h(x_N) \pi^{t_1}(y_1)...\pi^{t_M}(y_M)\Big]\vert 0\Big>_{\rm connected} =0 \nonumber
\end{eqnarray}
It is in this ``physical" connected-time-ordered product sense 
that the rigid global $SU(2)_L$ ``physical current" is conserved: 
the physical states, but not the BRST-invariant Lagrangian  (\ref{TotalEwSMLandauGauge}), 
obey the physical-current conservation equation 
(\ref{EwSMPhysicalAHMCurrentConservation}). 
It is this physical conserved current that generates our $SU(2)_L\otimes$BRST WTI.

The crucial observation for spontaneously broken $SU(2)_L\times U(1)_Y$ is that, just as in Section \ref{AbelianHiggsModel},
\begin{eqnarray}
\label{EwSMNGBSurfaceIntegralPrime}
&& \int_{LightCone}  dz \quad {\widehat {z}}^{LightCone;i} \Big< 0\vert T\Big[   \Big(\half \HVEV { \partial}^i {\vec \pi}\Big)(z)   \\
&&\quad \quad \times h(x_1)...h(x_N) \pi^{t_1}(y_1)...\pi^{t_M}(y_M)\Big]\vert 0\Big>_{\rm Connected} \neq 0\,. \nonumber
\end{eqnarray}
$\vec \pi$ is massless in 
Landau gauge, capable of carrying (along the light-cone) long-ranged pseudo-scalar forces out to the  very ends of the light-cone $(z^{LightCone}\to \infty)$, but not inside it.

Eq. (\ref{EwSMNGBSurfaceIntegralPrime}) shows that the spontaneously broken $SU(2)_L$  charge 
\begin{equation}
\label{EwSMCharge}
{\vec Q}_{L;2\otimes 1}(t)\equiv{ =}\int d^3z {\vec J}^0_{L;2\otimes 1}(t,{\vec z}) 
\end{equation}
is not conserved, even for connected time-ordered products, in
Landau gauge
\begin{eqnarray}
\label{EwSMChargeAHMTimeOrderedProductsNotConserved}
&&\Big< 0\vert T\Big[ \Big(  \frac{d}{dt}{\vec Q}_{L;2\otimes 1}(t) \Big)  \\
&&\quad \quad \times h(x_1)...h(x_N) \pi^{t_1}(y_1)...\pi^{t_M}(y_M)\Big]\vert 0\Big>_{\rm Connected}   \neq 0 \,.\nonumber
\end{eqnarray}

The classic 
proof of the Goldstone theorem \cite{Goldstone1961,Goldstone1962,Kibble1967} 
requires a conserved charge $\frac{d}{dt}{Q}=0$, 
so that proof fails 
\cite{Higgs1964,Englert1964,Guralnik1964,Kibble1967} for spontaneously broken gauge theories, 
allowing spontaneously broken electroweak $SU(2)_L\times U(1)_Y$ to generate mass-squared-gaps $M_W^2$ for the ${W}_\mu^\pm$, and $M_Z^2=M_W^2+M_B^2=\frac{1}{4}(g_1^2+g_2^2)\HVEV^2$ for the neutral current boson ${Z}_\mu$,
and avoid  massless particles apart from the photon ${A}_\mu$ in its observable physical spectrum. 
This is true, even in $R_\xi(\xi=0)$ 
Landau gauge, where there is a Goldstone theorem \cite{Guralnik1964,Kibble1967},
so $\tilde {\vec \pi}$ are derivatively coupled (hence massless) NGB, 
and where there is an LSS theorem \cite{LSS-3Proof}, so $\vec \pi$ is massless.

Massless $\vec \pi$ (not $\tilde{\vec \pi}$) is the basis  
of our pion-pole-dominance-based $SU(2)_{L-R}$ WTIs, derived in Appendices \ref{EwSMLightParticleCurrents} and \ref{EwSMTotalMasterEqWTI}, which give: 
relations among 1-$\phi$-I connected amputated 
$\phi$-sector Greens functions $\Gamma_{N,M}^{2\otimes 1}$ (\ref{EwSMGreensFWTI}); 
1-soft-pion theorems  (\ref{EwSMAdlerSelfConsistencyPrimePrime}, 
	\ref{EwSMAdlerSelfConsistency}, \ref{EwSMInternalTMatrix}); 
infra-red finiteness for $\mpisq =0$ (\ref{EwSMAdlerSelfConsistencyPrimePrime}, \ref{EwSMAdlerSelfConsistency});{}
an LSS theorem (\ref{EwSMTMatrixGoldstoneTheoremPrime}); 
vanishing 1-$\phi$-R connected amputated
on-shell $\phi$-sector T-Matrix elements $T^{2\otimes 1}_{N,M}$ 
(\ref{EwSMAdlerSelfConsistencyPrimePrime},  \ref{EwSMInternalTMatrix}); which all 
realize the full $SU(2)_L \otimes U(1)_Y\otimes BRST$ symmetry of this Section.

\subsection{Construction of the scalar-sector effective $SU(2)_L\times U(1)_Y$ Lagrangian 
from those $SU(2)_{L-R}$ WTIs that  govern connected amputated 1-$\phi$-I Greens functions}
\label{EwSMGreensFunctionsAHM}
We now consider
the global axial-vector isospin current ${\vec J}^{\mu}_{L-R;2\otimes 1}$, 
forming time-ordered amplitudes of
products of ${\vec  J}^\mu_{L-R;2\otimes 1}$ 
with N scalars (coordinates $x$, momenta $p$)   
and M pseudo-scalars (coordinates $y$, momenta $q$, isospin $t$) 
$\big< 0 \vert T\Big[ 
{\vec  J}_{L-R;2\otimes 1}^\mu(z)h(x_1)\cdots h(x_N) 
						\pi^{t_1}(y_1)\cdots \pi^{t_M}(y_M)
\Big]\vert 0\big>$
and examine the divergence of such connected amplitudes:
\bea
\label{EwSMMasterEquation}
&&\partial_\mu  \big< 0\vert 
T\Big[J_{L-R;2\otimes 1}^{\mu;t}(z) \\
&&\quad h(x_1)\cdots h(x_N)\pi^{t_1}(y_1)\cdots \pi^{t_M}(y_M)\Big]
		\vert 0 \big>_{Connected} \nonumber
\eea

In Appendix \ref{EwSMTotalMasterEqWTI} we derive $SU(2)_{L-R}$ ``pion-pole-dominance"  1-$\phi$-R
connected amputated T-Matrix  WTI (\ref{EwSMInternalTMatrix}) for SSB $SU(2)_L\times U(1)_Y$ from the divergence of the connected amplitudes considered above. 
Their solution is a tower of recursive $SU(2)_{L-R}$ WTI (\ref{EwSMGreensFWTI}) 
that govern 1-$\phi$-I  $\phi$-sector connected amputated Greens functions $\Gamma_{N,M}^{2\otimes 1}$.
For $\vec \pi$ with $CP=-1$, the result is 
precisely as (\ref{GreensWTIPrime}) for SSB $SU(2)$:
\begin{eqnarray}
	\label{EwSMGreensWTIPrime}
	&&\HVEV\Gamma_{N,M+1}^{2\otimes 1;t_1...t_Mt}(p_1 ...p_N;q_1...q_M0) \nonumber  \nonumber \\
	&&\quad \quad =\sum ^M_{m=1} \delta^{tt_m}\Gamma_{N+1,M-1}^{2\otimes 1;t_1...{\widehat {t_m}}...t_M}
(p_1...p_Nq_m;q_1...{\widehat {q_m}}...q_M) \nonumber \\
&&\quad \quad -\sum ^N_{n=1}\Gamma_{N-1,M+1}^{2\otimes 1;t_1...t_Mt}(p_1 ...{\widehat {p_n}}...p_N;q_1...q_Mp_n)\,,
\end{eqnarray} 
valid for $N,M \ge0$. 
On the left-hand-side of (\ref{EwSMGreensWTIPrime}) there are 
N renormalized $h$ external legs (coordinates x, momenta p), 
M renormalized ($CP=-1$) ${\vec \pi}$ external legs (coordinates y, momenta q, isospin t), 
and 1 renormalized soft external ${\vec \pi}(k_\mu=0)$  (coordinates z, momenta k).
``Hatted" fields with momenta $({\widehat {p_n}},{\widehat {q_m}})$ are omitted.

The rigid $SU(2)_{L-R}$ WTI 1-soft-pion theorems (\ref{EwSMGreensWTIPrime})
relate a 1-$\phi$-I Green's function with $(N+M+1)$ external fields 
(which include a zero-momentum ${\vec \pi}$), 
to two 1-$\phi$-I  Green's functions with $(N+M)$ external fields.%

The Green's functions $\Gamma_{N,M}^{2\otimes 1;t_1...t_M}(p_1...p_N;q_1...q_M)$ 
are not themselves gauge-independent. 
Furthermore, although 1-$\phi$-I, they are 1-(${\vec W}^\mu,B_\mu$)-R by cutting transverse ${\vec W}_\mu$ or $B_\mu$ gauge boson lines.

The 1-$\phi$-I ${\vec \pi}$  and $h$ inverse propagators are:
\begin{eqnarray}
\label{EwSMInversePropagators}
\Gamma_{0,2}^{2\otimes 1;t_1t_2}(;q,-q) &\equiv& \delta^{t_1t_2} \Gamma^{2\otimes 1}_{0,2}(;q,-q) \nonumber \\
\Gamma_{0,2}^{2\otimes 1}(;q,-q) &\equiv& \left[ \Delta_{\pi}^{2\otimes 1}(q^2) \right]^{-1}=-\HVEV\mpisq \nonumber \\
\Gamma^{2\otimes 1}_{2,0}(q,-q;) &\equiv& \left[ \Delta^{2\otimes 1}_{BEH}(q^2) \right]^{-1} 
\end{eqnarray}

We can now form the $\phi$-sector effective momentum space Lagrangian in Landau gauge.
All perturbative quantum loop corrections, to all-loop-orders 
and including all UVQD, log-divergent and finite  contributions, 
are included in the $\phi$-sector effective Lagrangian:
1-$\phi$-I Green's functions $\Gamma^{2\otimes 1;t_1...t_M}_{N,M}(p_1...p_N;q_1...q_M)$; 
wavefunction renormalizations;  
renormalized $\phi$-scalar propagators (\ref{EwSMpNGBPropagator},\ref{EwSMBEHPropagator}); 
the Brout-Englert-Higgs (BEH) VEV $\HVEV$ (\ref{EwSMHVEV}); 
all gauge boson and ghost propagators. 
This includes the full all-loop-orders renormalization of the $SU(2)_L\times U(1)_Y$ $\phi$-sector,
originating in quantum loops containing transverse virtual gauge bosons, $\phi$-scalars, anti-ghosts and ghosts:
${\vec W}^\mu ,B_\mu;h,{\vec \pi};{\bar {\vec \eta}},{\bar \eta}_B;{\vec \omega},\omega_B$ respectively. 
Because they arise entirely from global $SU(2)_{L-R}$ WTI, 
our results are independent of regularization-scheme \cite{KrausSiboldAHM}.

We want to classify operators arising in $SU(2)_L\times U(1)_Y$ loops,
and separate the finite operators
\footnote{
	In the $SU(2)_L\times U(1)_Y$ Standard Electro-weak Model, 
	there are finite operators that arise entirely from $SM$ degrees of freedom   
	that are crucially important for computing experimental observables. 
	The most familiar are the successful 1-loop high precision Standard Model predictions 
	for the top-quark from Z-pole physics \cite{LynnStuart1985,Kennedy1988,EXPOSTAR,
	Levinthal1990,ALEPH1991,Levinthal1992a,Levinthal1992b,LEPWorkingGroup1993} in 1984
	and from the $W^{\pm}$ mass \cite{Sirlin1980} in 1980, 
	as well as the 2-loop BEH mass from
	Z-pole physics \cite{LynnStuart1985,Verzegnassi1987,Kennedy1988,EXPOSTAR,Levinthal1992a,
		LEPSLCWorkingGroup1995}
	and from the $W^{\pm}$ mass \cite{Sirlin1980,Verzegnassi1987}. 
	Those precisely predicted the experimentally measured masses 
	of the top quark at FNAL \cite{Nobel1999}, 
	and of BEH scalar at CERN \cite{
		LynnStuart1985,Kennedy1988,EXPOSTAR,
		Levinthal1992a,LEPSLCWorkingGroup1995,
		Nobel2013}.
	These operators also include the high-precision electroweak S,T and U 
	\cite{Kennedy1988, PeskinTakeuchi, Ramond2004}).
	Such finite operators are not the point of this paper. 
}
from the divergent ones.
We focus on finite relevant operators, 
as well as quadratic and logarithmically divergent operators.

We classified the irrelevant operators \eqref{ignoredoperators} in Section \ref{GreensFunctionsAHM} and the classification is the same for the $SU(2)_{L}\otimes U(1)_{Y}$ case. 
The only difference now is that the light degrees of freedom include ${\vec W}^\mu ,B_\mu;h,{\vec \pi};{\bar {\vec \eta}},{\bar \eta}_B;{\vec \omega},\omega_B$. 
We ignore these operators here as well.

Such finite operators appear throughout the $SU(2)_{L-R}$ Ward-Takahashi IDs (\ref{EwSMGreensWTIPrime}): 
\begin{itemize}
\item $N+M \geq 5$ is  ${\cal O}_{2\otimes 1}^{1/ \Lambda^2; Irrelevant}$ and ${\cal O}_{2\otimes 1}^{Dim>4;Light}$;
\item the left hand side of  (\ref{EwSMGreensWTIPrime}) for $N+M=4$ is also  
${\cal O}_{2\otimes 1}^{1/ \Lambda^2; Irrelevant}$ and ${\cal O}_{2\otimes 1}^{Dim>4;Light}$;
\item  $N+M\leq 4$ operators ${\cal O}_{2\otimes 1}^{Dim\leq 4;NonAnalytic}$  appear in (\ref{EwSMGreensWTIPrime}). 
\end{itemize}

Finally, there are  $N+M\leq 4$ operators that are analytic in momenta. 
We expand these in powers of momenta, 
count the resulting dimension of each term in the operator Taylor-series, 
and ignore ${\cal O}_{2\otimes 1}^{Dim>4;Light}$ and ${\cal O}_{2\otimes 1}^{1/ \Lambda^2; Irrelevant}$ terms in that series. 

As we did in \eqref{FormSchwingerPotential} for the $SU(2)_{L}$ model we suppress gauge fields and form the
all-loop-orders renormalized scalar-sector effective Lagrangian 
with operator dimension less than or equal to $4$
for ($h, {\vec \pi}$) with CP=($1,-1$)

\begin{eqnarray}
\label{EwSMFormSchwingerPotential}
&& L^{Eff;2\otimes 1}_{\phi} =  \Gamma^{2\otimes 1}_{1,0}(0;)h +\frac{1}{2!} \Gamma^{2\otimes 1}_{2,0}(p,-p;)h^2 \nonumber \\
&&  \quad \quad + \frac{1}{2!} \Gamma_{0,2}^{2\otimes 1;t_1t_2}(;q,-q)\pi_{t_1}\pi_{t_2} +\frac{1}{3!} \Gamma^{2\otimes 1}_{3,0}(000;)h^3  \nonumber \\ 
&& \quad \quad + \frac{1}{2!} \Gamma_{1,2}^{2\otimes 1;t_1t_2}(0;00) h \pi_{t_1}\pi_{t_2} +\frac{1}{4!} \Gamma^{2\otimes 1}_{4,0}(0000;)h^4  \nonumber \\
&&  \quad \quad + \frac{1}{2!2!} \Gamma_{2,2}^{2\otimes 1;t_1t_2}(00;00) h^2 \pi_{t_1}\pi_{t_2}  \\
&& \quad \quad +  \frac{1}{4!}\Gamma_{0,4}^{2\otimes 1;t_1t_2t_3t_4}(;0000)\pi_{t_1}\pi_{t_2} \pi_{t_3}\pi_{t_4} + {\cal O}^{2\otimes 1}_{Ignore} \,. \nonumber 
\end{eqnarray} 

The connected amputated Green's function identities (\ref{EwSMGreensWTIPrime})
severely constrain the effective Lagrangian (\ref{EwSMFormSchwingerPotential}). 
For pedagogical clarity, we first separate out the isospin indices
\begin{eqnarray}
\label{EwSMIsospinIndices}
	\Gamma_{0,2}^{2\otimes 1;t_1t_2}(;q,-q) &\equiv& \delta^{t_1t_2} \Gamma^{2\otimes 1}_{0,2}(;q,-q)\,, \nonumber \\
	\Gamma_{1,2}^{2\otimes 1;t_1t_2}(-q;q0) &\equiv& \delta^{t_1t_2}\Gamma^{2\otimes 1}_{1,2}(-q;q0) \,,\nonumber  \\
	\Gamma^{2\otimes 1;t_1t_2}_{2,2}(00;00)&\equiv&\delta^{t_1t_2}\Gamma^{2\otimes 1}_{2,2}(00;00)\,, \\
	 \Gamma_{0,4}^{2\otimes 1;t_1t_2t_3t_4}(;0000)  &\equiv& \Gamma^{2\otimes 1}_{0,4}(;0000) \nonumber \\
	&\times& \left[ \delta^{t_1t_2}\delta^{t_3t_4} + \delta^{t_1t_3}\delta^{t_2t_4} + \delta^{t_1t_4}\delta^{t_2t_3} \right]\,. \nonumber 
\end{eqnarray}

The Ward-Takahashi IDs (\ref{EwSMGreensWTIPrime}) for Greens functions 
severely constrain the effective Lagrangian (\ref{EwSMFormSchwingerPotential}):
\begin{itemize}
\item WTI $N=0, M=1$
\begin{eqnarray}
\label{EwSMNM01}
\delta^{t_1t_2}\Gamma^{2\otimes 1}_{1,0}(q;)  &=& \HVEV \Gamma_{0,2}^{2\otimes 1;t_1t_2}(;q,-q)\,,\nonumber \\
\Gamma^{2\otimes 1}_{1,0}(0;) &=& \HVEV \Gamma^{2\otimes 1}_{0,2}(;00) \,,
\end{eqnarray}
since no momentum can run into the tadpoles.
\item WTI $N=1, M=1$
\begin{eqnarray}
\label{EwSMNM11}
\delta^{t_1t_2}\Gamma^{2\otimes 1}_{2,0}(-q,q;) &-& \Gamma_{0,2}^{2\otimes 1;t_1t_2}(;q,-q)  \nonumber \\
&=&\HVEV \Gamma_{1,2}^{2\otimes 1;t_1t_2}(-q;q0) \,, \nonumber \\
\Gamma^{2\otimes 1}_{2,0}(-q,q;) &-& \Gamma^{2\otimes 1}_{0,2}(;q,-q)   \\
&=&\HVEV \Gamma^{2\otimes 1}_{1,2}(-q;q0)  \nonumber \\
&=&\HVEV \Gamma^{2\otimes 1}_{1,2}(0;00)  + {\cal O}^{2\otimes 1}_{Ignore} \nonumber \,,\\ 
\Gamma^{2\otimes 1}_{2,0}(00;) &=& \Gamma^{2\otimes 1}_{0,2}(;00) + \HVEV \Gamma^{2\otimes 1}_{1,2}(0;00)\,.\nonumber 
\end{eqnarray}
\item WTI $N=2, M=1$
\begin{eqnarray}
\label{EwSMNM21}
\HVEV\Gamma^{2\otimes 1;t_1t_2}_{2,2}(00;00) &=& \delta^{t_1t_2}\Gamma^{2\otimes 1}_{3,0}(000;) -2\Gamma^{2\otimes 1;t_1t_2}_{1,2}(0;00) \,, \nonumber \\
\HVEV\Gamma^{2\otimes 1}_{2,2}(00;00) &=& \Gamma^{2\otimes 1}_{3,0}(000;) -2\Gamma^{2\otimes 1}_{1,2}(0;00) \,.
\end{eqnarray}
\item WTI $N=0, M=3$
\begin{eqnarray}
\label{EwSMNM03}
\HVEV\Gamma^{2\otimes 1;t_1t_2t_3t_4}_{0,4}(;0000) &=& \delta^{t_1t_2}\Gamma_{1,2}^{2\otimes 1;t_3t_4}(0;00)  \nonumber \\
&+& \delta^{t_1t_3}\Gamma_{1,2}^{2\otimes 1,t_2t_4}(0;00)  \\
&+& \delta^{t_1t_4}\Gamma_{1,2}^{2\otimes 1;t_2t_3}(0;00)\,, \nonumber \\
\HVEV\Gamma^{2\otimes 1}_{0,4}(;0000) &=& \Gamma^{2\otimes 1}_{1,2}(0;00)\,.\nonumber
\end{eqnarray}
\item WTI $N=1, M=3$
\begin{eqnarray}
\label{EwSMNM13}
\delta^{t_1t_2}\Gamma^{2\otimes 1;t_3t_4}_{2,2}(00;00) &+& \delta^{t_1t_3}\Gamma^{2\otimes 1;t_2t_4}_{2,2}(00;00) \nonumber \\
&+&\delta^{t_1t_4}\Gamma^{2\otimes 1;t_2t_3}_{2,2}(00;00)  \\
&-& \Gamma^{2\otimes 1;t_1t_2t_3t_4}_{0,4}(;0000) =0 \,, \nonumber \\
\Gamma^{2\otimes 1}_{2,2}(00;00)&=&\Gamma^{2\otimes 1}_{0,4}(;0000)\,. \nonumber
\end{eqnarray}
\item WTI $N=3, M=1$
\begin{eqnarray}
\label{EwSMNM31}
-\delta^{t_1t_2}\Gamma^{2\otimes 1}_{4,0}(0000;)&+&3\Gamma^{2\otimes 1;t_1t_2}_{2,2}(00;00)=0 \,, \\
-\Gamma^{2\otimes 1}_{4,0}(0000;)&+&3\Gamma^{2\otimes 1}_{2,2}(00;00)=0 \,.\nonumber
\end{eqnarray}
\end{itemize}

The quadratic and quartic coupling constants are defined in terms of 2-point and  4-point 1-$\phi$-I connected amputated GF
\begin{eqnarray}
\label{EwSMSchwingerWignerData}
 \Gamma^{2\otimes 1}_{0,4}(;0000)  &\equiv& -2 \lambda_{\phi}^2 \nonumber \\
\end{eqnarray}

The  all-loop-orders renormalized $\phi$-sector momentum-space effective Lagrangian (\ref{EwSMFormSchwingerPotential}) - constrained only by those $SU(2)_L\times U(1)_Y$ WTI governing Greens functions (\ref{EwSMGreensWTIPrime}) - 
may be written (for Wigner mode, the scale-invariant point and Goldstone mode)
\begin{eqnarray}
\label{EwSMLEffectiveSM}
&&L^{Eff;all-modes}_{2\otimes 1;Landau} = L^{Kinetic;Eff;all-modes}_{2\otimes 1;Landau} \nonumber \\
&& \quad \quad \qquad \qquad -V^{Eff;all-modes}_{2\otimes 1;Landau} + {\cal O}_{Ignore}^{2\otimes 1} \nonumber \\
&&L^{Kinetic;Eff;all-modes}_{2\otimes 1;Landau} \nonumber \\
&&\quad \quad \qquad \qquad =\half \Big( \Gamma^{2\otimes 1}_{0,2}(;p,-p) - \Gamma^{2\otimes 1}_{0,2}(;00) \Big) h^2 \nonumber \\
&&\quad \quad \qquad \qquad +\half \Big( \Gamma^{2\otimes 1}_{0,2}(;q,-q)- \Gamma^{2\otimes 1}_{0,2}(;00)  \Big) {\vec \pi}^2 \nonumber \\
&&V^{Eff;all-modes}_{2\otimes 1;Landau}= \mpisq \Big[ \frac{h^2 + {\vec \pi}^2}{2} +\HVEV h \Big] \nonumber \\
&&\quad \quad \qquad \qquad + \lambda_{\phi}^2 \Big[ \frac{h^2 + {\vec \pi}^2}{2} +\HVEV h \Big]^2
\end{eqnarray} 
with
finite non-trivial wavefunction renormalization 
\begin{eqnarray}
\label{EwSMWavefunctionA}
\Gamma^{2\otimes 1}_{0,2}(;q,-q)-\Gamma^{2\otimes 1}_{0,2}(;00) \sim q^2 
\end{eqnarray}{}

As in the $SU(2)_{L}$ model discussed in Section \ref{GreensFunctionsAHM},
the $\phi$-sector effective Lagrangian (\ref{EwSMLEffectiveSM}) 
has insufficient boundary conditions 
to distinguish among the three modes 
	\cite{Lee1970,Symanzik1970a,Symanzik1970b,Vassiliev1970}  
of the BRST-invariant Lagrangian $L_{2\otimes 1}$ in (\ref{EwSMLEffectiveSM}). 
The expression for the effective potential in various modes of the theory, 
but now for the $SU(2)_{L}\otimes U(1)_{Y}$ model, 
can be found in \eqref{WignerSIGoldstonePotentials}.
Eqn. (\ref{EwSMLEffectiveSM}) has exhausted the constraints 
(on the allowed terms in the $\phi$-sector effective Lagrangian) 
due to those $SU(2)_{L-R}$ WTIs 
that govern 1-$\phi$-I $\phi$-sector Green's functions $\Gamma^{2\otimes 1}_{N,M}$ 
(\ref{EwSMGreensWTIPrime}, \ref{GreensFWTI}). 
In order to provide boundary conditions 
that distinguish among the effective potentials in (\ref{WignerSIGoldstonePotentials}) in the $SU(2)_{L}\otimes U(1)_{Y}$ model, 
we must turn to the $SU(2)_{L-R}$ WTIs 
that govern $\phi$-sector 1-$\phi$-R T-Matrix elements $T^{2\otimes 1}_{N,M}$.

\subsection{The Lee-Stora-Symanzik (LSS) Theorem:
IR finiteness
and automatic tadpole renormalization}
\label{SectionLSSTheorem}

In strict obedience to K. Symanzik's edict ``...unless otherwise constrained by Ward Identities...", 
we now further constrain the allowed terms in the $\phi$-sector effective Lagrangian 
with  those $SU(2)_L\times U(1)_Y$ Ward-Takahashi identities 
that govern 1-$\phi$-R T-Matrix elements $T^{2\otimes 1}_{N,M}$.

In Appendix \ref{EwSMTotalMasterEqWTI}, we extend Adler's self-consistency condition  
(originally written for the global 
$SU(2)_L\times SU(2)_R$ \GMLfull Linear Sigma Model with PCAC \cite{Adler1965,AdlerDashen1968}), 
but now derived for the standard electroweak gauge theory in Landau gauge (\ref{EwSMAdlerSelfConsistency})
\begin{eqnarray}
\label{EwSMAdlerSelfConsistencyPrimePrime} 
&&\HVEV T^{2\otimes 1;tt_1...t_M}_{N,M+1}(p_1...p_N;0q_1...q_M) \\
&& \quad \quad \times (2\pi)^4\delta^4 \Big(\sum_{n=1}^N p_n +\sum_{m=1}^M q_m \Big) \Big\vert^{p_1^2 =p_2^2...=p_N^2=m_{BEH}^2}_{q_1^2 =q_2^2...=q_M^2=0} =0 \,.\nonumber
\end{eqnarray}
Eqn. (\ref{EwSMAdlerSelfConsistencyPrimePrime}) 
``asserts the absence of infrared (IR) divergences 
in the scalar-sector (of AHM) Goldstone mode  (in Lorenz gauge). 
Although individual Feynman diagrams are IR divergent, 
those IR divergent parts cancel exactly in each order of perturbation theory. 
Furthermore, the Goldstone mode amplitude must vanish in the soft-pion limit''
(B.W. Lee in \cite{Lee1970}).

It is crucial to note that the external states in $T_{N,M}$ are $N$ $h$'s and $M$ $\pi$'s,
 not $\tilde\pi$'s. We are working in the soft-$\pi$, not the soft-$\tilde\pi$ limit.

The $N=0,M=1$ case of (\ref{EwSMAdlerSelfConsistencyPrimePrime}) 
is the LSS theorem: 
\begin{eqnarray}
\label{EwSMTMatrixGoldstoneTheoremPrime}
\HVEV T^{2\otimes 1;t_1t_2}_{0,2}(;00)=0
\end{eqnarray}
As for \eqref{TMatrixGoldstoneTheoremPrime} in $SU(2)_L$,  
this looks like the Goldstone Theorem,
but, since it involves $\vec \pi$ (not $\tilde{\vec \pi}$), it is quite distinct.

Since the 2-point $T^{2\otimes 1;t_1t_2}_{0,2}$ is already 1-$\phi$-I
(cf. footnote 7), 
we may write the  LSS theorem 
as a further constraint on the 1-$\phi$-I Greens function 
and thus the $\vec \pi$ mass:
\begin{eqnarray}
\label{EwSMGFGoldstoneTheoremPrime}
\Gamma^{2\otimes 1;t_1t_2}_{0,2}\left(;00\right) &&=\delta^{t_1t_2}\Gamma^{2\otimes 1}_{0,2}\left(;00\right) \nonumber \\
\HVEV \Gamma^{2\otimes 1}_{0,2}\left(;00\right) &&=\HVEV \big[ \Delta^{2\otimes 1}_\pi (0) \big]^{-1} \nonumber \\ &&=\HVEV\mpisq = 0.
\end{eqnarray}

Evaluating the effective potential in (\ref{EwSMLEffectiveSM}) with $\HVEV \neq 0$, 
and then in the Kibble representation
\begin{eqnarray}
\label{EwSMPreLSSGoldstoneModePotential}
V^{Eff;pre-LSS-Goldstone}_{2\otimes 1;\phi;Landau} 
&=& \\
\frac{\mpisq}{2} \Big[ {\tilde H}^2 -\HVEV ^2\Big] 
&+&\frac{\lambda_{\phi}^2}{4} \Big[ {\tilde H}^2 -\HVEV ^2\Big]^2\,.\nonumber 
\end{eqnarray}
As expected, the NGB ${\tilde {\vec \pi}}$ have disappeared from the effective potential, 
have purely derivative couplings through their kinetic term, 
and obey the shift symmetry 
\bea
\label{EwSMShiftSymmetry}
{\tilde {\vec \pi}}\to {\tilde {\vec \pi}}+\HVEV {\vec \theta}+{\tilde {\vec \pi}}\times {\vec \theta}+{\cal O}({\vec \theta}^2)
\eea 
 for constant $\vec \theta$.
In other words, the Goldstone theorem is already properly enforced. 

As for the pre-LSS effective potential 
	\eqref{PreLSSGoldstoneModePotential} of $SU(2)_L$,
Eqn. (\ref{EwSMPreLSSGoldstoneModePotential}) 
appears to embrace a disaster: 
the term linear in $\phi^{\dagger} \phi -\half \HVEV ^2$  persists, 
destroying the symmetry of the famous ``Mexican hat",  
and implying that $SU(2)_L\times U(1)_Y$ is not actually in Goldstone mode. 
As for $SU(2)_L$,
the LSS theorem (\ref{EwSMTMatrixGoldstoneTheoremPrime})
(and not the Goldstone theorem) comes to the rescue,
forcing the electroweak gauge theory 
fully into its true Goldstone $\HVEV \neq 0$ mode
\begin{eqnarray}
\label{EwSMGoldstoneModePotential}
V^{Eff;LSS-Goldstone}_{2\otimes 1;\phi;Landau} &&= 
	\frac{\lambda_{\phi}^2}{4} \Big[ {\tilde H}^2-\HVEV^2\Big]^2 \nonumber \\
&&=\lambda_{\phi}^2 \Big[ \phi^{\dagger} \phi -\half \HVEV ^2\Big]^2  \\ &&= \lambda_{\phi}^2  
\left[ {\frac{h^2+{\vec \pi}^2}{2}} +\HVEV h\right] ^2 \nonumber
\end{eqnarray}

A central result of this paper is to recognize that,
in order to force (\ref{EwSMPreLSSGoldstoneModePotential}) 
to become (\ref{EwSMGoldstoneModePotential}),
the LSS theorem incorporates a ``new" on-shell T-Matrix  symmetry,
which is not a full symmetry of the BRST-invariant electroweak Lagrangian.
Electro-weak physics, but not its Lagrangian, 
has the $SU(2)_L\otimes U(1)_Y\otimes BRST$ symmetry of Section \ref{StandardModel} 
and Appendices \ref{EwSMLightParticleCurrents} and \ref{EwSMTotalMasterEqWTI},
a conserved current (\ref{EwSMPhysicalAHMCurrentConservation}), 
un-deformed WTIs governing connected amputated Green's functions 
	(\ref{EwSMGreensWTIPrime}), 
un-deformed WTIs governing connected amputated on-shell T-Matrix elements 
	(\ref{EwSMAdlerSelfConsistencyPrimePrime}), 
and an LSS theorem (\ref{EwSMTMatrixGoldstoneTheoremPrime}).

A crucial effect of the LSS theorem (\ref{EwSMTMatrixGoldstoneTheoremPrime}),
together with the  $N=0, M=1$ $SU(2)_L\times U(1)_Y$ Ward-Takahashi Greens function identity (\ref{EwSMGreensWTIPrime}), 
is to automatically eliminate tadpoles in (\ref{EwSMFormSchwingerPotential})
\begin{eqnarray}
\label{EwSMZeroTadpoles}
\Gamma^{2\otimes 1}_{1,0}(0;) &=& \HVEV \Gamma^{2\otimes 1}_{0,2}(;00) =0
\end{eqnarray}
so that separate tadpole renormalization is un-necessary.

Eq. (\ref{EwSMGoldstoneModePotential}) is the $\phi$-sector effective potential of spontaneously broken $SU(2)_L\times U(1)_Y$ gauge theory, in
Landau gauge, constrained by the LSS theorem%
\footnote{
	Imagine we suspected that $\vec\pi$ is not all-loop-orders massless in Landau 
	gauge in the SSB standard electroweak model, and simply/naively wrote a mass-squared $m_{\pi:Pole}^2$ 
	into the  $\vec\pi$ inverse-propagator
	\begin{eqnarray}
	\label{EwSMGoldstoneTheoremViolation}
 \left[ \Delta^{2\otimes 1}_{\pi}(0)  \right]^{-1}  &\equiv& -\mpisq  \\
&=&-m_{\pi ;Pole}^2\Bigg[ 1+ m_{\pi ;Pole}^2\int dm^2 \frac{\rho^{2\otimes 1}_{\pi}(m^2)}{m^2 } \Bigg]^{-1} \nonumber
	\end{eqnarray}
	However, the LSS theorem (\ref{EwSMGFGoldstoneTheoremPrime}) insists instead that
	\begin{eqnarray}
	\label{EwSMGoldstoneTheoremNGBMass}
	&&\left[ \Delta^{2\otimes 1}_{\pi}(0) \right]^{-1} \equiv -\mpisq =  \Gamma^{2\otimes 1}_{0,2}(;00)=0 \quad \quad
	\end{eqnarray} 
	The $\pi$-pole-mass vanishes exactly.
	\begin{eqnarray}
	\label{PionPoleMass}
	m_{\pi;Pole}^2 &=& \mpisq\Bigg[ 1- \mpisq\int dm^2 \frac{\rho^{2\otimes 1}_{\pi}(m^2)}{m^2 } \Bigg]^{-1} =0 \quad \quad 
	\end{eqnarray}
}.
The effective Lagrangian \eqref{EwSMLEffectiveSM}, 
with effective potential \eqref{EwSMGoldstoneModePotential}:
\begin{itemize}
\item is derived from the local BRST-invariant standard electroweak Lagrangian $L_{2\otimes 1}$ (\ref{LTotalEwSM});
\item includes all divergent
${\cal O}(\Lambda^2),{\cal O}(\ln \Lambda^2)$ and finite terms that arise 
to all perturbative loop-orders 
in the full $SU(2)_L\times U(1)_Y$ gauge theory, due to virtual gauge bosons, $\phi$ scalars, anti-ghosts and ghosts
($W^\pm_\mu,Z_\mu,A_\mu;h,{\vec \pi};{\bar {\vec \eta}},{\bar \eta}_B;{\bar {\vec \omega}},{\bar { \omega}}_B$ respectively);
\item obeys the LSS theorem  
	(\ref{EwSMTMatrixGoldstoneTheoremPrime},\ref{EwSMGFGoldstoneTheoremPrime}) 
and all other $SU(2)_L\times U(1)_Y$ Ward-Takahashi Green's function and T-Matrix identities;
\item obeys the Goldstone theorem in Landau gauge, 
having massless derivatively coupled NGBs, $\tilde{\vec \pi}$.
\item has its vacuum forced to be at $H=\HVEV$ ($\expval{\vec{\pi}}=0$)
by $SU(2)_L \otimes U(1)_Y \otimes BRST$ symmetry, 
obeying stationarity  \cite{ItzyksonZuber} of that true minimum;
\item preserves the theory's renormalizability and unitarity, 
which requires that wavefunction  renormalization
($\HVEV_{Bare}=\big[ Z^{\phi}_{2\otimes 1}\big]^{1/2}\HVEV$ 
	\cite{LSS-2,Bjorken1965,ItzyksonZuber}) 
forbids UVQD, relevant, or any other dimension-2 operator corrections to $\HVEV$.
\end{itemize}
The LSS theorem (\ref{EwSMTMatrixGoldstoneTheoremPrime}) has once again
caused all relevant operators to vanish!

The  Green's function Ward-Takahashi ID (\ref{EwSMGreensWTIPrime}) for $N=1,M=1$, 
constrained by the LSS theorem (\ref{EwSMGFGoldstoneTheoremPrime}), 
relates the BEH mass  to the coefficient of the $h{\vec \pi}^2$ vertex 
\begin{eqnarray}
\label{EwSMWTIHiggsMass}
\Gamma^{2\otimes 1}_{2,0}(00;)=\HVEV\Gamma^{2\otimes 1}_{1,2}(0;00) \,.
\end{eqnarray} 
Therefore, the BEH mass-squared
\begin{eqnarray}
\label{EwSMBEHMassAHMS}
m_{BEH}^2 = 2\lambda_{\phi}^2\HVEV^2
\end{eqnarray} 
arises entirely from SSB, 
as does (together with its electroweak decays) 
the resonance pole-mass-squared
\begin{eqnarray}
\label{EwSMBEHPoleMassPrime}
m^2_{BEH;Pole}  &=& 2\lambda_\phi^2 \HVEV^2\Big[ 1- 2\lambda_\phi^2 \HVEV^2 
	\int dm^2 \frac{\rho_{BEH}^{2\otimes 1}(m^2)}{m^2 - i\epsilon} \Big]^{-1} \nonumber \\
	&+& {\cal O}^{Ignore}_{2\otimes 1;\phi}.
\end{eqnarray}

\subsection{Comparison of results with the minimization procedure}

It's important to compare the results of our LSS theorem to those of the mainstream literature. For pedagogical simplicity, in this Subsubsection we suppress mention of vacuum energy and ${\cal O}^{Ignore}_{2\otimes 1;\phi}$. After renormalization, but before application of the LSS theorem, the effective potential \eqref{EwSMLEffectiveSM}, which is  derived entirely from Green's function WTIs, can be written 
\bea
\label{EwSMMainstreamPotential}
V_{2\otimes 1;\phi}^{Eff} &&=\Big( \mu_{\phi}^2+\lambda^2_\phi \HVEV^2 \Big)\Big(\frac{h^2+{\vec \pi}^2}{2}+\HVEV h\Big)\nonumber \\
&&+\lambda_{\phi}^2\Big(\frac{h^2+{\vec \pi}^2}{2}+\HVEV h\Big)^2 
\eea
where $V_{2\otimes 1;\phi}^{Eff} ,\phi,\mu_{\phi}^2,\lambda_{\phi}^2,\HVEV^2$ in (\ref{EwSMMainstreamPotential}) are all renormalized quantities. The vanishing of relevant operators in the effective $SU(2)_L\times U(1)_Y$ electroweak theory is therefore not itself controversial:

The literature minimizes (\ref{EwSMMainstreamPotential}) to find the vacuum:
\bea
\label{EwSMMinimizeVacuum}
&&\frac{\partial}{\partial h}V_{2\otimes 1;\phi}^{Eff}\Big\vert_{h=\vec{\pi}=0}  =\HVEV\Big( \mu_{\phi}^2+\lambda^2_\phi \HVEV^2 \Big) =0\,.
\eea
This is interpreted as a calculation of $\HVEV^2$
\bea
\label{EwSMMinimizeGoldstoneVacuumHiggsVEV}
 \HVEV^2 =-\frac{\mu_{\phi}^2}{\lambda^2_\phi}\,,
\eea
where, in renormalized $\mu^2_{\phi}$, 
UVQD and all other relevant contributions 
are regarded as having cancelled against a bare counter-term $\delta\mu^2_{\phi;Bare}$.

In contrast, in this paper we have never minimized a potential. 
Instead, we have derived a tower of Adler self-consistency conditions 
(\ref{EwSMAdlerSelfConsistencyPrimePrime},\ref{EwSMAdlerSelfConsistency}) 
in Landau gauge in Appendix \ref{EwSMTotalMasterEqWTI}, 
derived directly from the exact $SU(2)_L\times U(1)_Y$ symmetry obeyed by gauge-independent on-shell T-Matrix elements. 
One of these, the $N=0,M=1$ case, is the LSS theorem:
\bea
\label{EwSMLSSInsteadOfMinimize}
\HVEV\mpisq=\HVEV \Big( \mu_{\phi}^2+\lambda^2_\phi \HVEV^2 \Big) =0
\eea
which, for the SSB case, gives
\bea
\label{EwSMLSSInsteadOfMinimizeSSB}
\mpisq =\mu_{\phi}^2+\lambda^2_\phi \HVEV^2 =0.
\eea
The practical effect of \eqref{EwSMLSSInsteadOfMinimize}
is the same as minimization of the effective potential.
We therefore agree with the mainstream literature 
that all relevant operators vanish 
in the effective low-energy $SU(2)_L\times U(1)_Y$ theory,
but there is one vast difference. 
Although the mainstream minimization approach 
regards relevant operators as having cancelled, 
that cancellation is not protected by any symmetry. 
In contrast, relevant operators here vanish 
because of the exact $SU(2)_L\times U(1)_Y$ symmetry, 
represented by the LSS theorem, 
remnant in gauge-independent on-shell T-Matrix elements. 
Although $SU(2)_L\times U(1)_Y$ is not a symmetry of the BRST-invariant Lagrangian, 
it is exact for the physics of our new $SU(2)_L\otimes U(1)_Y\otimes BRST$ symmetry.

\section{Extension to 1-generation CP-conserving Standard Model with Dirac Neutrino Mass}
\label{nuDSM}

We finally extend the theory next to include fermion matter fields
-- the 3rd generation of Standard Model quarks and leptons, 
together with a right-handed $\tau$ neutrino with Dirac mass $m_{\nu_\tau}$. 
This conserves $CP$.

We also gauge $SU(3)_C$, 
adding to our model 8 QCD gluons $G_\mu^A$, 
so that the local/gauge group is $SU(3)_{C}\otimes SU(2)_{L} \otimes U(1)_{Y}$.  
As usual, the strong $CP$ problem arises in a term
$L_\theta = -\frac{\theta}{64\pi^2}\epsilon^{\mu\nu\lambda\sigma}G^A_{\mu\nu}G^A_{\lambda\sigma}$ 
induced by instantons, 
with the $CP$-violating parameter $\theta$. 
We artificially  set $\theta =0$ in order to conserve $CP$,
as $CP$ conservation is essential for our proofs of WTI 
(see eq.s (\ref{AxialCurrentMEven},\ref{VectorCurrentMOdd} and
	\ref{AxialVectorMasterEquation}). 

We call this model $\nu_{D}SM_{CP}$.
$\nu_{D}SM_{CP}$ then obeys various Ward-Takahashi identities.
As proved in \cite{LSS-2}, 
the form of the tower of $SU(2)_{L-R}$ Green's function WTI 
and that of the on-shell $SU(2)_{L-R}$ T-Matrix Adler self-consistency relations, 
are un-affected by the addition of 
quarks ($q^{c \dagger}_{3;L}= (t^{c \dagger},b^{c \dagger})_L, t^c_R, b^c_R$,
 with c the color quantum number)
 and leptons 
 ($l_{3;L}^\dagger=(\nu_{\tau}^\dagger,\tau^\dagger)_L, \nu_{\tau;R}, \tau_R$).
Neither are the form of 
the tower of $SU(2)_{L-R}$ Green's function WTI \eqref{EwSMGreensWTIPrime} 
and that of the on-shell $SU(2)_{L-R}$ T-Matrix 
	Adler self-consistency relations \eqref{EwSMAdlerSelfConsistencyPrimePrime}, 
derived from soft-pions, affected by the addition of gluons $G_\mu^A$.
Meanwhile, vector $SU(3)_C$ and $U(1)_{QED}$ WTI add nothing to our discussion of axial-vector soft-$\vec \pi$ theorems, 
just as conserved vector $SU(2)_{L+R}$ in (\ref{VectorCurrentMOdd}) 
adds nothing to the discussion of axial-vector soft-pions. 

The $CP$-conserving $SU(3)_C\otimes SU(2)_L \otimes U(1)_Y$ invariant Lagrangian (summed over color) studied in this section is then
\begin{eqnarray}
	\label{nuDSMLagrangian}
	L_{\nu_{D}SM_{CP}} &=& L_{2\otimes 1} + L_{QCD} + L_{l} + L_{Yuk} + L_{\nu_D} \nonumber \\ 
	L_{QCD} &=& -\frac{1}{2} \Tr (G^A_{\mu \nu}G^{A;\mu\nu}) + i\bar{q}^{c}_{3;L}\slashed{D}q_{3;L}^{c} \nonumber \\ &+& i\bar{b}^{c}_{R}\slashed{D}b_{R}^{c} + i\bar{t}^{c}_{R}\slashed{D}t_{R}^{c}  \\ 
	L_{l} &=& i\bar{l}_{3;L}\slashed{D}l_{3;L} + i\bar{\tau}_{R}\slashed{D}\tau_{R} + i\bar{\nu_\tau}_{R}\slashed{D}\nu_{\tau R} \nonumber 
	\\L_{Yuk} &=& y_{\tau}\bar{l}_{3;L}(i \sigma_{2}\phi^{*})\tau_R + y_{\tau}^{*}\bar{\tau}_{R}\phi^{T}(-i\sigma_{2})l_{3;L} \nonumber  
	\\ &+& y_b \bar{q}^{c}_{3;L}\phi b^c_R + y^{*}_b \bar{b}^{c}_{R}\phi^{\dagger} q^c_{3;L} \nonumber 
	\\ &+& y_t\bar{q}^{c}_{3;L}(i\sigma_{2}\phi^{*})t^{c}_R + y^{*}_t\bar{t}^{c}_{R}\phi^{T}(-i\sigma_{2})q^{c}_{3;L} \nonumber \\
	L_{\nu_D} &=& -y_{\nu}\bar{l}_{3;L}\phi \nu_{\tau;R}  -y^{*}_{\nu}\bar{\nu}_{\tau;R}\phi^{\dagger} l_{3;L}\,.\nonumber
\end{eqnarray}

In Landau gauge, the $SU(2)_{L}$ sub-current is
\begin{eqnarray}
\label{nuDSMLCurrent}
\vec{J}^{\mu}_{L,v_{D}SM_{CP}} &=& \\
&&\!\!\!\!\!\!\!\!\!\!
\vec{J}^{\mu}_{L+R,v_{D}SM_{CP}} + \vec{J}^{\mu}_{L-R,v_{D}SM_{CP}} + \vec{J}^{\mu}_{L,2\otimes 1} \,,\nonumber
\end{eqnarray}
with fermion contributions
\begin{eqnarray} 
2\vec{J}^{\mu}_{L+R,v_{D}SM_{CP}} &=& \sum_{i}  {\bar l}_{3;i}    \gamma^{\mu}\vec{t} l_{3;i} +  \sum_{i,c}  {\bar q}^c_{3;i}  \gamma^{\mu}\vec{t}q_{3;i}^{c} 
\\ 2\vec{J}^{\mu}_{L-R,v_{D}SM_{CP}} &=& \sum_{i}  {\bar l}_{3;i}\gamma^{\mu}\gamma^5\vec{t} l_{3;i} +  \sum_{i,c}    {\bar q}^c_{3;i}  \gamma^{\mu}\gamma^5\vec{t}q_{3;i}^{c}
\,. \nonumber
\end{eqnarray}
(Here colors $c=r,w,b$, quark flavors  $q_{3;i}^{c}=t^c,b^c$, 
and lepton flavors $l_{3;i}=\nu_{\tau}, \tau$.)
$\vec{J}^{\mu}_{L,2\otimes 1}$ is given in \eqref{EwSMIsospinCurrentPrime}.

Since $CP$ is conserved, the fermionic parts of the current are separately conserved for matrix elements
\begin{eqnarray}
\label{nuDSMLCurrentDivergence}
\left\langle\partial_{\mu} \vec{J}^{\mu}_{L+R,v_{D}SM_{CP}}\right\rangle &=& 0 \nonumber \\
\left\langle\partial_{\mu} \vec{J}^{\mu}_{L-R,v_{D}SM_{CP}}\right\rangle&=& 0
\end{eqnarray}
in the sense of (\ref{AxialCurrentMEven},\ref{VectorCurrentMOdd}). 
We focus on the global isospin current $\vec{J}^{\mu}_{L,v_{D}SM_{CP}}$ 
and examine the divergence of amplitudes such as
\bea
\label{nuDSMConnectedAmplitude}
&&\partial_\mu  \big< 0\vert 
T\Big[J_{L;\nu_D SM_{CP} }^{\mu;t}(z) \\
&&\quad h(x_1)\cdots h(x_N)\pi^{t_1}(y_1)\cdots \pi^{t_M}(y_M)\Big]
		\vert 0 \big>_{Connected}\,. \nonumber
\eea
We now write down the WTI for the $\nu_{D}SM_{CP}$ model, one concerning on-shell T-Matrix elements
\begin{eqnarray}
\label{AdlerSCnuDSM}
&&\expval{H}T^{\nu_D SM_{CP};tt_1 t_2\ldots t_M}_{N,M+1}(p_1 \cdots p_N;0q_1 \ldots q_M)  \\ 
&&\times(2\pi)^4 \delta^{4}\Big(\sum\limits_{n=1}^N p_{n} + \sum\limits_{m=1}^M q_{m}\Big)\Big|_{q_1^2 = q_2^2 \ldots = q_M^2 = 0}^{p_1^2 = p_2^2 \ldots = p_M^2 = m_{BEH}^2} =0\,, \nonumber
\end{eqnarray}
and the other relating 1-$\phi$-I 
(but one $\vec{W}^{\mu}, B^{\mu}, G_A^{\mu}$ reducible) 
connected, amputated Greens functions $\Gamma_{N,M}^{\nu_{D}SM_{CP}}$:
\begin{eqnarray}
\label{WTInuDSM}
&\expval{H}&\Gamma_{N,M+1}^{\nu_{D}SM_{CP};t_1 \ldots t_M t}(p_1\cdots p_{N}; q_1\cdots q_M0)
\\ &=&  \sum\limits_{m=1}^{M} \delta^{t t_m}\Gamma_{N+1,M-1}^{\nu_{D}SM_{CP};t_1 \ldots \widehat{t_m} \ldots t_M t} \nonumber \\
&& \qquad\qquad (p_1\cdots p_{N}q_m; q_1 \cdots \widehat{q_m} \cdots q_M) \nonumber
\\&-&\sum\limits_{n=1}^{N}\Gamma_{N-1,M+1}^{\nu_{D}SM_{CP};t_1 \ldots t_M t}(p_1 \cdots \widehat{p_n}  \cdots p_{N};q_1 \cdots q_M p_n) \,.\nonumber
\end{eqnarray}
These WTI are mathematically identical 
to the ones in (\ref{EwSMGreensWTIPrime},\ref{EwSMAdlerSelfConsistencyPrimePrime}), 
but now include all fermion and boson loops, 
to all perturbative loop-orders, 
from virtual
$q^{c}_L$, $t^c_R$, $b^c_R$, $l_{3;L}$, $\nu_{\tau;R}$, $\tau_R$,
$G_\mu^A$, ${\vec W}_\mu$, $B_\mu$, $h$, and ${\vec \pi}$, 
and the ghosts and anti-ghosts associated with the gauge bosons.

The LSS theorem for $\nu_D SM$ is the $N=0,M=1$ case of \eqref{AdlerSCnuDSM}
\begin{eqnarray}
\label{nuDSMLSS}
\expval{H}T_{0,2}^{\nu_D SM_{CP}}(;00) = 0\,,
\end{eqnarray}
forcing $m_{\pi}^{2}=0$, 
and this is the $SU(3)_{C}\otimes SU(2)_{L}\otimes U(1)_{Y}$ analogue 
of \eqref{EwSMTMatrixGoldstoneTheoremPrime}.

The WTI \eqref{WTInuDSM} and \eqref{nuDSMLSS} severely constrain the effective Lagrangian. 
Considering only operators with $N+M \leq 4$ in \eqref{WTInuDSM}, 
and using the LSS theorem \eqref{nuDSMLSS}, 
the scalar-sector effective potential becomes
\begin{eqnarray}
V^{Eff;Goldsone}_{\nu_D SM_{CP};Landau} 
= \lambda_{\phi}^{2}\Big[\frac{h^2 + \pi^2}{2} + \expval{H}h \Big]^2.
\end{eqnarray}
As the WTI \eqref{WTInuDSM} and the LSS theorem  \eqref{nuDSMLSS} are mathematically the same as their analogues in $SU(2)_L \otimes U(1)_Y$ theory, 
all the results and conclusions obtained there apply to 
the $CP$-conserving $\nu_D SM_{CP}$ model.

\section{OG, BWL and GDS: This research, viewed through the prism of mathematical rigor demanded by Raymond Stora}
\label{Stora}
Raymond Stora regarded vintage QFT as incomplete,
fuzzy in its definitions, and primitive in technology. For
example, he worried about whether the off-shell T-matrix
could be mathematically rigorously defined to exist in
Lorenz gauge: e.g. without running into some IR subtlety.
The Adler self-consistency conditions proved here guarantee
the IR finiteness of the scalar-sector on-shell T-matrix.

Raymond contributed hugely to this work, especially Section \ref{StoraSymmetry}. His insight formed part of the foundation of this paper. Because of his contribution, his intentions were to be a co-author on this paper but he passed away before this manuscript reached is final form. He was a perfectionist, as his son Dr. Thierry Stora explained \lq \lq However, he was always very critical in accepting to publish scientific results, which often took years before he accepted that they would be submitted to a peer reviewed journal."

Although he agreed on the correctness of the results
presented here, Raymond might complain that we fall short
of a strict mathematically rigorous proof (according to his
exacting mathematical standards). He reminded us that
much has been learned about quantum field theory, via
modern path integrals, in the recent $\sim$45 years. In the time
up to his passing, he was intent on improving this work by
focusing on the following three issues:

\begin{itemize}
\item{properly defining and proving the Lorenz-gauge
results presented here with modern path integrals;}
\item{tracking our central results directly to SSB, via
BRST methods, in an arbitrary manifestly IR finite
't Hooft $R_{\xi}$ gauge, i.e. proving to his satisfaction that
they are not an artifact of Lorenz gauge;}
\end{itemize}
Any errors, wrong-headedness, misunderstanding, or misrepresentation
appearing in this paper are solely our fault.

\medskip
\section{Conclusion}
\label{Conclusions}
$SU(2)_L$, $SU(2)_L\times U(1)_Y$ and $\nu_D SM_{CP}$ physics 
(e.g. on-shell T-Matrix elements) 
have more symmetry than their BRST-invariant Lagrangians, including
\begin{itemize}
\item a new tower of  rigid SSB $SU(2)_{L-R}$ 1-soft-$\pi$ WTIs 
governing relations among Green's functions;
\item a new tower of rigid SSB $SU(2)_{L-R}$  WTIs 
that forces 1-soft-$\pi$ on-shell T-Matrix elements to vanish,
and represents the on-shell behavior of their newly identified 
$SU(2)_{L}\otimes$BRST, $SU(2)_{L}\otimes U(1)_Y\otimes$BRST 
or $SU(3)_c\otimes SU(2)_{L}\otimes U(1)_Y\otimes$BRST
symmetry, respectively.  
\item A new Lee-Stora-Symanzik theorem. 
\end{itemize}

Our results are ubiquitous to gauge theories 
that spontaneously break a certain Lie algebraic structure group, $\cal G$,
at a low scale, say for definiteness $m_{Weak}$. 
We have proved global ${\cal G}\otimes$BRST symmetry and motivated: 
its 2 associated towers of 1-soft-pion WTI
(one for Green's functions, and another for on-shell T-Matrix elements, 
including the LSS theorem);
severe WTI constraints on the $m_{Weak}$-scalars' effective potential.

Such $SU(2)_{L}\otimes U(1)_Y \otimes \text{BRST}$ symmetry 
may be the reason why the Standard Model 
viewed as an effective low-energy weak-scale theory, 
and augmented by classical General Relativity and neutrino mixing,
is the most experimentally and observationally successful and accurate 
known theory of Nature.
That ``Core  Theory" \cite{WilcekCoreTheory} 
has no known experimental or observational counter-examples.

\section{Acknowledgments}
\label{Acknowledgements}
BWL thanks: 
Jon Butterworth and the Department of Physics and Astronomy at University College London for support as a UCL Honorary Senior Research Associate;
Chris Pope, the George P. and Cynthia W. Mitchell Center for Fundamental Physics and Astronomy, 
and Texas A\&M University,  
for support and hospitality during the  2010-2011 academic year, when this work began.
GDS thanks the CERN Theory group for their support and hospitality in 2012-13.
OG and GDS are partially supported by CWRU grant DOE-SC0009946.  

\bibliography{SU2L-Proof_16Feb2018_arxiv_v2.bib}
\bibliographystyle{unsrt}

\appendix

\section{Landau-gauge reduction of $SU(2)_L$  Ward-Takahashi identities to the $SU(2)_{L-R}$ axial-vector WTI governing ${\vec J}_{L-R;Schwinger}^\mu$}
\label{GlnuDSMWTIs}

We focus on the global isospin current ${\vec J}^{\mu}_{L}$. 
We are interested in global-symmetric relations 
among 1-scalar-particle-irreducible (1-$\phi$-I) connected amputated Green's functions (GF),
and 1-scalar-particle-Reducible (1-$\phi$-R) connected amputated T-Matrix elements, 
with external $\phi$ scalars, 
Analysis is done in terms of the exact renormalized interacting $SU(2)_L$ fields, 
which asymptotically become the in/out states, i.e. free fields for physical S-Matrix elements.

We form time-ordered amplitudes of
products of ${\vec  J}^\mu_{L}$, 
with N scalars (coordinates x, momenta p), 
and M pseudo-scalars (coordinates y, momenta q, isospin t):
$\big< 0 \vert T\Big[ 
{\vec  J}_{L}^\mu(z)h(x_1)\cdots h(x_N) 
						\pi^{t_1}(y_1)\cdots \pi^{t_M}(y_M)
\Big]\vert 0\big>$.
Here $h=H-\HVEV$ and ${\vec \pi}$ are all-loop-orders renormalized fields, 
normalized so that
$\left\langle0\vert\pi^i(0)\vert\pi^j\right\rangle=\delta^{ij}$.

In order to form a Master Equation, whose solutions are the $SU(2)_L$ WTI, 
we examine the divergence of such connected amplitudes:
\bea
\label{SMMasterEquation}
&&\partial_\mu  \big< 0\vert 
T\Big[\Big(J_{L}^{\mu;t}(z) \Big)\\
&&\quad \times h(x_1)\cdots h(x_N)\pi^{t_1}(y_1)\cdots \pi^{t_M}(y_M)\Big]
		\vert 0 \big>_{Connected}\,. \nonumber
\eea

\subsection{Right-hand side of Master Equation}
\label{RHS}

Making use of  current conservation \eqref{SMIsospinCurrent},
\bea
\label{CurrentConservationPrime}
\partial_{\mu} {\vec J}^{\mu}_{L}&=&\half M_W\Big[ H \partial_\beta{\vec W}^\beta+ {\vec \pi} \times \partial_\beta{\vec W}^\beta \Big] ,
\eea
G. 't Hooft's gauge-fixing \cite{tHooft1971}  condition (\ref{GaugeConditionsPrimePrime})
\begin{eqnarray}
\label{GaugeConditionsPrimePrimePrime}
&&\big< 0\vert T\Big[ \Big( \partial_{\mu}{\vec W}^{\mu}(z)\Big)  \\
&&\quad \times h(x_1)...h(x_N)\pi_{t_1}(y_1)...\pi_{t_M}(y_M)\Big]\vert 0\big>_{\rm connected}  =0 \,, \nonumber
\end{eqnarray}
and the equal-time commutation relations (\ref{SMIsospinCurrent})
\bea
\delta(z_0-x_0) \left[\Big( {\vec J}^0_{L}-{\vec J}^0_{Schwinger} \Big)(z)  ,h(x)\right] &=& 0 \\
\delta(z_0-x_0) \left[\Big( {\vec J}^0_{L}-{\vec J}^0_{Schwinger} \Big)(z)  ,{\vec \pi}(x)\right] &=& 0\,, \quad \quad\nonumber
\eea
a short calculation reveals
\bea
\label{MasterRHS1}
&&\partial_\mu  \big< 0\vert 
	T\Big[J_{L}^{\mu;t}(z) h(x_1)\cdots h(x_N) \nonumber\\
		&&\quad\quad\quad\quad\quad\quad \times \pi^{t_1}(y_1)\cdots \pi^{t_M}(y_M)\Big]
		\vert 0 \big>_{Connected} \nonumber \\
&&= \big< 0\vert 
	T\Big[ \Big(\partial_\mu J_{L}^{\mu;t}(z)\Big)  \\
&&\quad\quad\quad\times h(x_1)\cdots h(x_N) \pi^{t_1}(y_1)\cdots \pi^{t_M}(y_M)\Big]
		\vert 0 \big>_{Connected}  \nonumber\\
&& \quad+ \sum_{n=1}^{N} \big< 0\vert 
	T\Big[ h(x_1)\cdots h(x_{n-1})\delta(z^0-x_n^0)\Big[ J_{L}^{0;t}(z),h(x_n)\Big] \nonumber \\
&&\quad\quad\quad\times h(x_{n+1})\cdots h(x_N) \pi^{t_1}(y_1)\cdots \pi^{t_M}(y_M)\Big]
		\vert 0 \big>_{Connected}  \nonumber\\
&& \quad+ \sum_{m=1}^{M} \big< 0\vert 
	T\Big[ h(x_1)\cdots h(x_N)\pi^{t_1}(y_1)\cdots \pi^{t_{m-1}}(y_{m-1})\nonumber \\
&& \quad\quad\quad\times \delta(z^0-y_m^0)\Big[ J_{L}^{0;t}(z),\pi^{t_m}(y_m)\Big]\nonumber\\
&& \quad\quad\quad\times \pi^{t_{m+1}}(y_{m+1})\cdots \pi^{t_M}(y_M)\Big]
		\vert 0 \big>_{Connected} \,.\nonumber\\
&& = \sum_{n=1}^{N} \big< 0\vert 
	T\Big[ h(x_1)\cdots h(x_{n-1})
	\delta(z^0-x_n^0)\Big[ J_{Schwinger}^{0;t}(z),h(x_n)\Big] \nonumber \\
&&\quad\quad\quad\times h(x_{n+1})\cdots h(x_N) \pi^{t_1}(y_1)\cdots \pi^{t_M}(y_M)\Big]
		\vert 0 \big>_{Connected}  \nonumber\\
&& \quad+ \sum_{m=1}^{M} \big< 0\vert 
	T\Big[ h(x_1)\cdots h(x_N)\pi^{t_1}(y_1)\cdots \pi^{t_{m-1}}(y_{m-1})\nonumber \\
&& \quad\quad\quad\times \delta(z^0-y_m^0)\Big[ J_{Schwinger}^{0;t}(z),\pi^{t_m}(y_m)\Big] \nonumber \\
&&\quad\quad\quad\times \pi^{t_{m+1}}(y_{m+1})\cdots \pi^{t_M}(y_M)\Big]
		\vert 0 \big>_{Connected} \nonumber
\eea
in Landau gauge.
Here we have N external renormalized scalars $h=H-\HVEV$ (coordinates x), 
and M external ($CP=-1$) renormalized pseudo-scalars ${\vec \pi}$ (coordinates y, isospin $t$). 
We have also thrown away a sum of $M$ terms, proportional to $\HVEV$,
that corresponds entirely to disconnected graphs.

\subsection{Left-hand side of Master Equation: Surface terms}
\label{SurfaceLHS}

We begin by studying the surface integral of 
the global $SU(2)_L$ current \eqref{AHMCurrentPrime} in Landau gauge 
\begin{eqnarray}
\label{SMIsospinCurrentPrime}
{\vec J}^{\mu}_{L} &=& {\vec J}^{\mu}_{L;Schwinger} + {\vec {\cal J}}^{\mu}_{L}  \\
{\vec {\cal J}}^{\mu}_{L} &=&-\frac{1}{4} g_2 {\vec W}^{\mu}\left[ H^2+{\vec \pi}^2\right] +  {\vec W}^{\mu \nu} \times {\vec W}_{\nu} \nonumber \\
&-& \lim_{\xi \to 0} \frac{1}{\xi}\left[{\vec W}^{\mu} \times \partial_\beta{\vec W}^{\beta} \right] - \partial^\mu {\bar {\vec \eta}}\times {\vec \omega} \nonumber
\end{eqnarray}

In order to transform (\ref{SMMasterEquation}) into a surface term (and crucially,  to later form 1-soft-pion $SU(2)_L$ WTI), we Fourier-transform the current with far-infra-red ultra-soft momentum.
We use Stokes theorem, where $ {\widehat {z}_\mu}^{3-surface}$ is a unit vector normal to the $3$-surface. The time-ordered product constrains the $3$-surface to lie on, or inside, the light-cone. 

In this subsection, we will prove that
\begin{eqnarray}
\label{SMSurfacePionPoleDominance}
0&=&\lim_{k_\lambda \to 0} \int d^4z e^{ikz} \partial_{\mu} \Big< 0\vert T\Big[ \Big( 
{\vec {\cal J}}^{\mu}_{L} (z)  \Big) \nonumber \\
&&\quad \quad \times h(x_1)...h(x_N) \pi^{t_1}(y_1)...\pi^{t_M}(y_M)\Big]\vert 0\Big>_{Connected} \nonumber \\
&& =\int d^4z \partial_{\mu} \Big< 0\vert T\Big[ \Big( 
{\vec {\cal J}}^{\mu}_{L} (z)  \Big)   \\
&&\quad \quad \times h(x_1)...h(x_N) \pi^{t_1}(y_1)...\pi^{t_M}(y_M)\Big]\vert 0\Big>_{Connected} \nonumber \\
&& =\int_{3-surface\to\infty}
\!\!\!\!\!\!\!\!\!\!\!\!\!\!\!\!\!\!\!\!\!\!\!\!\!\!
d^3z \quad {\widehat {z}_\mu}^{3-surface} \Big< 0\vert T\Big[ \Big( 
{\vec {\cal J}}^{\mu}_{L} (z)  \Big)  \nonumber \\
&&\quad \quad \times h(x_1)...h(x_N) \pi^{t_1}(y_1)...\pi^{t_M}(y_M)\Big]\vert 0\Big>_{Connected}\,. \nonumber 
\end{eqnarray}

\subsubsection{$SU(2)_L$ gauge fields' kinetic and interaction}
We ignore the surface integral of the appropriate term in (\ref{SMIsospinCurrentPrime})
\begin{eqnarray}
\label{SMGaugesSurface}
0&=&
\lim_{k_\lambda \to 0} \int d^4z e^{ikz}\partial_{\mu} 
\Big< 0\vert T\Big[ \Big(  {\vec W}^{\mu \nu}\times {\vec W}_\nu\Big)(z) \\
&&\quad \quad \times h(x_1)...h(x_N) \pi^{t_1}(y_1)...\pi^{t_M}(y_M)\Big]\vert 0\Big>_{Connected} \nonumber \\
&& \quad =\int_{3-surface\to\infty}  
\!\!\!\!\!\!\!\!\!\!\!\!\!\!\!\!\!\!\!\!\!\!\!\!\!\!
d^3z \quad {\widehat {z}_\mu}^{3-surface} \Big< 0\vert T\Big[ \Big(  {\vec W}^{\mu \nu}\times {\vec W}_\nu \Big)(z)  \nonumber \\
&&\quad \quad \times h(x_1)...h(x_N) \pi^{t_1}(y_1)...\pi^{t_M}(y_M)\Big]\vert 0\Big>_{Connected} \nonumber 
\end{eqnarray}
because each term in the current contains at least one massive ${\vec W}^\beta$
\begin{eqnarray}
\label{KineticCurrentGaugeFields}
 {\vec W}^{\mu \nu}\times {\vec W}_\nu
 &=& 
-{\vec W}_\nu \times \partial^\mu{\vec W}^{ \nu}+{\vec W}_\nu \times \partial^\nu{\vec W}^{ \mu} \nonumber \\
&+&g_2\Big[ {\vec W}^\mu \big({\vec W}^{ \nu}\cdot {\vec W}_{ \nu}\big)-{\vec W}^\nu \big({\vec W}^{ \mu}\cdot {\vec W}_{ \nu}\big)\Big]\,.
\nonumber
\end{eqnarray}
${\vec W}^\beta$ are massive  in spontaneously broken $SU(2)_L$. 
Propagators connecting ${\vec W}^\beta$ 
from points on $z^{3-surface}\to \infty$ 
to the localized interaction points $(x_1...x_N;y_1...y_M)$ 
must stay inside the light-cone, 
but die off exponentially with mass $M_W^2\neq 0$
and are incapable of carrying information to the 3-surface at infinity. 

\subsubsection{Ghosts and anti-ghosts}
\label{GhostSurfaceTerm}
The surface integral of the appropriate term in (\ref{SMIsospinCurrentPrime})
\bea
\label{GhostSurfaceTerms}
&&\int_{3-surface\to\infty}  
\!\!\!\!\!\!\!\!\!\!\!\!\!\!\!\!\!\!\!\!\!\!\!\!\!\!
d^3z \quad {\widehat {z}_\mu}^{3-surface} 
	\Big< 0\vert T\Big[ \Big(\partial^\mu {\bar {\vec \eta}}\times {\vec \omega} \Big)(z)\\
&&\quad \quad \times h(x_1)...h(x_N) \pi^{t_1}(y_1)...\pi^{t_M}(y_M)\Big]\vert 0\Big>_{Connected} 
\!\!\!=0 \nonumber 
\eea
because ghosts and anti-ghosts are not physical external asymptotic in-states and out-states for the S-Matrix.

\subsection{LHS of Master Equation: Connected amplitudes linking the $\phi$-sector with external currents}
\label{AmplitudesLHS}

Connected momentum-space amplitudes, 
with $N$ external BEHs, $M$ external $\vec \pi$s,
and a current ${\vec J}_L^\mu$, 
are defined in terms of $\phi$-sector connected time-ordered products
\begin{eqnarray}
\label{CurrentConnectedAmplitudes} 
&&i{{\cal G}}_{\mu;N,M}^{t;t_1...t_M}
\Big( { J}^{t}_{L;\beta};p_1...p_N;q_1...q_M \Big) \nonumber \\
&&\qquad\quad\times (2\pi)^4\delta^4 \Big(k+\sum_{n=1}^N p_n +\sum_{m=1}^M q_m \Big)  \\
&& \quad \equiv\int d^4z e^{ikz} \prod_{n=1}^N\int d^4x_n e^{ip_nx_n} \prod_{m=1}^M\int d^4y_m e^{iq_my_m}  \nonumber \\
&&\qquad\quad \times 
\big< 0\vert T\Big[ \Big( { J}^{t}_{L;\mu} (z)\Big)  \nonumber  \\
&&\qquad \quad \times h(x_1)...h(x_N)\pi^{t_1}(y_1)...\pi^{t_M}(y_M)\Big]\vert 0\big>_{Connected}\nonumber
\end{eqnarray}
These appear throughout the proof of the WTI\footnote{
	Our $i{{\cal G}}_{\mu;N,M}^{t;t_1...t_M}$ for the $SU(2)_L$ gauge theory 
	is in direct analogy with B.W. Lee's $i{{ G}}_{\mu}^{t;t_1...t_M}$ 
	for global $SU(2)_L\times SU(2)_R$ \cite{Lee1970}.
},  
so that
\begin{eqnarray}
\label{CurrentDivergenceConnectedAmplitudes} 
&&-k^\mu{{\cal G}}_{\mu;N,M}^{t;t_1...t_M}\Big({ J}^{t}_{L;\beta};p_1...p_N;q_1...q_M\Big) \nonumber \\
&&\qquad\quad\times (2\pi)^4\delta^4 \Big(k+\sum_{n=1}^N p_n +\sum_{m=1}^M q_m \Big)  \\
&& \quad \equiv\int d^4z e^{ikz} \prod_{n=1}^N\int d^4x_n e^{ip_nx_n} \prod_{m=1}^M\int d^4y_m e^{iq_my_m}  \nonumber \\
&&\qquad\quad \times 
\partial_\mu^z 
\big< 0\vert T\Big[ \Big( { J}^{t}_{L;\mu} (z)\Big)  \nonumber  \\
&&\qquad \quad \times h(x_1)...h(x_N)\pi^{t_1}(y_1)...\pi^{t_M}(y_M)\Big]\vert 0\big>_{Connected}\nonumber
\end{eqnarray}

\subsubsection{Gauge-invariant scalar-sector Lagrangian}
\label{ScalarInvariantLagrangian}

The contribution of the appropriate term in (\ref{SMIsospinCurrentPrime}) 
to the LHS of the Master Equation vanishes
\begin{eqnarray}
\label{GaugeScalarConnectedAmplitudes} 
&&-k^\mu{{\cal G}}_{\mu;N,M}^{t;t_1...t_M}\Big( \Big[-\frac{1}{4}g_2 {\vec W}_\mu \Big(H^2+{\vec \pi}^2 \Big) \Big];p_1...p_N;q_1...q_M\Big) \nonumber \\
&& \qquad \quad \sim-\frac{1}{4}g_2k^\mu {\vec W}_\mu (k)\Big(\cdot\cdot\cdot \Big)(k^2)\\
&&\qquad =0\nonumber\,,
\end{eqnarray}
because the gauge condition (\ref{GaugeConditionsPrimePrime}) obeyed by the states reads, in momentum-space
\begin{eqnarray}
\label{MomentumGaugeConditionsW}
k^\mu {\vec W}_\mu (k) =0\,.
\end{eqnarray}

\subsubsection{Gauge fixing Lagrangian $L^{GaugeFix;Landau}$}
\label{GaugeFixingLagrangian}

The contribution of the appropriate term in (\ref{SMIsospinCurrentPrime}) to the LHS of the Master Equation vanishes
\begin{eqnarray}
\label{GaugeFixingConnectedAmplitudes} 
&&-k^\mu{{\cal G}}_{\mu;N,M}^{t;t_1...t_M}\Big( \Big[ -\lim_{\xi \to 0} \frac{1}{\xi} {\vec W}_\mu\times \partial_\nu {\vec W}^\nu \Big]; p_1...p_N;q_1...q_M\Big) \nonumber \\
&&\qquad \sim \lim_{\xi \to 0} \frac{1}{\xi} k^\mu {\vec W}_\mu (k)\times \Big(i k_\nu {\vec W}^\nu (k)  \Big)  \\
&&\qquad =0 \nonumber\,.
\end{eqnarray}
because of the momentum-space gauge condition (\ref{MomentumGaugeConditionsW}).

Two factors of the gauge condition in ({\ref{GaugeFixingConnectedAmplitudes}) make it sufficiently convergent, like $L^{GaugeFix;Landau}=-\lim_{\xi \to 0} \frac{1}{2\xi} \Big( \partial_\mu{\vec W}^\mu+\xi M_W{\vec \pi} \Big)^2$, as $\xi\to 0$.
But ${\vec W}^\mu$ are also massive 
and incapable of carrying information to the 3-surface at infinity, 
so ({\ref{GaugeFixingConnectedAmplitudes}) dies exponentially. 

\subsubsection{Total $SU(2)_L$ Lagrangian $L^{Landau}$}
\label{TotalLagrangianPrime}
\bea
\label{MasterLHSPrime}
&&\lim_{k_\lambda \to 0} \int d^4z e^{ikz} \partial_\mu  \big< 0\vert 
	T\Big[J_{L}^{\mu;t}(z) h(x_1)\cdots h(x_N) \nonumber \\
&&\quad\times \pi^{t_1}(y_1)\cdots \pi^{t_M}(y_M)\Big]
		\vert 0 \big>_{Connected} \\
&&=\lim_{k_\lambda \to 0} \int d^4z e^{ikz} \partial_\mu  \big< 0\vert 
	T\Big[J_{Schwinger}^{\mu;t}(z) h(x_1)\cdots h(x_N) \nonumber \\
&&\quad\times \pi^{t_1}(y_1)\cdots \pi^{t_M}(y_M)\Big]
		\vert 0 \big>_{Connected}  \nonumber\,.
\eea

\subsection{Vector ${\vec  J}_{L+R;Schwinger}^\mu$ for $M$ even,\\ Axial-vector ${\vec  J}_{L-R;Schwinger}^\mu$ for $M$ odd}
\label{SchwingerMaster}

Divide the remainder of 
	the global $SU(2)_L$ isospin current (\ref{SMIsospinCurrentPrime}), 
i.e. the part which survives Subsections \ref{RHS}, \ref{SurfaceLHS} and
	\ref{AmplitudesLHS}, 
namely ${\vec J}^{\mu}_{Schwinger}$,
into vector $SU(2)_{L+R}$ and axial-vector $SU(2)_{L-R}$ parts:
\begin{eqnarray}
\label{TotalCurrent}
 {\vec J}^{\mu}_{Schwinger}&=& {\vec J}^{\mu}_{L+R;Schwinger}+ {\vec J}^{\mu}_{L-R;Schwinger}\nonumber\\ 
{\vec J}^{\mu}_{L+R;Schwinger}&=&\half{\vec \pi}\times \partial^\mu {\vec \pi}  \\
 {\vec J}^{\mu}_{L-R;Schwinger}&=& \half\Big({\vec \pi} \partial^\mu H-H\partial^\mu {\vec \pi}\Big) \nonumber\,.
\end{eqnarray}

The classical equations of motion show that none of these sub-currents is conserved  
\bea
\label{DivergenceSMCurrent}
\partial_{\mu} {\vec J}^{\mu}_{L+R;Schwinger}&\neq&0\,; \nonumber\\
\quad\partial_{\mu} {\vec J}^{\mu}_{L-R;Schwinger}&\neq&0\,;  \\
\partial_{\mu} {\vec J}^{\mu}_{L;Schwinger}&\neq&0\,.\nonumber
\eea

We are interested only in models where CP is conserved, 
so that amplitudes connecting the $\phi$-sector 
with the total $SU(2)_L$ isospin current 
(\ref{SMIsospinCurrentPrime}) conserve $CP$. 
Therefore, on-shell and off-shell connected amputated 
T-matrix elements and Green's functions 
of an odd number of $(CP=-1)$ $\vec \pi$s and their derivatives, are zero.

$SU(2)_{L+R}$ and $SU(2)_{L-R}$ are not here 
	strictly applicable sub-groups of $SU(2)_L$; 
but, while the global vector current ${\vec J}^{\mu}_{L+R;Schwinger}$ 
transforms as an even number of $\vec \pi$s, 
the global axial-vector current ${\vec J}^{\mu}_{L-R;Schwinger}$ 
transforms as an odd number of $\vec \pi$s.
Therefore, for $M$ even, the axial-vector current and its divergence obey
\begin{eqnarray}
\label{AxialCurrentMEven}
&&\Big< 0\vert T\Big[ \Big(  {\vec J}^{\mu}_{L-R;Schwinger} (z)\Big)    \\
&&\quad \times h(x_1)...h(x_N) \pi^{t_1}(y_1)...\pi^{t_M}(y_M)\Big]\vert 0\Big>_{Connected} =0 \,,
\nonumber \\
&& \Big< 0\vert T\Big[ \Big( \partial_{\mu} {\vec J}^{\mu}_{L-R;Schwinger}(z)\Big)  \nonumber \\
&&\quad \times h(x_1)...h(x_N)  \pi^{t_1}(y_1)...\pi^{t_M}(y_M)\Big]\vert 0\Big>_{Connected}=0 \,.
\nonumber
\end{eqnarray}

Meanwhile, for odd M, the vector current
\begin{eqnarray}
\label{VectorCurrentMOdd}
&&\Big< 0\vert T\Big[ \Big(  {\vec J}^{\mu}_{L+R;Schwinger} (z)\Big)    \\
&&\quad \times h(x_1)...h(x_N) \pi^{t_1}(y_1)...\pi^{t_M}(y_M)\Big]\vert 0\Big>_{Connected} =0 \,,
\nonumber \\
&& \Big< 0\vert T\Big[ \Big( \partial_{\mu} {\vec J}^{\mu}_{L+R;Schwinger}(z)\Big)  \nonumber \\
&&\quad \times h(x_1)...h(x_N)  \pi^{t_1}(y_1)...\pi^{t_M}(y_M)\Big]\vert 0\Big>_{Connected}=0 \,.
\nonumber
\end{eqnarray}
This, plus $CP$ conservation, 
allows us to write two sets of $SU(2)_L$ Ward-Takahashi identities 
governing the $\phi$-sector of $SU(2)_L$: 
one for the vector current ${\vec  J}_{L+R;Schwinger}^\mu$ based on even $M$, 
and one for the axial-vector current ${\vec  J}_{L-R;Schwinger}^\mu$ based on odd $M$.
 In this paper, 
 we focus on those strong constraints placed 
 on the scalar-sector of the $SU(2)_L$, by the axial-vector WTI.

\subsection{Axial-vector Master Equation in ${\vec  J}_{L-R;Schwinger}^\mu$}
\label{AxialVectorSchwingerMaster}
We now assemble the axial-vector Master Equation, from which we derive our ``1-soft-$\pi$" $SU(2)_{L-R}$ WTI.
The LHS is from (\ref{MasterLHSPrime}), 
the RHS from \eqref{MasterRHS1}:
\begin{eqnarray}
\label{AxialVectorMasterEquation}
&&\lim_{k_\lambda \to 0} \int d^4z e^{ikz} \int d^4z e^{ikz}\partial_{\mu} 
\Big< 0\vert T\Big[ \Big( {\vec J}^{\mu}_{L-R;Schwinger}(z) \Big) \nonumber  \\
&&\quad \quad \times h(x_1)...h(x_N) \pi^{t_1}(y_1)...\pi^{t_M}(y_M)\Big]\vert 0\Big>_{Connected} \nonumber\\
&& =\lim_{k_\lambda \to 0}\int d^4z e^{ikz}\sum_{n=1}^{n=N} \big< 0\vert 
	T\Big[ h(x_1)\cdots h(x_{n-1}) \\
&& \quad\times \delta(z^0-x_n^0)\Big[ J_{L-R;Schwinger}^{0;t}(z),h(x_n)\Big] \nonumber \\
&&\quad\times h(x_{n+1})\cdots h(x_N) \pi^{t_1}(y_1)\cdots \pi^{t_M}(y_M)\Big]
		\vert 0 \big>_{Connected}  \nonumber\\
&& + \lim_{k_\lambda \to 0}\int d^4z e^{ikz}\sum_{m=1}^{m=M} \big< 0\vert 
	T\Big[ h(x_1)\cdots h(x_N)\nonumber \\
&&\quad\times \pi^{t_1}(y_1)\cdots \pi^{t_{m-1}}(y_{m-1})\nonumber \\
&& \quad\times \delta(z^0-y_m^0)\Big[ J_{L-R;Schwinger}^{0;t}(z),\pi^{t_m}(y_m)\Big] \nonumber \\
&&\quad\times \pi^{t_{m+1}}(y_{m+1})\cdots \pi^{t_M}(y_M)\Big]
		\vert 0 \big>_{Connected}\,,\nonumber
\end{eqnarray}
written in terms of the physical states of the complex scalar $\phi$.
Here we have N external renormalized scalars $h=H-\HVEV$ (coordinates x, momenta p), 
and M external ($CP=-1$) renormalized pseudo-scalars ${\vec \pi}$ (coordinates y, momenta q, isospin t). 

Eq. (\ref{AxialVectorMasterEquation}) is true for any M, odd or even. 
It is derived from (\ref{SMMasterEquation}) for $M$ odd in 
(\ref{MasterRHS1},\ref{SMSurfacePionPoleDominance},\ref{MasterLHSPrime}).
It is also satisfied trivially, because of (\ref{AxialCurrentMEven}), for $M$ even.

This paper is based on the de facto conservation of $ {\vec J}^{\mu}_{L-R;Schwinger}$, 
for $CP$-conserving on-shell and off-shell 
connected amputated  Green's functions and T-matrix elements in the $\phi$-sector, 
in the 1-soft-$\pi$ limit (\ref{AxialVectorMasterEquation}) of $SU(2)_L$.

\section{$SU(2)_L$ Ward-Takahashi identities}
\label{DerivationWTIAHM}

The purpose of this Appendix is to derive two towers of quantum $SU(2)_{L-R}$ Ward-Takahashi identities, which exhaust the information content of (\ref{AxialVectorMasterEquation}), and severely constrains the dynamics (i.e. the connected time-orderd products) of the physical states 
of the spontaneously broken bosonic $SU(2)_L$. 

For  pedagogical completeness, we repeat certain results found elsewhere the paper.

1) The axial-vector Master Equation (\ref{AxialVectorMasterEquation}),
from which we derive our $SU(2)_{L-R}$ WTI, is derived in Subsection \ref{AxialVectorSchwingerMaster}. 

2) We employ Vintage QFT and canonical quantization,
imposing equal-time commutators on the exact renormalized fields, 
yielding at space-time points $y, z$:
\begin{eqnarray}
\label{EqTimeCommAHM}
&& \delta(z_0-y_0)\left[ {J}^{t;0}_{L-R;Schwinger}(z),H(y)\right] \nonumber \\
&& \qquad =-\half i{\pi}^t(y)\delta^4(z-y) \\
&& \delta(z_0-y_0)\left[ {J}^{t_1;0}_{L-R;Schwinger}(z),\pi^{t_2}(y)\right] \nonumber \\
&&\qquad= \half i\delta^{t_1t_2}H(y)\delta^4(z-y) \nonumber
\end{eqnarray} 
Field normalization follows from the non-trivial commutators
\begin{eqnarray}
\label{ZeroEqTimeCommAHM}
  \delta(z_0-y_0)\left[ \partial^0 H(z),H(y)\right] &=& -i\delta^4(z-y)  \\
 \delta(z_0-y_0)\left[ \partial^0 \pi^{t_1} (z),\pi^{t_1}(y)\right] &=& -i\delta^{t_1t2}\delta^4(z-y) \nonumber\,.
\end{eqnarray} 

3) As appropriate to our study of the massless $\vec \pi$, 
we use pion-pole dominance to derive 1-soft-pion theorems, 
and form the surface integral
\begin{eqnarray}
\label{SurfacePionPoleDominanceAppendixA}
&&\lim_{k_\lambda \to 0} \int d^4z e^{ikz} \partial_{\mu} \Big< 0\vert T\Big[ 
\bigg(  \half \Big( {\vec \pi}\partial^\mu h - h \partial^\mu {\vec \pi}\Big) (z)\bigg)  
\nonumber \\
&&\quad \quad \times h(x_1)...h(x_N) \pi^{t_1}(y_1)...\pi^{t_M}(y_M)\Big]\vert 0\Big>_{Connected} \nonumber \\
&& =\int d^4z \partial_{\mu} \Big< 0\vert T\Big[ 
\bigg(  \half \Big( {\vec \pi}\partial^\mu h - h \partial^\mu {\vec \pi}\Big) (z)\bigg)    \\
&&\quad \quad \times h(x_1)...h(x_N) \pi^{t_1}(y_1)...\pi^{t_M}(y_M)\Big]\vert 0\Big>_{Connected} \nonumber \\
&& =\int_{3-surface\to \infty}  
\!\!\!\!\!\!\!\!\!\!\!\!\!\!\!\!\!\!\!\!\!\!\!\!
d^3z \quad {\widehat {z}_\mu}^{3-surface} \Big< 0\vert T\Big[ 
\bigg(  \half \Big( {\vec \pi}\partial^\mu h - h \partial^\mu {\vec \pi}\Big) (z)\bigg)  
 \nonumber \\
&&\quad \quad \times h(x_1)...h(x_N) \pi^{t_1}(y_1)...\pi^{t_M}(y_M)\Big]\vert 0\Big>_{Connected} \nonumber \\
&& =0 \nonumber\,,
\end{eqnarray}
where we have used Stokes theorem, 
and $ {\widehat {z}_\mu}^{3-surface}$ is a unit vector normal to the $3$-surface. 
The time-ordered product constrains the $3$-surface to lie on, 
or inside, the light-cone. 

At a given point on the surface of a large enough 4-volume $\int d^4z$ (i.e. the volume of all space-time): all fields are asymptotic in-states and out-states, properly quantized as free fields, with each field species orthogonal to the others,
and  they are evaluated at equal times, making time-ordering un-necessary on the
3-surface at infinity.
The surface integral (\ref{SurfacePionPoleDominanceAppendixA}) vanishes because  $h$ is massive in spontaneously broken $SU(2)_L$, 
with $m_{BEH}^2\neq 0$. 
Propagators connecting $h$ 
from points on the 3-surface at infinity 
to the localized interaction points $(x_1...x_N;y_1...y_M)$ 
must stay inside the light-cone, 
die off exponentially with mass,
and are incapable of carrying information that far. 

It is very important for pion-pole-dominance,  and the $SU(2)_{L-R}$ WTI here, that this argument fails for the remaining term in ${\vec J}^\mu_{L-R;Schwinger}$ in (\ref{AxialVectorMasterEquation}):
\begin{eqnarray}
\label{NGBSurfaceIntegral}
&& \int_{3-surface\to \infty} 
\!\!\!\!\!\!\!\!\!\!\!\!\!\!\!\!\!\!\!\!\!\!\!\!
 d^3z \quad {\widehat {z}_\mu}^{3-surface}  \Big< 0\vert T\Big[  \Big(\half\HVEV \partial^\mu {\vec \pi}\Big)(z)   \\
&&\quad \quad \times h(x_1)...h(x_N) \pi^{t_1}(y_1)...\pi^{t_M}(y_M)\Big]\vert 0\Big>_{Connected} \neq 0 \nonumber\,.
\end{eqnarray}
$\vec \pi$ are massless in SSB $SSU(2)_L$ in Landau gauge, 
capable of carrying (along the light-cone) l
ong-ranged pseudo-scalar forces out to the 2-surface  
$(z^{\mathrm 2-surface}\to \infty)$,
i.e. to the very ends of the light-cone (but not inside it).
That masslessness is the basis of our pion-pole-dominance-based $SU(2)_{L-R}$ WTIs, which give 1-soft-pion theorems 
(\ref{SoftPionLimitPropagator}), infra-red finiteness for $\mpisq =0$ (\ref{AdlerSelfConsistency}),  and an LSS theorem.

4) Using (\ref{EqTimeCommAHM}) in (\ref{AxialVectorMasterEquation}) 
on the right-hand-side, 
and (\ref{SurfacePionPoleDominanceAppendixA}) in 
(\ref{AxialVectorMasterEquation}) on the left-hand-side, 
we rewrite the Master Equation
\begin{eqnarray}
	\label{MasterEquation}
	&&\lim_{k_\lambda\to 0}\int d^4z e^{ikz}\Big\{ -\HVEV\partial_{\mu}^z \big< 0\vert T\Big[  \big(\partial^\mu\pi^{t}(z)\big)   \\
	&& \quad \times h(x_1)...h(x_N) \pi^{t_1}(y_1)...\pi^{t_M}(y_M)\Big]\vert 0\big>_{Connected} \nonumber \\
	&&\quad - \sum_{m=1}^M  i\delta^4(z-y_m) \delta^{tt_m}\big< 0\vert T\Big[ h(z) h(x_1)...h(x_N) \nonumber \\
	&&\quad \quad \quad \quad \times  \pi^{t_1}(y_1)...{\widehat {\pi^{t_m} (y_m)}}...\pi^{t_M}(y_M)\Big]\vert 0\big>_{Connected} \nonumber \\
	&&\quad + \sum_{n=1}^N   i\delta^4(z-x_n) \big< 0\vert T\Big[ h(x_1)...{\widehat {h(x_n)}}...h(x_N) \nonumber \\
	&&\quad \quad \quad \quad \times  \pi^{t}(z){\pi}^{t_1}(y_1)...\pi^{t_M}(y_M)\big]\vert 0\big>_{Connected}\Big\} \nonumber \\
	&& \quad =0 \,,\nonumber
\end{eqnarray}
where the ``hatted" fields ${\widehat {h(x_n)}}$ and ${\widehat {\pi^{t_m} (y_m)}}$ 
are to be removed, 
and we have suppressed the $k_\mu -z_\mu$ Fourier transform 
for clarity of presentation. 
We have also thrown away a sum of $M$ terms, proportional to $\HVEV$,
which corresponds entirely to disconnected graphs.

5) Connected momentum-space amplitudes, 
with $N$ external BEHs and $M$ external $\pi$s, 
are defined in terms of $\phi$-sector connected time-ordered products
\begin{eqnarray}
\label{Amplitudes} 
&&iG_{N,M}^{t_1...t_M}(p_1...p_N;q_1...q_M)(2\pi)^4\delta^4 \Big(\sum_{n=1}^N p_n +\sum_{m=1}^M q_m \Big) \nonumber \\
&& \quad =\prod_{n=1}^N\int d^4x_n e^{ip_nx_n} \prod_{m=1}^M\int d^4y_m e^{iq_my_m}  \\
&&\quad \times \big< 0\vert T\Big[ h(x_1)...h(x_N) \pi^{t_1}(y_1)...\pi^{t_M}(y_M)\Big]\vert 0\big>_{Connected} \nonumber
\end{eqnarray}

The Master Equation (\ref{MasterEquation}) can then be re-written
\begin{eqnarray}
\label{AmplitudeIdentity} 
&&\lim_{k_\lambda\to 0}\Big\{i\HVEV k^2 G_{N,M+1}^{tt_1...t_M}(p_1...p_N;kq_1...q_M) \nonumber \\
&&\quad -\sum_{n=1}^N G_{N-1,M+1}^{tt_1...t_M}(p_1...{\widehat {p_n}}...p_N;(k+p_n)q_1...q_M) \nonumber \\
&&\quad +\sum_{m=1}^M \delta^{tt_m}G_{N+1,M-1}^{t_1...{\widehat {t_m}}...t_M}((k+q_m)p_1...p_N;q_1...{\widehat {q_m}}...q_M) \Big\}\nonumber \\
&& \quad =0
\end{eqnarray}
with the ``hatted" momenta $({\widehat {p_n}},{\widehat {q_m}})$ and isospin ${\widehat {t_m}}$ removed  in (\ref{AmplitudeIdentity}), and an overall momentum conservation factor of $(2\pi)^4\delta^4 \Big(k+\sum_{n=1}^N p_n +\sum_{m=1}^M q_m \Big)$. 

6) Special cases of (\ref{Amplitudes}) are the BEH and $\vec \pi$ propagators
\begin{eqnarray}
\label{ConnectedAmplitudePropagators} 
iG_{2,0}(p_1,-p_1;)&=&i\int \frac{d^4p_2}{(2\pi)^4}  G_{2,0}(p_1,p_2;)\nonumber \\
&=&\int d^4x_1 e^{ip_1x_1} \big< 0\vert T \Big[ h(x_1)h(0)\Big]\vert 0 \big> \nonumber \\
&\equiv& i\Delta_{BEH}(p_1^2) \nonumber \\
iG_{0,2}^{t_1t_2}
(;q_1,-q_1)&=&i\int \frac{d^4q_2}{(2\pi)^4}  G_{0,2}^{t_1t_2}(;q_1,q_2) \nonumber \\
&=&\int d^4y_1 e^{iq_1y_1}\big< 0\vert T \Big[ \pi^{t_1} (y_1)\pi^{t_2}(0)\Big]\vert 0\big> \nonumber \\
&\equiv& i\delta^{t_1t_2}\Delta_{\pi}(q_1^2)
\end{eqnarray}

7) In what follows, isospin indices will become increasingly cumbersome. We therefore  adopt B.W. Lee's \cite{Lee1970} convention of supressing isospin indices, allowing momenta to implicitly carry them.

8) $\phi$-sector connected amputated 1-$(h,\pi)$-Reducible (1-$\phi$-R) transition matrix (T-matrix): With an overall momentum conservation factor 
$(2\pi)^4\delta^4 \Big(\sum_{n=1}^N p_n +\sum_{m=1}^M q_m \Big)$, the $\phi$-sector connected amplitudes are related to $\phi$-sector connected amputated T-matrix elements
\begin{eqnarray}
\label{TMatrix} 
&&G_{N,M}(p_1...p_N;q_1...q_M)  \\
&&\qquad \equiv\prod_{n=1}^N\Big[i\Delta_{BEH}(p_n^2)\Big] \prod_{m=1}^M\Big[i\Delta_{\pi}(q_m^2)\Big] \nonumber\\
&&\qquad\quad \times T_{N,M}(p_1...p_N;q_1...q_M) \nonumber 
\end{eqnarray}
so that the Master Equation (\ref{MasterEquation}) can be written
\begin{eqnarray}
\label{TMatrixIdentity} 
&&\lim_{k_\lambda\to 0}\Big\{i\HVEV k^2 \Big[i\Delta_\pi(k^2)\Big]T_{N,M+1}(p_1...p_N;kq_1...q_M) \nonumber \\
&&\quad -\sum_{n=1}^N T_{N-1,M+1}(p_1...{\widehat {p_n}}...p_N;(k+p_n)q_1...q_M) \nonumber \\
&&\quad \times \Big[i\Delta_\pi((k+p_n)^2)\Big] \Big[i\Delta_{BEH}(p_n^2)\Big]^{-1} \\
&&\quad +\sum_{m=1}^M T_{N+1,M-1}((k+q_m)p_1...p_N;q_1...{\widehat {q_m}}...q_M) \nonumber \\
&&\quad \times \Big[i\Delta_{BEH}((k+q_m)^2)\Big] \Big[i\Delta_{\pi}(q_m^2)\Big]^{-1} \Big\} \nonumber \\
&& \quad =0\,,\nonumber
\end{eqnarray}
with the ``hatted" momenta $({\widehat {p_n}},{\widehat {q_m}})$ removed  in (\ref{TMatrixIdentity}), and an overall momentum conservation factor of $(2\pi)^4\delta^4 \Big(k+\sum_{n=1}^N p_n +\sum_{m=1}^M q_m \Big)$.

9) Consider the ``soft-pion limit" 
\begin{eqnarray}
\label{SoftPionLimitPropagator}
\lim_{k_\mu \to 0} k^2\Delta_\pi (k^2) =1 \,,
\end{eqnarray}
where the $\vec \pi$ is hypothesized to be all-loop-orders massless
\begin{eqnarray}
\label{NGBPropagator}
\Delta_{\pi}(k^2) &=& \frac{1}{k^2
+ i\epsilon} + \int dm^2 \frac{\rho^{\pi}_{AHM}(m^2)}{k^2-m^2 + i\epsilon}
\end{eqnarray} 
in the K$\ddot a$ll$\acute e$n-Lehman representation \cite{Bjorken1965}. 
The Master Equation (\ref{MasterEquation}) then becomes
\begin{eqnarray}
\label{SoftPionTMatrixID} 
&&-\HVEV T_{N,M+1}(p_1...p_N;0q_1...q_M) \nonumber \\
&&\quad =\sum_{n=1}^N T_{N-1,M+1}(p_1...{\widehat {p_n}}...p_N;p_nq_1...q_M) \nonumber \\
&&\quad \times \Big[i\Delta_\pi(p_n^2)\Big] \Big[i\Delta_{BEH}(p_n^2)\Big]^{-1} \nonumber \\
&&\quad -\sum_{m=1}^M T_{N+1,M-1}((k+q_m)p_1...p_N;q_1...{\widehat {q_m}}...q_M) \nonumber \\
&&\quad \times \Big[i\Delta_{BEH}(q_m^2)\Big] \Big[i\Delta_{\pi}(q_m^2)\Big]^{-1} 
\end{eqnarray}
in the 1-soft-pion limit.
As usual the ``hatted" momenta $({\widehat {p_n}},{\widehat {q_m}})$ and associated fields are removed  in (\ref{SoftPionTMatrixID}), and an overall momentum conservation factor  $(2\pi)^4\delta^4 \Big(\sum_{n=1}^N p_n +\sum_{m=1}^M q_m \Big)$ applied.

The set of 1-soft-pion theorems (\ref{SoftPionTMatrixID}) have the form 
\begin{eqnarray}
\HVEV T_{N,M+1}\sim T_{N-1,M+1} - T_{N+1,M-1}
\end{eqnarray}
relating, by the addition of a zero-momentum pion, an $N+M+1$-point function to $N+M$-point functions.

10) The Adler self-consistency relations, 
for the $SU(2)_L$ gauge theory 
rather than global $SU(2)_L \times SU(2)_R$ \cite{Adler1965,AdlerDashen1968}, 
are obtained from (\ref{SoftPionTMatrixID}) 
by putting the remainder of the particles on mass-shell
\begin{eqnarray}
\label{AdlerSelfConsistency} 
&&\HVEV T_{N,M+1}(p_1...p_N;0q_1...q_M)\nonumber \\
&& \quad \quad \times (2\pi)^4\delta^4 \Big(\sum_{n=1}^N p_n +\sum_{m=1}^M q_m \Big) \Big\vert^{p_1^2 =p_2^2...=p_N^2=m_{BEH}^2}_{q_1^2 =q_2^2...=q_M^2=0}  \nonumber \\
&& \quad \quad =0 \,.
\end{eqnarray}
This guarantees the infra-red (IR) finiteness of the $\phi$-sector on-shell T-matrix in SSB $SU(2)_L$ gauge theory in $R_\xi (\xi=0)$ 
Landau gauge, with massless $\vec \pi$ in the 1-soft-pion limit.

11) 1-$(h,\pi)$ Reducibility (1-$\phi$-R) and 1-$(h,\pi)$ Irreducibility (1-$\phi$-I): With some exceptions, the $\phi$-sector connected amputated transition matrix $T_{N,M}$ can be cut in two by cutting an internal $h$ or $\vec \pi$ line, and are designated 1-$\phi$-R. In contrast, the $\phi$-sector connected amputated Green's functions $\Gamma_{N,M}$ are defined to be 1-$\phi$-I,
 i.e. they cannot be cut apart by cutting an internal $h$ or $\vec \pi$ line.
\begin{eqnarray}
\label{1SPReducibility}
T_{N,M} = \Gamma_{N,M} + (1-\phi-R)
\end{eqnarray}

Both $T_{N,M}$ and $\Gamma_{N,M}$ 
	are 1-$({\vec W}_\mu)$-Reducible (1-${\vec W}_\mu$-R)\footnote{The set of 1-$\phi$-I graphs includes all the 1-P-I graphs and infinitely many more. Kraus and Sibold\cite{KrausSiboldAHM}, and Grassi\cite{Grassi1999} derived WTIs for 1-P-I graphs for the AHM and for general gauge theories with non-semi-simple gauge groups respectively. As we consider a different set of Green's functions, the WTIs derived here are fundamentally different from those derived in \cite{KrausSiboldAHM,Grassi1999}. We are interested only in processes involving external scalars and in the scalar-sector effective potential; to calculate those, 1-$\phi$-I graphs are the correct set to consider.} 
i.e. they can be cut in two by cutting an internal transverse ${\vec W}_\mu$ photon line. 

12) The special 2-point functions $T_{0,2}(;q,-q)$ and
$T_{2,0}(p,-p;)$, and the 3-point vertex  $T_{1,2}(q;0,-q)$, are 1-$\phi$-I,
i.e. they are not 1-$\phi$-R. 
They are therefore equal to the corresponding 1-$\phi$-I 
connected amputated Green's functions. 
The 2-point functions
\begin{eqnarray}
\label{TMatrix2PointA}
T_{2,0}(p,-p;)&=&\Gamma_{2,0}(p,-p;)=\big[\Delta_{BEH}(p^2)\big]^{-1} \nonumber \\
T_{0,2}(;q,-q)&=&\Gamma_{0,2}(;q,-q)=\big[\Delta_{\pi}(q^2)\big]^{-1}  
\end{eqnarray}
are related to the $(1h,2\pi)$ 3-point $h{\vec \pi}^2$ vertex 
\begin{eqnarray}
\label{3PointVertex}
T_{1,2}(p;q,-p-q) = \Gamma_{1,2}(p;q,-p-q) 
\end{eqnarray}
by a 1-soft-pion theorem (\ref{SoftPionTMatrixID})
\begin{eqnarray}
\label{TMatrix2and3Point}
&&\HVEV T_{1,2}(q;0,-q) -T_{2,0}(q,-q;)+T_{0,2}(;q,-q)  \\
&&\quad \quad =\HVEV T_{1,2}(q;0,-q) -\big[\Delta_{BEH}(q^2)\big]^{-1} +\big[\Delta_{\pi}(q^2)\big]^{-1} \nonumber \\
&&\quad \quad =\HVEV \Gamma_{1,2}(q;0,-q) -\Gamma_{2,0}(q,-q;)+\Gamma_{0,2}(;q,-q) \nonumber \\
&&\quad \quad =\HVEV \Gamma_{1,2}(q;0,-q) -\big[\Delta_{BEH}(q^2)\big]^{-1} +\big[\Delta_{\pi}(q^2)\big]^{-1} \nonumber \\
&&\quad \quad =0 \,.\nonumber
\end{eqnarray}

13) The Lee-Stora-Symanzik (LSS) theorem,
in spontaneously broken $SU(2)_L$ in $R_\xi (\xi=0)$ Landau gauge, 
is a special case of that SSB gauge theory's Adler self-consistency relations 
(\ref{AdlerSelfConsistency})
\begin{eqnarray}
\label{AppendixGoldstoneTheorem} 
\HVEV T_{0,2}(;00)&=&0 \nonumber \\
\HVEV \Gamma_{0,2}(;00)&=&0  \\
\HVEV \big[ \Delta_\pi (0) \big]^{-1} &=&-\HVEV\mpisq=0\,,\nonumber
\end{eqnarray}
proving that $\vec \pi$ is massless. 
That all-loop-orders renormalized masslessness 
is protected/guarranteed by the global $SU(2)_L$ symmetry 
of the physical states of the gauge theory after spontaneous symmetry breaking.

14) 
Figure \ref{fig:LeeFig10} illustrates $T_{External;N,M+1}$,
a  $\phi$-sector T-Matrix with one soft ${\vec \pi} (q_\mu =0)$ attached to an external-leg, showing that
\begin{eqnarray}
\label{ExternalLegTMatrix}
&&\HVEV T_{N,M+1}^{ExternalLeg}(p_1...p_N;0q_1...q_M) \nonumber \\
&&\quad \quad =\sum_{n=1}^N \Big[ i\HVEV\Gamma_{1,2}(p_n,0,-p_n)\Big] \Big[ i\Delta_\pi (p_n^2)\Big]  \nonumber \\
&&\quad \quad \times T_{N-1,M+1}(p_1...{\widehat{p_n}}...p_N;p_nq_1...q_M) \nonumber \\
&&\quad \quad +\sum_{m=1}^M \Big[ i\HVEV\Gamma_{1,2}(q_m,0,-q_m)\Big] \Big[ i\Delta_{BEH} (q_m^2)\Big]  \nonumber \\
&&\quad \quad \times T_{N+1,M-1}(q_mp_1...p_N;q_1....{\widehat{q_m}}...q_M) \nonumber \\
&&\quad \quad =\sum_{n=1}^N \Big(1-\Big[ i\Delta_\pi (p_n^2)\Big]\Big[ i\Delta_{BEH} (p_n^2)\Big]^{-1} \Big) \nonumber \\
&&\quad \quad \times T_{N-1,M+1}(p_1...{\widehat{p_n}}...p_N;p_nq_1...q_M) \nonumber \\
&&\quad \quad -\sum_{m=1}^M \Big(1-\Big[ i\Delta_{BEH} (q_m^2)\Big]\Big[ i\Delta_\pi (q_m^2)\Big]^{-1} \Big)  \nonumber \\
&&\quad \quad \times T_{N+1,M-1}(q_mp_1...p_N;q_1....{\widehat{q_m}}...q_M)  
\end{eqnarray}
when we employ (\ref{TMatrix2and3Point}). 
Now separate
\begin{eqnarray}
\label{DefineInternalTMatrixA}
&&T_{N,M+1}(p_1...p_N;0q_1...q_M) \nonumber \\
&&\quad \quad =T_{External;N,M+1}(p_1...p_N;0q_1...q_M) \nonumber \\
&&\quad \quad +T_{Internal;N,M+1}(p_1...p_N;0q_1...q_M)  
\end{eqnarray}
so that 
\begin{eqnarray}
\label{InternalTMatrix}
&&\HVEV T_{Internal;N,M+1}(p_1...p_N;0q_1...q_M) \nonumber \\
&&\quad \quad =\sum_{m=1}^M T_{N+1,M-1}(q_mp_1...p_N;q_1....{\widehat{q_m}}...q_M)  \nonumber \\
&&\quad \quad -\sum_{n=1}^N T_{N-1,M+1}(p_1...{\widehat{p_n}}...p_N;p_nq_1...q_M) \quad \quad 
\end{eqnarray}

\begin{figure}
\centering
\includegraphics[width=1\hsize,trim={0cm 5cm 0cm 5.5cm},clip]{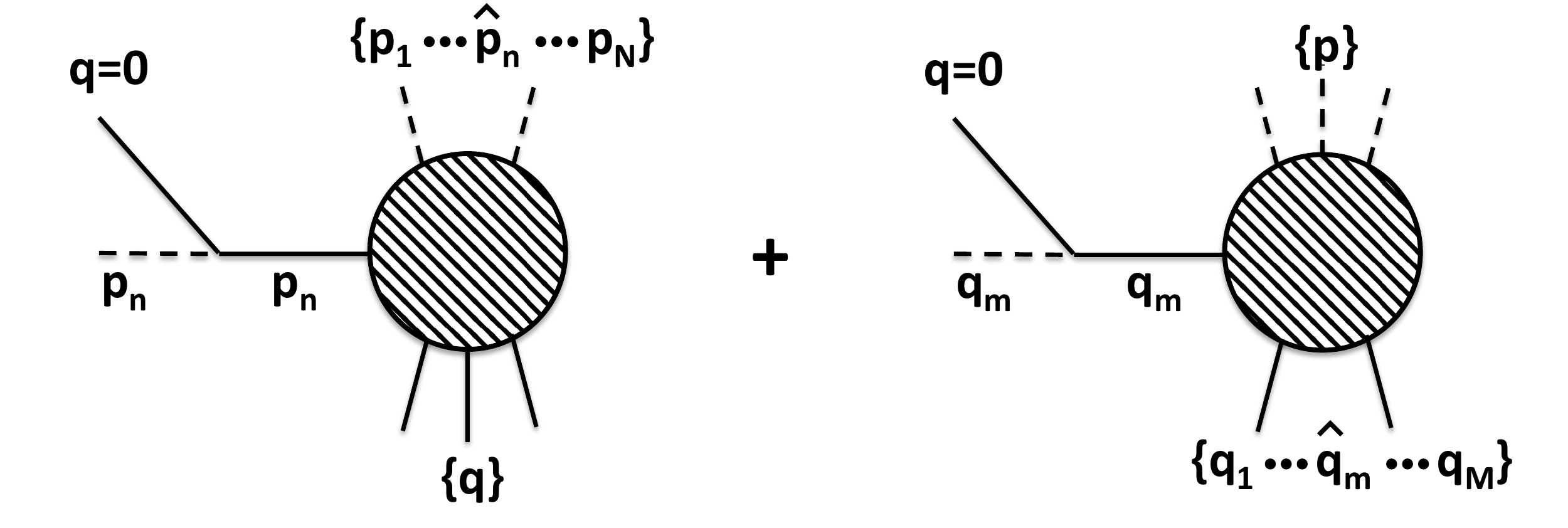}
\caption{$T_{N,M+1}^{Internal}$: Hashed circles are 1-$\phi$-R $T_{N,M}$, solid lines $\pi$, dashed lines $h$. One (zero-momentum) soft pion is attached to an external leg in all possible ways. $T_{N,M}$ is 1-$A^\mu$-R by cutting an $\vec{W}^\mu$ line. Fig. \ref{fig:LeeFig10}  is the $SU(2)_L$ analogy of B.W. Lee's Figure 10 \cite{Lee1970}.}
\label{fig:LeeFig10} 
\end{figure}

15) Removing the 1-$\phi$-R  graphs from both sides of (\ref{InternalTMatrix}) 
yields the recursive $SU(2)_{L-R}$ WTI  
for  1-$\phi$-I connected amputated Green's functions $\Gamma_{N,M}$: 
\begin{eqnarray}
\label{GreensFWTI}
&&\HVEV \Gamma_{N,M+1}(p_1...p_N;0q_1...q_M)  \\
&&\quad \quad =\sum_{m=1}^M \Gamma_{N+1,M-1}(q_mp_1...p_N;q_1....{\widehat{q_m}}...q_M)  \nonumber \\
&&\quad \quad -\sum_{n=1}^N \Gamma_{N-1,M+1}(p_1...{\widehat{p_n}}...p_N;p_nq_1...q_M) \,.{\quad \quad}\nonumber
\end{eqnarray}

B.W. Lee \cite{Lee1970} gave an inductive proof 
for the corresponding recursive $SU(2)_L \times SU(2)_R$ WTI 
in the global Gell-Mann Levy model with PCAC \cite{GellMannLevy1960}. 
Specifically, he proved  that, 
given the global $SU(2)_L \times SU(2)_R$ analogy of (\ref{InternalTMatrix}), 
the global $SU(2)_L \times SU(2)_R$ analogy of (\ref{GreensFWTI}) follows. 
He did this by examining the explicit reducibility/irreducibility 
of the various  Feynman graphs involved. 

That proof also works for SSB $SU(2)_{L-R}$, 
thus establishing our tower of 1-$\phi$-I connected amputated Green's functions' 
recursive $SU(2)_{L-R}$ WTI (\ref{GreensFWTI}) 
or the local $SU(2)_L$ gauge theory. 

Rather than including the lengthy proof here, 
we paraphrase \cite{Lee1970} as follows: 
(\ref{TMatrix2and3Point}) shows that (\ref{GreensFWTI}) is true for $(n=1,m=1)$. 
Assume it is true for all $(n,m)$ such that $n+m < N+M$.
Consider (\ref{InternalTMatrix}) for $n=N$,$m=M$. 
The two classes of graphs contributing to $T_{N,M+1}^{Internal}(p_1...p_N;0q_1...q_M)$
are displayed in Figure \ref{fig:LeeFig11}. \

\begin{figure}
	\centering
	\includegraphics[width=1\hsize,trim={0cm 4cm 0cm 3cm},clip]{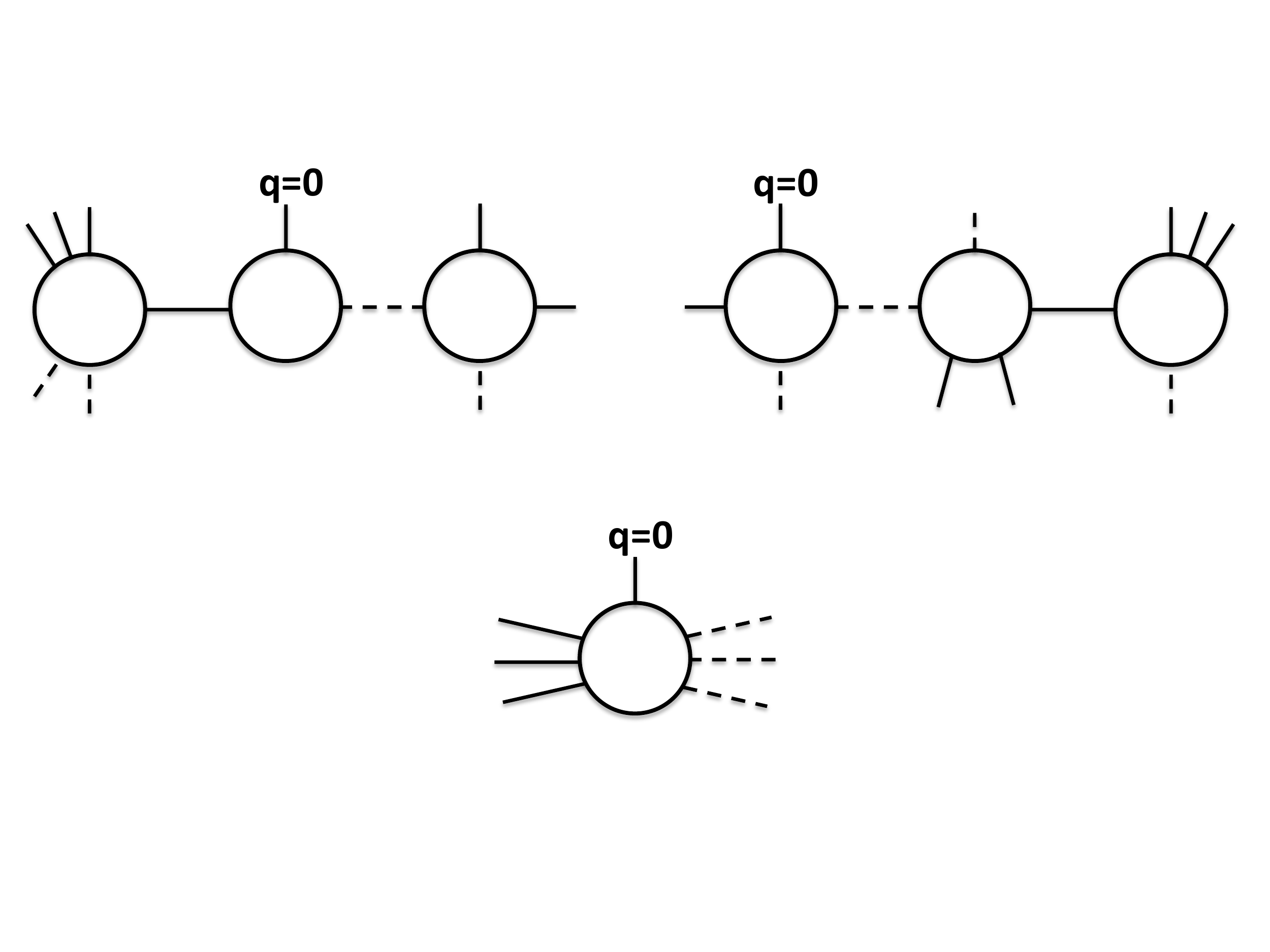}
	\caption{
	Circles are 1-$\phi$-I $\Gamma_{n,m}$, solid lines $\pi$, dashed lines $h$, with $n+m<N+M$. 1 (zero-momentum) soft pion emerges in all possible ways from the  connected amputated Green's functions. $\Gamma_{n,m}$ is 1-$\vec{W}^\mu$-R by cutting an $\vec{W}^\mu$ line. Fig. \ref{fig:LeeFig11}  is the $SU(2)_L$ analogy of B.W. Lee's Figure 11 \cite{Lee1970}. 
	}
	\label{fig:LeeFig11} 
\end{figure}

The top graphs in Figure \ref{fig:LeeFig11} are 1-$\phi$-R. 
For $(n,m;n+m<N+M)$, we may use (\ref{GreensFWTI}), 
for those 1-$\phi$-I Green's functions $\Gamma_{n,m}$ that contribute to 
(\ref{InternalTMatrix}),
to show that the contributions of 1-$\phi$-R graphs 
to both sides of  (\ref{InternalTMatrix}) are identical.

The bottom graphs in  Figure \ref{fig:LeeFig11} 
are 1-$\phi$-I, and so already obey (\ref{GreensFWTI}). 

16) The LSS theorem makes tadpoles vanish. 
\begin{eqnarray}
\label{Tadpoles}
&&\big<0\vert h(x=0)\vert0\big>_{Connected} 
	= i \Big[i\Delta_{BEH}(0)\Big]\Gamma_{1,0}(0;)\,,\quad
\end{eqnarray}
but the $N=0,M=1$ case of (\ref{GreensFWTI}) reads
\begin{eqnarray}
\label{GoldstoneTadpoles}
\Gamma_{1,0}(0;)&=&  \HVEV \Gamma_{0,2}(;00) = 0 \,,
\end{eqnarray}
where we used (\ref{AppendixGoldstoneTheorem}). 
Tadpoles therefore all vanish automatically, 
and separate tadpole renormalization is unnecessary.
Since we can choose the origin of coordinates anywhere we like
\begin{eqnarray}
\label{GoldstoneTadpolesVanishExtended}
\big<0\vert h(x)\vert0\big>_{Connected} &=& 0\,.
\end{eqnarray}

17) Renormalized $\HVEV$ obeys
\begin{eqnarray}
\label{HVEV}
\big<0\vert H(x)\vert0\big>_{Connected} &=&\big<0\vert h(x)\vert0\big>_{Connected} +\HVEV \nonumber \\
&=& \HVEV \nonumber \\
\partial_\mu \HVEV &=&0\,.
\end{eqnarray}

18) Benjamin W. Lee's 
1970 Cargese summer school lectures' \cite{Lee1970} proof of $\phi$-sector WTI 
focussed on the global $SU(2)_L \times SU(2)_R$ Gell-Mann Levy theory with PCAC, 
but it also gave a detailed pedagogical account 
of the appearance of the Goldstone theorem and massless Nambu-Goldstone bosons 
in global theories. We include a translation guide in Table 1.

\begin{table*}[t]
	\centering
	\caption{Derivation of Ward-Takahashi IDs}
	\begin{tabular}{c|c|c}
	\hline
	Property&This paper&B.W.Lee \cite{Lee1970}\\
	\hline
	Lagrangian invariant&BRST&global group\\
	structure group&$SU(2)_{L}$&$SU(2)_L\!\times\!SU(2)_R$\\
	rigid/global group&$SU(2)_{L}$&$SU(2)_L\!\times\!SU(2)_R$\\
	local/gauge group&$SU(2)_L$& \sout{\phantom{long}}\\
	global currents&${\vec J}_{L}^\mu$&${\vec V}^\mu ;{\vec A}^\mu$ \\
	PCAC&no&yes\\
	current divergence&$\sim\partial_\beta {\vec W}^\beta$&$0; f_\pi\mpisq{\vec \pi}$\\
	gauge fixing&$\partial_\beta {\vec W}^\beta=0$&\sout{\phantom{long}}\\
	gauge&Landau&\sout{\phantom{long}}\\
	ghosts ${\vec {\bar \eta}},{\vec \omega}$&isospin&\sout{\phantom{long}}\\
	ghosts ${\bar \omega}^A,\omega^A$&color&\sout{\phantom{long}}\\
	conserved current&physical states&Lagrangian \\
	physical states&$h,{\vec \pi},{\vec W}_\mu$&$s,{\vec \pi}$\\
	unphysical states&ghosts&\sout{\phantom{long}} \\
	interaction&weak&strong \\
	fields&$H,{\vec \pi},{\vec W}_\mu,{\bar{\vec \eta}},{\vec \omega}$&$\sigma,{\vec \pi}$\\
	BEH scalar&$h=H-\HVEV$&$s=\sigma-\expval{\sigma}$\\
	VEV&$\HVEV$&$\expval{\sigma}=v=f_\pi$\\
	loop particles& physical \& ghosts &$s,{\vec \pi}$\\
	renormalization&all-loop-orders&all-loop-orders\\
	Amplitudes:&&\\
	&\sout{\phantom{long}}&G\\
	&$G_{N,M}$&H\\
	&${\cal G}_\mu^{N,M}$&$G_\mu$  \\
	no pion pole&\sout{\phantom{long}}&$\bar H$\\
	T-Matrix&$T_{N,M}$&$T$\\
	1-$\phi$&$h,{ \pi}$&$s,{\vec \pi}$\\ 
	$\phi$-sector $T_{N,M}$&1-$\phi$-R, 1-$\vec W$-R, 1-$\Phi$-R&1-$\phi$-R\\
	$\phi$-sector GF&$\Gamma_{N,M}$&$\Gamma_{N,M}$\\ 
	connected $\Gamma_{N,M}$&amputated&amputated\\
	connected $T_{N,M}$&amputated&amputated\\
	$\Gamma_{N,M}$&1-$\phi$-R, 1-$\vec W$-R, 1-$\Phi$-R&1-$\phi$-I\\
	external leg $\vec \pi$&$T_{External;N,M+1}$&$T_1$\\
	internal $\vec\pi$&$T_{Internal;N,M+1}$&$T_2$\\
	explicit breaking&\sout{\phantom{long}}&$\epsilon=-f_\pi\mpisq$\\
	pseudo-NGB mass-squared&$\mpisq$&$\mpisq$\\
	LSS Theorem&$\HVEV\Gamma^{t_1t_2}_{0,2}(;00)=0$&$f_\pi\Gamma^{t_1t_2}_{0,2}(;00)=\epsilon\delta^{t_1t_2}=0$\\
	LSS Theorem&1-D line&1-D boundary of 2-D quarter-plane\\
	SSB&Goldstone Mode&Goldstone Mode\\
	NGB after SSB&${\tilde{\vec \pi}}$&Implied ${\tilde{\vec \pi}}$\\
	BEH propagator&$\Delta_{BEH}$&$\Delta_\sigma$\\
	transverse propagator&$\Delta_{W}^{\mu \nu}$&\sout{\phantom{long}}\\
	pion propagator&$\Delta_{\pi}$&$\delta^{t_i t_j}\Delta_{\pi}$
	\end{tabular}
\end{table*}

\clearpage
\section{$R_\xi$-gauge renormalized $SU(2)_L$ for gauge bosons, complex scalar doublet, ghosts and anti-ghosts} 
\label{LightParticleCurrents}

In this Appendix, we construct, in an arbitrary $R_\xi$ gauge, 
the $SU(2)_L$ isospin current ${\vec J}^\mu_L$, 
together with its divergence 
and equal-time commutators with fields.

Thirteen renormalized bosons and ghosts appear in $SU(2)_L$:
a scalar $H$, three pseudo-scalars $\vec \pi$, 
three isospin gauge fields ${\vec W}_\mu$, 
three isospin ghosts $\vec \omega$, 
and three anti-ghosts $\bar {\vec \eta}$.
\bea
\label{SMBosonsGhosts}
\{\Phi_i\}&=&\{ H, {\vec \pi}; {\vec W}_{\mu},  {\vec \omega}, {\bar {\vec \eta}};  
 \}; \quad i=1,13
\eea

\subsection{$SU(2)_L$ classical current and its divergence}

{We form the classical current}
\begin{eqnarray}
\label{IncrementalCurrents}
-{\hat J}^{\mu}_{L} &\equiv& -\big(g_2{\vec \Omega} \big)\cdot {\vec J}^{\mu}_{L} \\
&=& \sum_i^{Particles} \Big(\frac{\partial}{\partial(\partial_{\mu}\Phi_i)} L\Big) \Big(\delta_{SU(2)_L}\Phi_i\Big) \nonumber
\end{eqnarray}

{and its divergence}
\begin{eqnarray}
\label{IncrementalDivergences}
-\partial_{\mu} {\hat J}^{\mu}_{L} &\equiv& -\big(g_2{\vec \Omega} \big)\cdot \partial_\mu{\vec J}^{\mu}_{L} \nonumber \\
&=& \sum_i^{Particles} \Big[ \Big(\frac{\partial}{\partial(\partial_{\mu}\Phi_i)}  L\Big)
\Big(\partial_{\mu} \delta_{SU(2)_L}\Phi_i\Big) \nonumber \\
&&  \qquad \qquad +\Big(\partial_\mu \frac{\partial}{\partial(\partial_{\mu}\Phi_i)}  L\Big)\Big(\delta_{SU(2)_L}\Phi_i\Big) \Big] \nonumber \\
&=& \sum_i^{Particles} \Big[ \Big(\frac{\partial}{\partial(\partial_{\mu}\Phi_i)}  L\Big)
\Big(\delta_{SU(2)_L}\partial_\mu\Phi_i\Big) \nonumber \\
&&  \qquad \qquad +\Big(\frac{\partial}{\partial( \Phi_i)}  L\Big)\Big(\delta_{SU(2)_L}\Phi_i\Big) \Big] \nonumber \\
&=&\delta_{SU(2)_L}L_{SU(2)_L}  \\
&=& \delta_{SU(2)_L}\Big(L_{SU(2)_L}^{GaugeFix; R_\xi}
	+L_{SU(2)_L}^{Ghost;R_\xi} \Big)\nonumber \\
&=& \delta_{SU(2)_L} s\Big[{\bar {\vec \eta}} \cdot \Big({\vec F}^W + \half \xi {\vec b} \Big)\Big] \nonumber \\
&=& -\frac{1}{\xi}{\vec F}^W \cdot \delta_{SU(2)_L}{\vec F}^W \nonumber \\
&=& -\half M_W\Big[ H{\vec F}^W+{\vec \pi} \times {\vec F}^W  \Big]\cdot\big(g_2\Omega \big) \nonumber \\
\partial_{\mu} {\vec J}^{\mu}_{L}&=& \half M_W\Big[ H{\vec F}^W+{\vec \pi} \times {\vec F}^W  \Big] \nonumber\,.
\end{eqnarray} 
We have used the equations of motion (EOM) for certain $\{\Phi_i\}$;
in particular,
the Lagrange-multiplier field $\vec b$ is constrained,
\bea
\label{bconstraint}
{\vec b}&=& -\frac{1}{\xi} {\vec F}_W \,,
\eea
and the ${\vec \omega}$ ghost EOM is
\bea
\label{GhostEofM}
0&=& s{\vec F}_W  \\
&=& \partial^2{\vec \omega}+g_2\partial_\mu\Big({\vec W}^\mu \times {\vec \omega}\Big)
+\xi\frac{M_W^2}{\HVEV} \Big( H {\vec \omega}+{\vec \pi} \times {\vec \omega}\Big)\,.
\nonumber
\eea
Eq. (\ref{GhostEofM}) are used 
in the construction of the classical current ${\vec J}_L^\mu$
and its divergence $\partial_\mu{\vec J}_L^\mu$, 
but not in the quantum mechanics 
of the Lagrangian (\ref{LagrangianAHMRxiGauge},\ref{tHooftGaugeFixing}). 

\subsection{Equal-time field commutators}
\label{Equal-timeFieldCommutators}
We normalize the $\phi$ field so that
\begin{eqnarray}
\label{FieldNormalization}
\delta(z_0-y_0)\left[ \partial^0H(z),H(y)\right] &=& -i\delta^4(z-y) \\
\delta(z_0-y_0)\left[ \partial^0\pi^i(z),{\pi}^j(y)\right]  &=& -i \delta^{ij}\delta^4({ z}-{ y})\nonumber
\end{eqnarray}
with similar normalizations for the other fields, 
as appropriate to gauge bosons and fermionic ghosts.
Except for such momentum-field quantum conditions, 
all other equal-time commutators  vanish
\begin{eqnarray}
\label{VanishingFieldCommutators}
\delta(z_0-y_0)\left[ \Phi_i(z),\Phi_j(y)\right] &=& 0, \quad \forall i,j \\
\delta(z_0-y_0)\left[ \partial^0\Phi_i(z),\Phi_j(y)\right]  &=& 0,
\quad \forall i,j, \mathrm{with~} i\neq j\nonumber\,,
\end{eqnarray}
with anti-commutators as appropriate for ghosts.

\subsection{$SU(2)_L$ currents and commutators in $R_{\xi}$ Gauge}
\label{CurrentsDivergences}
\subsubsection{Global $SU(2)_L$ Schwinger model \cite{Schwinger1957}}
\label{SchwingerModel}
It is useful to remember some results from 
the global $SU(2)_L$ Schwinger model \cite{Schwinger1957},
with Lagrangian
\begin{eqnarray}
\label{SchwingerLagrangian}
L_{Schwinger} &=&\vert \partial_{\mu}\phi \vert ^2 - V\\
V &=& \mu_\phi^2 (\phi^{\dagger}\phi) + \lambda^2_\phi (\phi^{\dagger}\phi)^2 \nonumber
\,.
\end{eqnarray}
Consider the current
\begin{eqnarray}
\label{SchwingerJLeft}
&&{\vec J}^{\mu}_{Schwinger} = \half {\vec \pi}\times \partial^\mu {\vec \pi} + \half\Big( {\vec \pi}\partial^\mu H -H \partial^\mu {\vec \pi}\Big) \,.
\end{eqnarray}
Some useful commutators are
\begin{eqnarray}
&&\delta(z^0-y^0)\left[ {\vec J}_{Schwinger}^{0}(z),H(y)\right]\nonumber \\
 &&\qquad \qquad=-\half i\delta^4(z-y) {\vec \pi}(z) \\
&& \delta(z^0-y^0)\left[ { J}_{Schwinger}^{0;t}(z),{ \pi^{t_1}}(y)\right] \nonumber \\
&&\qquad \qquad=\half i \delta^4(z-y)
\times \Big( \epsilon^{tt_1t_2} \pi^{t_2} (z)+ \delta^{tt_1}H(z) \Big) \nonumber 
\,.
\end{eqnarray}

\subsubsection{Gauge-invariant scalar Lagrangian $L^\phi$}
\label{ScalarLagrangian}
The gauge-invariant scalar Lagrangian is
\begin{eqnarray}
\label{Lphi}
L^\phi &=&\vert D_{\mu}\phi \vert ^2 - V \\
D_{\mu}\phi &=& \left[\partial_{\mu} + ig_2W_{\mu}\right]\phi \nonumber \\
W_{\mu}&=& {\vec t} \cdot {\vec W}_{\mu} \equiv \half{\vec \sigma} \cdot {\vec W}_{\mu}\nonumber
\end{eqnarray}
Consider the current
\begin{eqnarray}
\label{ScalarLagrangianJLeft}
{\vec J}^{\mu;\phi}_L&=& {\vec J}^{\mu}_{Schwinger} + {\vec {\cal J}}_L^{\mu;\phi}\\
{\vec {\cal J}}^{\mu;\phi}_{L} &=&-\frac{1}{4} g_2 {\vec W}^{\mu}\left[ H^2+{\vec \pi}^2\right]\,. \nonumber 
\end{eqnarray}
Some useful commutators are
\begin{eqnarray}
\delta(z^0-y^0)&&\left[ {\vec J}_{L}^{0;\phi}(z),H(y)\right] \nonumber \\
&=&\delta(z^0-y^0)\left[  {\vec J}_{Schwinger}^{0}(z),H(y)\right] \nonumber \\
 \delta(z^0-y_1^0)&&\left[ { J}_{L}^{0;t;\phi}(z),{ \pi}^{t_1}(y_1)\right] \nonumber \\
&=& \delta(z^0-y_1^0)\left[  { J}_{Schwinger}^{0;t}(z),{ \pi}^{t_1}(y_1)\right]\,. \nonumber
\end{eqnarray}

\subsubsection{Gauge-invariant isospin Lagrangian $L^W$}
\label{IsospinLagrangian}
The gauge-invariant isospin Lagrangian is
\begin{eqnarray}
\label{LIsospin}
L^W&=&-\half Tr \left(W_{\mu \nu}W^{\mu \nu}\right) = -\frac{1}{4}{\vec W}_{\mu \nu}\cdot {\vec W}^{\mu \nu} \quad \\
W_{\mu \nu}&=&\partial_{\mu}W_{\nu}-\partial_{\nu}W_{\mu} +ig_2\left[W_{\mu},W_{\nu}\right] \nonumber \\
W_{\mu}&=& {\vec t} \cdot {\vec W}_{\mu} \equiv \half{\vec \sigma} \cdot {\vec W}_{\mu}\nonumber \\
{\vec W}_{\mu \nu}&=&\partial_{\mu} {\vec W}_{\nu}-\partial_{\nu} {\vec W}_{\mu} -g_2{\vec W}_{\mu} \times {\vec W}_{\nu} \nonumber 
\end{eqnarray} 
with Pauli matrices $\vec \sigma$.
Consider the current
\begin{eqnarray}
\label{IsospinJLeft}
{\vec J}^{\mu;W}_{L} &=&  {\vec W}^{\mu \nu} \times {\vec W}_{\nu} .
\end{eqnarray}
Useful commutators are:
\begin{eqnarray}
\delta(z^0-y^0)\left[ {\vec J}_{L}^{0;W}(z),H(y)\right] &=&0 \nonumber \\
\delta(z^0-y^0)\left[ {\vec J}_{L}^{0;W}(z),{\vec \pi}(y)\right] &=&0 
\end{eqnarray} 

\subsubsection{Isospin  gauge-fixing Lagrangian $L^{R_{\xi}}$}
\label{IsospinGaugeFixingCurrent}
The isospin  gauge-fixing Lagrangian is 
\begin{eqnarray}
\label{IsospinGaugeFixing}
L^{R_{\xi};GaugeFix}&=&-\frac{1}{2\xi}\left[\partial_{\beta}{\vec W}^{\beta}+\xi M_W{\vec \pi}\right]^2 \\
M_W&=&\half g_2\HVEV;  \nonumber \\
{\vec F}_W &=& \partial_{\beta}{\vec W}^{\beta}+\xi M_W {\vec \pi}\nonumber\,.
\end{eqnarray} 
Consider the current
\begin{eqnarray}
\label{LIsospinGaugeFixingIsospinCurrents}
{\vec J}^{\mu ;R_{\xi};GaugeFix}_L&=-& \frac{1}{\xi}\left[{\vec W}^{\mu} \times {\vec F}_W \right]  
\end{eqnarray}
Useful commutators are
\begin{eqnarray}
\delta(z_0-y_0)&&\left[ {\vec J}^{0 ;R_{\xi};GaugeFix}_L(z),H(y)\right] =0 \nonumber \\
\delta(z_0-y_0)&&\left[{\vec J}^{0 ;R_{\xi};GaugeFix}_L(z),{\vec \pi}(y)\right] =0 \nonumber
\end{eqnarray} 

\subsubsection{$R_{\xi}$ ghost-sector Lagrangian} 
\label{IsospinGhostsCurrent}
The $R_{\xi}$ ghost-sector Lagrangian is
\begin{eqnarray}
\label{LGhost}
L^{ R_\xi}_{Ghosts} &=& \partial^\mu {\bar {\vec \eta}}\cdot \Big[ \partial_\mu {\vec \omega} +g_2  {\vec W}_{\mu} \times {\vec \omega} \Big]\nonumber \\
&-& \xi\frac{M_W^2}{\HVEV} {\bar {\vec \eta}}\cdot \Big[ H {\vec { \omega}} +{\vec \pi} \times {\vec { \omega}}\Big]  \nonumber \\
&+& \text{surface}\, \text{terms}
\end{eqnarray}
Consider the current
\begin{eqnarray}
\label{LGhostIsospinCurrent}
{\vec J}^{\mu;R_\xi}_{L;Ghosts} &=&- \partial^\mu {\bar {\vec \eta}}\times {\vec \omega}
\end{eqnarray}
Useful commutators are
\begin{eqnarray}
0&=&\delta(z_0-y_0)\left[ {\vec J}^{0;R_\xi}_{L;Ghosts} (z),H(y)\right] \nonumber \\
0&=&\delta(z_0-y_0)\left[ {\vec J}^{0;R_\xi}_{L;Ghosts} (z),{\vec \pi}(y)\right] \nonumber
\end{eqnarray} 

\subsection{Total $SU(2)_L$ current and commutators}
\label{TotalLandauGaugeSM}
The total $SU(2)_L$ current is therefore
\begin{eqnarray}
\label{SMIsospinCurrent}
{\vec J}^{\mu}_{L} &=& {\vec J}^{\mu}_{L;Schwinger} + {\vec {\cal J}}^{\mu}_{L} \nonumber \\
{\vec J}^{\mu}_{Schwinger} &=& \half {\vec \pi}\times \partial^\mu {\vec \pi} + \half\Big( {\vec \pi}\partial^\mu H -H \partial^\mu {\vec \pi}\Big) \nonumber \\
{\vec {\cal J}}^{\mu}_{L} &=&-\frac{1}{4} g_2 {\vec W}^{\mu}\left[ H^2+{\vec \pi}^2\right] \nonumber \\
&+&  {\vec W}^{\mu \nu} \times {\vec W}_{\nu} \nonumber \\
&-& \lim_{\xi \to 0} \frac{1}{\xi}\left[{\vec W}^{\mu} \times {\vec F}_W(0) \right] \nonumber \\
&-& \partial^\mu {\bar {\vec \eta}}\times {\vec \omega} \nonumber 
\end{eqnarray}
with
\begin{eqnarray}
{\vec F}_W &=& \partial_{\beta}{\vec W}^{\beta} +\xi M_W {\vec \pi}\nonumber \\
M_W&=&\frac{1}{2}g_2\HVEV \nonumber
\end{eqnarray}
Its divergence is
\begin{eqnarray}
\partial_{\mu} {\vec J}^{\mu}_{L}&=&\half M_W\left[{\vec \pi} \times {\vec F}_W +H {\vec F}_W \right] 
\end{eqnarray}
Useful commutators are
\begin{eqnarray}
 \delta(z_0-y_0)&&\left[ {\vec J}_{L}^{0}(z)- {\vec J}_{Schwinger}^{0}(z),H(y)\right]=0 \quad \nonumber \\
 \delta(z_0-y_0)&&\left[ {\vec J}_{L}^{0}(z)- {\vec J}_{Schwinger}^{0}(z),{\vec \pi}(y)\right]=0 \quad \nonumber \\
\delta(z^0-y^0)&&\left[ {\vec J}_{Schwinger}^{0}(z),H(y)\right]\nonumber \\
&&=-\half i\delta^4(z-y) {\vec \pi}(z)  \\
 \delta(z^0-y^0)&&\left[ { J}_{Schwinger}^{0;t}(z),{ \pi^{t_1}}(y)\right] \nonumber \\
&&=\half i \delta^4(z-y)
\times \Big( \epsilon^{tt_1t_2} \pi^{t_2} (z)+ \delta^{tt_1}H(z) \Big) \nonumber
\end{eqnarray} 

\subsection{Total $SU(2)_L$ Lagrangian}
\label{TotalSU2LandauGauge}
The total $SU(2)_L$ Lagrangian is thus
\begin{eqnarray}
\label{LTotalEwSM}
L&=&L^W+L^\phi +L_{GaugeFix}^{R_{\xi}^{W}} + L_{Ghosts}^{R_{\xi}^{W}}
\end{eqnarray}
with
\begin{eqnarray}
L^W&=&-\half Tr \left(W_{\mu \nu}W^{\mu \nu}\right)  \nonumber \\
W_{\mu}&=& {\vec t} \cdot {\vec W}_{\mu} \equiv \half{\vec \sigma} \cdot {\vec W}_{\mu}\nonumber \nonumber \\
W_{\mu \nu}&=&\partial_{\mu}W_{\nu}-\partial_{\nu}W_{\mu} +ig_2\left[W_{\mu},W_{\nu}\right] \nonumber \\
L^\phi&=&\vert D_{\mu}\phi \vert ^2 - V \nonumber \\
V &=& \mu^2_\phi (\phi^{\dagger}\phi) + \lambda^2_\phi (\phi^{\dagger}\phi)^2 \nonumber \\
D_{\mu}\phi &=& \left[\partial_{\mu} + ig_2W_{\mu}  \right]\phi  \\
L_{GaugeFix}^{R_{\xi}^{W}} &+& L_{Ghosts}^{R_{\xi}^{W}} \nonumber \\
&=&s\Big( {\bar {\vec \eta}} \cdot \Big[ {\vec F}_W +\half \xi {\vec b}_W\Big] \nonumber\\
{\vec F}_W &=& \partial_{\beta}{\vec W}^{\beta} +\xi M_W {\vec \pi} \nonumber \\
M_W&=&\half g_2 \HVEV  \nonumber
\end{eqnarray}

\section{$R_\xi$ gauge renormalized $SU(2)_L\times U(1)_Y$ 
for gauge bosons, a complex scalar doublet, ghosts and anti-ghosts}
\label{EwSMLightParticleCurrents}

In this Appendix, we construct the $SU(2)_L\times U(1)_Y$ 
isospin sub-current ${\vec J}^\mu_{L;2\otimes 1}$, 
together with its divergence, 
for bosons, ghosts and anti-ghosts in an arbitrary $R^{2\otimes 1}_\xi$ gauge. 
The Appendix is pedagogically complete, 
and repeats certain results found in the body of the paper. 

The notation ``$2\otimes 1$" is short-hand for $SU(2)_L\times U(1)_Y$.

Sixteen renormalized bosons and ghosts appear in $SU(2)_L\times U(1)_Y$:
one scalar $H$, three pseudo-scalars $\vec \pi$, 
three isospin gauge fields ${\vec W}_\mu$, 
three isospin ghosts $\vec \omega$, 
three anti-ghosts $\bar {\vec \eta}$, 
a hypercharge gauge field $B_\mu$, 
a hypercharge ghost $\omega_B$ and anti-ghost ${\bar \eta}_B$.
\bea
\label{SMBosonsGhostSMs}
\{\Phi_i\}&=&\{ H, {\vec \pi}; {\vec W}_{\mu},  {\vec \omega}, {\bar {\vec \eta}};B_\mu , \omega_B, {\bar \omega}_B 
 \}.  i=1,16
\eea

\subsection{Global $SU(2)\times U(1)_Y$ and BRST transformations}
\label{TransformationsSM}

Global BRST transformations \cite{BecchiRouetStora,Tyutin1975,Tyutin1976,Nakanashi1966,Lautrup1967,Weinberg1995} 
$s_{2\otimes 1}$ appear in \eqref{EwSMBRSTTransformations}.

Anomaly-free un-deformed rigid/global $\delta_{SU(2)_L\times U(1)_Y}$ 
transformations of fields by constant $\vec \Omega$ and $\Omega_B$ 
appear in \eqref{EwSMU(1)Transformations}.

The global BRST and $SU(2)_L\times U(1)_Y$ transformation sets  
\eqref{EwSMBRSTTransformations} and \eqref{EwSMU(1)Transformations}
commute 
\bea
\label{SU2BRSTFieldCommutatorsSM}
\Big[ \delta_{2\otimes 1}, s_{2\otimes 1} \Big] &=& 0
\eea

\subsection{$SU(2)_L$ sub-current and its divergence}
\label{IsospinCurrentAndDivergence}

Because $SU(2)_L$ is an applicable sub-group of $SU(2)_L\times U(1)_Y$,
\bea
\label{CommutatorsSU(2)}
\Big[ \delta_{SU(2)_L}, s_{2\otimes 1} \Big]&=& 0
\eea
as in \eqref{SU2LBRSTcommuteonL2x1}, 
and we can form the classical $SU(2)_L$ sub-current
\begin{eqnarray}
\label{IncrementalCurrentsSM}
-{\hat J}^{\mu}_{L;2\otimes1} &\equiv& -\big(g_2{\vec \Omega} \big)\cdot {\vec J}^{\mu}_{L;2\otimes1} \\
&=& \sum_i^{Particles} \Big(\frac{\partial}{\partial(\partial_{\mu}\Phi_i)} L_{2\otimes1}\Big) \Big(\delta_{SU(2)_L}\Phi_i\Big) \nonumber
\end{eqnarray}

and its divergence
\begin{eqnarray}
\label{IncrementalDivergencesSM}
&&-\partial_{\mu} {\hat J}^{\mu}_{L;2\otimes1} \equiv -\big(g_2{\vec \Omega} \big)\cdot \partial_\mu{\vec J}^{\mu}_{L;2\otimes1} \nonumber \\
&&\quad = \sum_i^{Particles} \Big[ \Big(\frac{\partial}{\partial(\partial_{\mu}\Phi_i)}  L\Big)
\Big(\partial_{\mu} \delta_{SU(2)_L}\Phi_i\Big) \nonumber \\
&&  \qquad \qquad +\Big(\partial_\mu \frac{\partial}{\partial(\partial_{\mu}\Phi_i)}  L\Big)\Big(\delta_{SU(2)_L}\Phi_i\Big) \Big] \nonumber \\
&&\quad = \sum_i^{Particles} \Big[ \Big(\frac{\partial}{\partial(\partial_{\mu}\Phi_i)}  L\Big)
\Big(\delta_{SU(2)_L}\partial_\mu\Phi_i\Big) \nonumber \\
&&  \qquad \qquad +\Big(\frac{\partial}{\partial( \Phi_i)}  L\Big)\Big(\delta_{SU(2)_L}\Phi_i\Big) \Big] \nonumber \\
&&\quad =\delta_{SU(2)_L}L_{2\otimes 1} \nonumber \\
&&\quad = \delta_{SU(2)_L}\Big(L_{GaugeFix}^{R^{2\otimes 1}_\xi}
	+L_{Ghost}^{R^{2\otimes 1}_\xi} \Big)\nonumber
\eea
\bea
&&\quad =\delta_{SU(2)_L}s_{2\otimes 1}\Big( {\bar {\vec \eta}} \cdot \Big[ {\vec F}_W +\half \xi {\vec b}_W\Big] +{\bar { \eta}}_B \Big[ {F}_B +\half \xi {b}_B\Big]\Big)\nonumber \\
&&\quad =s_{2\otimes 1}\Big( {\bar {\vec \eta}} \cdot \Big[ \delta_{SU(2)_L}{\vec F}_W \Big] +{\bar { \eta}}_B \Big[\delta_{SU(2)_L} {F}_B \Big]\Big)\nonumber \\
&&\quad =\delta_{SU(2)_L}s_{2\otimes 1} \Big( {\bar {\vec \eta}} \cdot {\vec F}_W  +{\bar { \eta}}_B {F}_B \Big)\nonumber \\
&&\quad =\delta_{SU(2)_L}\Big( \Big[ s_{2\otimes 1}{\bar {\vec \eta}} \Big]\cdot {\vec F}_W +{\bar {\vec \eta}} \cdot \Big[s_{2\otimes 1} {\vec F}_W \Big]\Big)\nonumber \\
&&\qquad \qquad +\delta_{SU(2)_L}\Big( \Big[s_{2\otimes 1} {\bar { \eta}}_B\Big] {F}_B +{\bar { \eta}}_B \Big[s_{2\otimes 1}{F}_B \Big]\Big)\nonumber \\
&&\quad =\delta_{SU(2)_L}\Big(  {\vec b} \cdot {\vec F}_W +b_B {F}_B \Big)\nonumber \\
&&\quad = -\frac{1}{\xi} {\vec F}_W \cdot \delta_{SU(2)_L}{\vec F}_W -\frac{1}{\xi} {F}_B \delta_{SU(2)_L}{F}_B \nonumber \\
&&\quad = - \Big( g_2{\vec \Omega} \Big) \cdot \Big\{ \frac{1}{2} M_W \Big[ H{\vec F}_W +{\vec \pi}\times {\vec F}_W\Big] \nonumber \\
&& \qquad \qquad+\frac{1}{2} M_B {F}_B \Big( -\pi_2,\pi_1,H\Big)\Big\}  \nonumber 
\end{eqnarray}
Thus
\begin{eqnarray}
\label{JuL2x1divergence}
&&\partial_\mu {\vec J}^\mu_{L;2\otimes 1} =  \frac{1}{2} M_W \Big[ H{\vec F}_W +{\vec \pi}\times {\vec F}_W\Big] 
\\&& \qquad \qquad
+\frac{1}{2} M_B {F}_B \Big( -\pi_2,\pi_1,H\Big) \nonumber
\end{eqnarray} 

To arrive at \eqref{JuL2x1divergence}, 
we have used the EOM for $\{\Phi_i\}$.
In particular,
the Lagrange-multiplier fields ${\vec b},b_B$ are constrained, 
\bea
\label{2x1bconstraints}
{\vec b}&=& -\frac{1}{\xi} {\vec F}_W; \qquad  {b}_B= -\frac{1}{\xi} { F}_B\\
\eea
and the ${\vec \omega},\omega_B$ ghost EOM are
\bea
\label{GhostEofMSM}
s_{2\otimes 1}{\vec F}_W&=& \partial^2{\vec \omega}+g_2\partial_\mu\Big({\vec W}^\mu \times {\vec \omega}\Big)
+\xi\frac{M_W^2}{\HVEV} \Big( H {\vec \omega}+{\vec \pi} \times {\vec \omega}\Big) \nonumber \\
&+& \xi \frac{M_B M_W}{\HVEV} \Big(-\pi_2,\pi_1,H\Big)\omega_B \\
&=&0 \nonumber \\
s_{2\otimes 1}{F}_B&=& \partial^2{\omega}_B
+\xi\frac{M_B^2}{\HVEV}  H { \omega}_B 
+\xi \frac{M_B M_W}{\HVEV} \Big[H{\vec \omega} +{\vec \pi}\times {\vec \omega}\Big]_3\nonumber \\
&=&0 \nonumber
\eea
Eq. (\ref{GhostEofMSM}) are used in the construction of the classical current ${\vec J}_L^\mu$
and its divergence $\partial_\mu{\vec J}_L^\mu$, 
but not in the quantum mechanics of the Lagrangian (\ref{LTotalEwSM}). 

\subsection{$SU(2)_L$ sub-current and commutators}
\label{CurrentsDivergencesEwSM}

The global $SU(2)_L$ Schwinger model \cite{Schwinger1957}, 
its relevant current,
the divergence of that current, and relevant commutators are found in
Subsubsection \ref{SchwingerModel}.

The gauge-invariant isospin Lagrangian $L^W$, 
its relevant current,
the divergence of that current, and relevant commutators are found in
Subsubsection \ref{IsospinLagrangian}.

The isospin  gauge-fixing Lagrangian $L^{R_\xi}_{GaugeFix}$ 
its relevant current,
the divergence of that current, and relevant commutators are found in
 Subsubsection \ref{IsospinGaugeFixingCurrent}.

The isospin ghost Lagrangian $L^{R_\xi}_{Ghosts}$, 
its relevant current,
the divergence of that current, 
and relevant commutators are found in Subsubsection \ref{IsospinGhostsCurrent}}

\subsubsection{Gauge-invariant hypercharge Lagrangian $L^B_{2\otimes 1}$}
\label{IsospinLagrangianEwSM}}
The gauge-invariant hypercharge Lagrangian is
\begin{eqnarray}
\label{LHyperchargeEwSM}
L^B_{2\otimes 1} &=&-\frac{1}{4}B_{\mu\nu}  B^{\mu\nu}  \\
B_{\mu\nu}&=& \partial_\mu B_\nu-\partial_\nu B_\mu  \nonumber 
\end{eqnarray}
Its relevant current is
\bea
{\vec J}^B_{L;2\otimes 1} &=&0 
\eea
and so has commutators
\bea
\delta(z^0-y^0)&&\left[ {\vec J}_{L;2\otimes 1}^{0;B}(z),H(y)\right]=0 \nonumber \\
\delta(z^0-y^0)&&\left[ {\vec J}_{L;2\otimes 1}^{0;B}(z),{\vec \pi}(y)\right]=0 \,.
\eea

\subsubsection{Gauge-invariant scalar Lagrangian $L^\phi_{2\otimes 1}$}
\label{ScalarLagrangianEwSM}
The gauge-invariant scalar Lagrangian is
\begin{eqnarray}
\label{LphiEwSM}
L^\phi_{2\otimes 1} &=&\vert D^{2\otimes 1}_{\mu}\phi \vert ^2 - V \nonumber \\
V &=& \mu^2 (\phi^{\dagger}\phi) + \lambda^2_\phi (\phi^{\dagger}\phi)^2  \\
D^{2\otimes 1}_{\mu}\phi &=& \left[\partial_{\mu} + ig_2W_{\mu}+i{\tilde e}B_\mu\right]\phi \nonumber \\
W_{\mu}&=& {\vec t} \cdot {\vec W}_{\mu} \equiv \half{\vec \sigma} \cdot {\vec W}_{\mu}\nonumber \\
{\tilde e}&=&\half Y_\phi g_1 =-\half g_1 \nonumber
\end{eqnarray}
Consider the current
\begin{eqnarray}
\label{ScalarLagrangianJLeft}
{\vec J}^{\mu;\phi}_{L;2\otimes 1}&=& {\vec J}^{\mu}_{Schwinger} + {\vec {\cal J}}_{L;2\otimes 1}^{\mu;\phi}  \\
{\vec {\cal J}}^{\mu;\phi}_{L;2\otimes 1} &=&-\frac{1}{4} g_2 {\vec W}^{\mu}\left[ H^2+{\vec \pi}^2\right] \nonumber \\
&+& \half g_1 B^{\mu}\Big(\pi_1\pi_3-\pi_2 H, \quad \pi_2\pi_3+\pi_1 H,  \nonumber \\
  &&\qquad \qquad\half(H^2 +\pi_3^2-\pi_1^2-\pi_2^2)\Big) \nonumber \
\end{eqnarray}
and the commutators
\begin{eqnarray}
\delta(z^0-y^0)&&\left[ {\vec J}_{L;2\otimes 1}^{0;\phi}(z),H(y)\right] \nonumber \\
&=&\delta(z^0-y^0)\left[  {\vec J}_{Schwinger}^{0}(z),H(y)\right]  \\
 \delta(z^0-y_1^0)&&\left[ { J}_{L;2\otimes 1}^{0;t;\phi}(z),{ \pi}^{t_1}(y_1)\right] \nonumber \\
&=& \delta(z^0-y_1^0)\left[  { J}_{Schwinger}^{0;t}(z),{ \pi}^{t_1}(y_1)\right] \nonumber
\end{eqnarray}

\subsubsection{Global $SU(2)_L\times U(1)_Y$ gauge-fixing, ghosts and anti-ghosts}
\label{GaugeFixingGhostsEwSM}
The global $SU(2)_L\times U(1)_Y$ Lagrantian for gauge-fixing, ghosts and anti-ghosts is
\begin{eqnarray}
\label{GaugeFixingGhosts}
&& L_{GaugeFix}^{R_\xi} + L_{Ghosts}^{R_\xi} \\
&& \quad \quad =s_{2\otimes 1}\Big( {\bar {\vec \eta}} \cdot \Big[ {\vec F}_W +\half \xi {\vec b}_W\Big] +{\bar { \eta}}_B \Big[ {F}_B +\half \xi {b}_B\Big]\Big)\nonumber \\
&& {\vec F}_W= \partial_\mu {\vec W}^\mu + \xi M_W {\vec \pi}\nonumber \\
&& {F}_B= \partial_\mu {B}^\mu + \xi M_B {\pi}_3\nonumber 
\end{eqnarray}
Consider the current
\bea
&& {\vec J}_{L;GaugeFix}^{\mu;R_\xi} + {\vec J}_{L;Ghosts}^{\mu;R_\xi}
= {\vec J}_{L;GaugeFix}^{\mu;R_\xi^{W}} + {\vec J}_{L;Ghosts}^{\mu;R_\xi} 
\eea
and its commutators
\bea
 &&\delta(z_0-y_0)\left[ {\vec J}_{L;GaugeFix}^{0;R_\xi}(z),H(y)\right]=0 \quad \nonumber \\
 &&\delta(z_0-y_0)\left[ {\vec J}_{L;GaugeFix}^{0;R_\xi}(z),{\vec \pi}(y)\right]=0 \quad \nonumber \\
&&\delta(z_0-y_0)\left[ {\vec J}_{L;Ghosts}^{0;R_\xi}(z),H(y)\right]=0 \quad \nonumber \\
 &&\delta(z_0-y_0)\left[ {\vec J}_{L;Ghosts}^{0;R_\xi}(z),{\vec \pi}(y)\right]=0 \quad \nonumber
\end{eqnarray}

\subsection{Total $SU(2)_L$ current and comutators}
\label{TotalLandauGaugeEwSM}
The total $SU(2)_L$ current is
\begin{eqnarray}
\label{EwSMIsospinCurrent}
{\vec J}^{\mu}_{L;2\otimes 1} &=& {\vec J}^{\mu}_{L;Schwinger} + {\vec {\cal J}}^{\mu}_{L;2\otimes 1}  \\
{\vec J}^{\mu}_{Schwinger} &=& \half {\vec \pi}\times \partial^\mu {\vec \pi} + \half\Big( {\vec \pi}\partial^\mu H -H \partial^\mu {\vec \pi}\Big) \nonumber\\
{\vec {\cal J}}^{\mu}_{L;2\otimes 1} &=&  {\vec W}^{\mu \nu} \times {\vec W}_{\nu} \nonumber \\
&+& \half g_1 B^{\mu}\Big(\pi_1\pi_3-\pi_2 H, \quad \pi_2\pi_3+\pi_1 H,  \nonumber \\
&& \qquad \qquad \half(H^2 +\pi_3^2-\pi_1^2-\pi_2^2)\Big) \nonumber \\
&-& \frac{1}{4} g_2 {\vec W}^{\mu}\left[ H^2+{\vec \pi}^2\right]  \nonumber \\
&-& \lim_{\xi \to 0} \frac{1}{\xi}\left[{\vec W}^{\mu} \times {\vec F}_W \right] \nonumber \\
&-& \partial^\mu {\bar {\vec \eta}}\times {\vec \omega} \nonumber \\
\partial_{\mu} {\vec J}^{\mu}_{L;2\otimes 1}&=& \half M_B F_B  \left( -\pi_2,\pi_1,H\right) \nonumber \\
&+&\half M_W\left[H {\vec F}_W+{\vec \pi} \times {\vec F}_W \right]  \nonumber
\eea
Relevant commutators are
\bea 
\delta(z_0-y_0)&&\left[ {\vec J}_{L;2\otimes 1}^{0}(z)- {\vec J}_{L;Schwinger}^{0}(z),H(y)\right]=0 \quad \nonumber \\
 \delta(z_0-y_0)&&\left[ {\vec J}_{L;2\otimes 1}^{0}(z)- {\vec J}_{L;Schwinger}^{0}(z),{\vec \pi}(y)\right]=0 \quad \nonumber \\
\delta(z^0-y^0)&&\left[ {\vec J}_{Schwinger}^{0}(z),H(y)\right]\nonumber \\
&=&-\half i\delta^4(z-y) {\vec \pi}(z) \nonumber \\
\delta(z^0-y^0_1)&&\left[ { J}_{Schwinger}^{0;t}(z),{ \pi^{t_1}}(y_1)\right] \nonumber \\
&=&\half i \delta^4(z-y_1)
\times \Big( \epsilon^{tt_1t_2} \pi^{t_2} (z)+ \delta^{tt_1}H(z) \Big) \nonumber 
\eea
with
\bea
{\vec F}_W &=& \partial_{\beta}{\vec W}^{\beta} +\xi M_W {\vec \pi}\nonumber \\
{F}_B &=& \partial_{\beta}{B}^{\beta} +\xi M_B { \pi}_3  \\
M_W&=&\half g_2 \HVEV; \qquad M_B = {\tilde e}\HVEV \nonumber \\
  {\tilde e} &=& \half Y_\phi g_1 = -\half g_1; \qquad M_W^2+M_B^2=M_Z^2; \nonumber 
\end{eqnarray} 

\subsection{Total $SU(2)_L\times U(1)_Y$ Lagrangian in $R_{\xi}$ gauge}
\label{TotalEwSMLandauGauge}
The total $SU(2)_L\times U(1)_Y$ Lagrangian in $R_{\xi}$ gauge is
\begin{eqnarray}
\label{LTotalEwSM}
L_{2\otimes 1}&=&L^W+L^B_{2\otimes 1} +L^\phi_{2\otimes 1} +L_{GaugeFix}^{R_{}^{2\otimes 1}} + L_{Ghosts}^{R_{}^{2\otimes 1}} 
\eea
with
\bea
L^W&=&-\half Tr \left(W_{\mu \nu}W^{\mu \nu}\right)  \nonumber \\
W_{\mu}&=& {\vec t} \cdot {\vec W}_{\mu} \equiv \half{\vec \sigma} \cdot {\vec W}_{\mu}\nonumber \nonumber \\
W_{\mu \nu}&=&\partial_{\mu}W_{\nu}-\partial_{\nu}W_{\mu} +ig_2\left[W_{\mu},W_{\nu}\right] \nonumber \\
L^B_{2\otimes 1} &=&-\frac{1}{4}B_{\mu\nu}  B^{\mu\nu}  \\
B_{\mu\nu}&=& \partial_\mu B_\nu-\partial_\nu B_\mu  \nonumber \\
L^\phi_{2\otimes 1} &=&\vert D^{2\otimes 1}_{\mu}\phi \vert ^2 - V \nonumber \\
V &=& \mu^2 (\phi^{\dagger}\phi) + \lambda^2_\phi (\phi^{\dagger}\phi)^2 \nonumber \\
D^{2\otimes 1}_{\mu}\phi &=& \left[\partial_{\mu} + ig_2W_{\mu} +i{\tilde e}B_\mu \right]\phi \nonumber \\
L_{GaugeFix}^{R_{}^{2\otimes 1}} &+& L_{Ghosts}^{R_{}^{2\otimes 1}} \nonumber \\
&=&s_{2\otimes 1}\Big( {\bar {\vec \eta}} \cdot \Big[ {\vec F}_W +\half \xi {\vec b}_W\Big] +{\bar { \eta}}_B \Big[ {F}_B +\half \xi {b}_B\Big]\Big)\nonumber\,.
\eea
Here
\bea
{\vec F}_W &=& \partial_{\beta}{\vec W}^{\beta} +\xi M_W {\vec \pi} \nonumber \\
{F}_B &=& \partial_{\beta}{B}^{\beta} +\xi M_B \pi_3 \\
M_W&=&\half g_2 \HVEV; \qquad M_B = {\tilde e}\HVEV \nonumber \\
  {\tilde e} &=& \half Y_\phi g_1 = -\half g_1; \qquad M_W^2+M_B^2=M_Z^2\,. \nonumber 
\end{eqnarray}

\section{$SU(2)_L\times U(1)_Y$ Master equation and WTIs in Landau gauge}
\label{EwSMTotalMasterEqWTI}
We focus on the global isospin current ${\vec J}^{\mu}_{L;2\otimes 1}$. 
As usual, we form time-ordered amplitudes of
products of ${\vec  J}^\mu_{L;2\otimes 1}$, 
with N scalars  
and M pseudo-scalars 
$\big< 0 \vert T\Big[ 
\Big({\vec  J}_{L;2\otimes 1}^\mu(z)\Big) h(x_1)\cdots h(x_N) 
						\pi^{t_1}(y_1)\cdots \pi^{t_M}(y_M)
\Big]\vert 0\big>$
and examine the divergence of such connected amplitudes:
\bea
\label{EwSMMasterEquation}
&&\partial_\mu  \big< 0\vert 
T\Big[ \Big( J_{L;2\otimes 1}^{\mu;t}(z)\Big) \\
&&\quad \times h(x_1)\cdots h(x_N)\pi^{t_1}(y_1)\cdots \pi^{t_M}(y_M)\Big]
		\vert 0 \big>_{Connected} \nonumber
\eea

\subsection{Right-hand side (RHS) of Master Equation}
\label{EwSMRHS}

Making use of  current conservation (\ref{EwSMIsospinCurrent}),
\bea
\label{EwSMCurrentConservationPrime}
\partial_{\mu} {\vec J}^{\mu}_{L;2\otimes 1}&=&\half M_W\Big[ H \partial_\beta{\vec W}^\beta+ {\vec \pi} \times \partial_\beta{\vec W}^\beta \Big] \nonumber \\
&+&\half M_B \partial_\beta B^\beta  \left( -\pi_2,\pi_1,H\right) 
\eea
G. 't Hooft's gauge-fixing \cite{tHooft1971}  conditions \eqref{GaugeConditionsPrimePrimePrime}, 
\begin{eqnarray}
\label{EwSMGaugeConditions}
&&\big< 0\vert T\Big[ \Big( \partial_{\mu}{\vec W}^{\mu}(z)\Big) \nonumber \\
&&\quad \times h(x_1)...h(x_N)\pi_{t_1}(y_1)...\pi_{t_M}(y_M)\Big]\vert 0\big>_{\rm connected} \nonumber \\
&&\quad =0 \nonumber \\
&&\big< 0\vert T\Big[ \Big( \partial_{\mu}{B}^{\mu}(z)\Big) \nonumber \\
&&\quad \times h(x_1)...h(x_N)\pi_{t_1}(y_1)...\pi_{t_M}(y_M)\Big]\vert 0\big>_{\rm connected} \nonumber \\
&&\quad =0
\end{eqnarray}
and the equal-time commutation relations (\ref{EwSMIsospinCurrent})
\bea
\delta(z_0-x_0) \left[\Big( {\vec J}^0_{L;2\otimes 1}-{\vec J}^0_{Schwinger} \Big)(z)  ,h(x)\right] &=& 0 \nonumber\\
\delta(z_0-x_0) \left[\Big( {\vec J}^0_{L;2\otimes 1}-{\vec J}^0_{Schwinger} \Big)(z)  ,{\vec \pi}(x)\right] &=& 0 \quad \quad
\eea
a short calculation reveals
\bea
\label{EwSMMasterRHS1}
&&\partial_\mu  \big< 0\vert 
	T\Big[J_{L;2\otimes 1}^{\mu;t}(z) h(x_1)\cdots h(x_N) \nonumber \\
&&\quad\times \pi^{t_1}(y_1)\cdots \pi^{t_M}(y_M)\Big]
		\vert 0 \big>_{Connected}  \nonumber\\
&&= \big< 0\vert 
	T\Big[ \Big(\partial_\mu J_{L;2\otimes 1}^{\mu;t}(z)\Big) \nonumber \\
&&\quad\times h(x_1)\cdots h(x_N) \pi^{t_1}(y_1)\cdots \pi^{t_M}(y_M)\Big]
		\vert 0 \big>_{Connected}  \\
&& + \sum_{n=1}^{n=N} \big< 0\vert 
	T\Big[ h(x_1)\cdots h(x_{n-1})\nonumber \\
&& \quad\times \delta(z^0-x_n^0)\Big[ J_{L;2\otimes 1}^{0;t}(z),h(x_n)\Big] \nonumber \\
&&\quad\times h(x_{n+1})\cdots h(x_N) \pi^{t_1}(y_1)\cdots \pi^{t_M}(y_M)\Big]
		\vert 0 \big>_{Connected}  \nonumber\\
&& + \sum_{m=1}^{m=M} \big< 0\vert 
	T\Big[ h(x_1)\cdots h(x_N)\pi^{t_1}(y_1)\cdots \pi^{t_{m-1}}(y_{m-1})\nonumber \\
&& \quad\times \delta(z^0-y_m^0)\Big[ J_{L;2\otimes 1}^{0;t}(z),\pi^{t_m}(y_m)\Big] \nonumber \\
&&\quad\times \pi^{t_{m+1}}(y_{m+1})\cdots \pi^{t_M}(y_M)\Big]
		\vert 0 \big>_{Connected} \nonumber
\eea
\bea
\label{EwSMMasterRHS2}
&& = \sum_{n=1}^{n=N} \big< 0\vert 
	T\Big[ h(x_1)\cdots h(x_{n-1})\nonumber \\
&& \quad\times \delta(z^0-x_n^0)\Big[ J_{Schwinger}^{0;t}(z),h(x_n)\Big] \nonumber \\
&&\quad\times h(x_{n+1})\cdots h(x_N) \pi^{t_1}(y_1)\cdots \pi^{t_M}(y_M)\Big]
		\vert 0 \big>_{Connected}  \nonumber\\
&& + \sum_{m=1}^{m=M} \big< 0\vert 
	T\Big[ h(x_1)\cdots h(x_N)\pi^{t_1}(y_1)\cdots \pi^{t_{m-1}}(y_{m-1})\nonumber \\
&& \quad\times \delta(z^0-y_m^0)\Big[ J_{Schwinger}^{0;t}(z),\pi^{t_m}(y_m)\Big] \nonumber \\
&&\quad\times \pi^{t_{m+1}}(y_{m+1})\cdots \pi^{t_M}(y_M)\Big]
		\vert 0 \big>_{Connected}  \nonumber
\eea
in Landau gauge.
Here we have N external renormalized scalars $h=H-\HVEV$ (coordinates x), 
and M external ($CP=-1$) renormalized pseudo-scalars ${\vec \pi}$ (coordinates y, isospin $t$). 
We have also thrown away a sum of $M$ terms, proportional to $\HVEV$,
that corresponds entirely to disconnected graphs.

\subsection{Left-hand side (LHS) of Master Equation: Surface terms}
\label{EwSMSurfaceLHS}

We begin by studying the surface integral of 
the global $SU(2)_{L;2\otimes 1}$ sub-current (\ref{EwSMIsospinCurrent}) of the $SU(2)_L\times U(1)_Y$ standard electroweak model in Landau gauge 
\begin{eqnarray}
\label{EwSMIsospinCurrentPrime}
{\vec J}^{\mu}_{L;2\otimes 1} &=& {\vec J}^{\mu}_{L;Schwinger} + {\vec {\cal J}}^{\mu}_{L;2\otimes 1} \nonumber \\
{\vec {\cal J}}^{\mu}_{L;2\otimes 1} &=&-\frac{1}{4} g_2 {\vec W}^{\mu}\left[ H^2+{\vec \pi}^2\right]  \\
&+& \half g_1 B^{\mu}\Big(\pi_1\pi_3-\pi_2 H, \quad \pi_2\pi_3+\pi_1 H,  \nonumber \\
&& \qquad \qquad \half(H^2 +\pi_3^2-\pi_1^2-\pi_2^2)\Big) \nonumber \\
&+&  {\vec W}^{\mu \nu} \times {\vec W}_{\nu} 
- \lim_{\xi \to 0} \frac{1}{\xi}\left[{\vec W}^{\mu} \times \partial_\beta{\vec W}^{\beta} \right] \nonumber \\
&-& \partial^\mu {\bar {\vec \eta}}\times {\vec \omega} \nonumber
\end{eqnarray} 

In order to transform the LHS of  (\ref{EwSMMasterEquation}) into a surface term (and crucially,  to later form 1-soft-pion $SU(2)_{L;2\otimes 1}$ WTI), we Fourier-transform the current with far-infra-red ultra-soft momentum.
We use Stokes theorem, where $ {\widehat {z}_\mu}^{3-surface}$ is a unit vector normal to the $3$-surface. The time-ordered product constrains the $3$-surface to lie on, or inside, the light-cone. 

In this and the next subsections, we will prove that
\begin{eqnarray}
\label{EwSMSurfacePionPoleDominance}
&&\lim_{k_\lambda \to 0} \int d^4z e^{ikz} \partial_{\mu} \Big< 0\vert T\Big[ \Big( 
{\vec {\cal J}}^{\mu}_{L;2\otimes 1} (z)  \Big) \nonumber \\
&&\quad \quad \times h(x_1)...h(x_N) \pi^{t_1}(y_1)...\pi^{t_M}(y_M)\Big]\vert 0\Big>_{Connected} \nonumber \\
&& =\int d^4z \partial_{\mu} \Big< 0\vert T\Big[ \Big( 
{\vec {\cal J}}^{\mu}_{L;2\otimes 1} (z)  \Big)   \\
&&\quad \quad \times h(x_1)...h(x_N) \pi^{t_1}(y_1)...\pi^{t_M}(y_M)\Big]\vert 0\Big>_{Connected} \nonumber \\
&& =\int_{3-surface\to\infty}  
\!\!\!\!\!\!\!\!\!\!\!\!\!\!\!\!\!\!\!\!\!\!\!\!\!\!\!
d^3z \quad {\widehat {z}_\mu}^{3-surface} \Big< 0\vert T\Big[ \Big( 
{\vec {\cal J}}^{\mu}_{L;2\otimes 1} (z)  \Big)  \nonumber \\
&&\quad \quad \times h(x_1)...h(x_N) \pi^{t_1}(y_1)...\pi^{t_M}(y_M)\Big]\vert 0\Big>_{Connected} \nonumber \\
&&=0\nonumber\,.
\end{eqnarray}

\subsubsection{$SU(2)_L$ gauge fields' kinetic and interaction}
The surface integral of the 3rd term in ${\vec {\cal J}}_\mu$ in (\ref{EwSMIsospinCurrentPrime})
\begin{eqnarray}
\label{EwSMGaugesSurface}
&&\int_{3-surface\to\infty}  
\!\!\!\!\!\!\!\!\!\!\!\!\!\!\!\!\!\!\!\!\!\!\!\!\!\!\!
d^3z \quad {\widehat {z}_\mu}^{3-surface}  \Big< 0\vert T\Big[ \Big(  {\vec W}^{\mu \nu}\times {\vec W}_\nu \Big)(z)   \\
&&\quad \quad \times h(x_1)...h(x_N) \pi^{t_1}(y_1)...\pi^{t_M}(y_M)\Big]\vert 0\Big>_{Connected} \nonumber \\
&&\quad=0 \nonumber\,,
\end{eqnarray}
because each term in the current contains at least one massive 
${\vec W}^+_\beta$ or ${\vec W}^-_\beta$\,.
\begin{eqnarray}
\label{EwSMKineticCurrentGaugeFields}
 {\vec W}^{\mu \nu}\times {\vec W}_\nu
 &=& 
-{\vec W}_\nu \times \partial^\mu{\vec W}^{ \nu}+{\vec W}_\nu \times \partial^\nu{\vec W}^{ \mu}  \\
&+&g_2\Big[ {\vec W}^\mu \big({\vec W}^{ \nu}\cdot {\vec W}_{ \nu}\big)-{\vec W}^\nu \big({\vec W}^{ \mu}\cdot {\vec W}_{ \nu}\big)\Big]
\nonumber
\end{eqnarray}
${\vec W}^\pm_\beta$ are massive  in spontaneously broken $SU(2)_L\times U(1)_Y$. Propagators connecting ${\vec W}^\pm_\beta$,
from points on the 3-surface at infinity 
to the localized interaction points $(x_1...x_N;y_1...y_M)$, 
must stay inside the light-cone, but die off exponentially with mass,
$M_W^2\neq 0$. 
They are incapable of carrying information that far. 

The surface integral of the last term in ${\vec {\cal J}}_\mu$ in (\ref{EwSMIsospinCurrentPrime}), i.e. ghosts and anti-ghosts, is shown to vanish in Subsubsection \ref{GhostSurfaceTerm}.

\subsection{LHS of Master Equation:  Connected amplitudes linking the $\phi$-sector with external currents}
\label{EwSMAmplitudesLHS}

Connected momentum-space amplitudes, with $N$ external BEHs, $M$ external $\vec \pi$s,
and a current ${\vec J}_{L;2\otimes 1}^\mu$, 
are defined in terms of $\phi$-sector connected time-ordered products
\begin{eqnarray}
\label{EwSMCurrentConnectedAmplitudes} 
&&i{{\cal G}}_{\mu;N,M}^{t;t_1...t_M}
\Big( { J}^{t}_{L;2\otimes 1;\mu};p_1...p_N;q_1...q_M \Big) \nonumber \\
&&\qquad\quad\times (2\pi)^4\delta^4 \Big(k+\sum_{n=1}^N p_n +\sum_{m=1}^M q_m \Big)  \\
&& \quad \equiv\int d^4z e^{ikz} \prod_{n=1}^N\int d^4x_n e^{ip_nx_n} \prod_{m=1}^M\int d^4y_m e^{iq_my_m}  \nonumber \\
&&\qquad\quad \times 
\big< 0\vert T\Big[ \Big( { J}^{t}_{L;2\otimes 1;\mu} (z)\Big)  \nonumber  \\
&&\qquad \quad \times h(x_1)...h(x_N)\pi^{t_1}(y_1)...\pi^{t_M}(y_M)\Big]\vert 0\big>_{Connected}\nonumber
\end{eqnarray}
These appear throughout the proof of the WTI, so that
\begin{eqnarray}
\label{EwSMCurrentDivergenceConnectedAmplitudes} 
&&-k^\mu{{\cal G}}_{\mu;N,M}^{t;t_1...t_M}\Big({ J}^{t}_{L;2\otimes 1;\mu};p_1...p_N;q_1...q_M\Big) \nonumber \\
&&\qquad\quad\times (2\pi)^4\delta^4 \Big(k+\sum_{n=1}^N p_n +\sum_{m=1}^M q_m \Big)  \\
&& \quad \equiv\int d^4z e^{ikz} \prod_{n=1}^N\int d^4x_n e^{ip_nx_n} \prod_{m=1}^M\int d^4y_m e^{iq_my_m}  \nonumber \\
&&\qquad\quad \times 
\partial_\mu^z 
\big< 0\vert T\Big[ \Big( { J}^{t}_{L;2\otimes 1;\mu} (z)\Big)  \nonumber  \\
&&\qquad \quad \times h(x_1)...h(x_N)\pi^{t_1}(y_1)...\pi^{t_M}(y_M)\Big]\vert 0\big>_{Connected}\nonumber
\end{eqnarray}

\subsubsection{Gauge-invariant scalar-sector Lagrangian}
\label{EwSMScalarInvariantLagrangian}

The contribution of 
the 1st term in ${\vec {\cal J}}_\mu$ in (\ref{EwSMIsospinCurrentPrime}) 
to the LHS of the Master Equation vanishes
\begin{eqnarray}
\label{EwSMGaugeScalarConnectedAmplitudes} 
&&-k^\mu{{\cal G}}_{\mu;N,M}^{t;t_1...t_M}\Big( \Big[-\frac{1}{4}g_2 {W}^t_\mu \Big(H^2+{\vec \pi}^2 \Big) \Big];p_1...p_N;q_1...q_M\Big) \nonumber \\
&& \qquad \quad \sim-\frac{1}{4}g_2k^\mu {W}^t_\mu (k)\Big(\cdot\cdot\cdot \Big)(k^2)\nonumber\\
&&\qquad =0
\end{eqnarray}
because the isospin gauge conditions (\ref{EwSMGaugeConditions}) obeyed by the states read, in momentum-space
\begin{eqnarray}
\label{MomentumGaugeConditions}
k^\mu {\vec W}_\mu (k) =0
\end{eqnarray}

The contribution of the 2nd term in ${\vec {\cal J}}_\mu$ in (\ref{EwSMIsospinCurrentPrime}) to the LHS of the Master Equation vanishes
\begin{eqnarray}
\label{EwSMGaugeScalarConnectedAmplitudes} 
&&-k^\mu{{\cal G}}_{\mu;N,M}^{t;t_1...t_M}\Big( \Big[\half g_1 B^{\mu}\Big(\pi_1\pi_3-\pi_2 H,\nonumber \\
&& \qquad \qquad \qquad \pi_2\pi_3+\pi_1 H, \quad \half(H^2 +\pi_3^2-\pi_1^2-\pi_2^2)\Big)^t \Big]\nonumber \\
&& \qquad \quad\quad\quad\quad  ;p_1...p_N;q_1...q_M\Big) \nonumber \\
&& \qquad \quad \sim\frac{1}{2}g_1k^\mu {B}_\mu (k)\Big(\cdot\cdot\cdot \Big)^t(k^2)\nonumber\\
&&\qquad =0
\end{eqnarray}
because the hypercharge gauge condition (\ref{EwSMGaugeConditions}) obeyed by the states reads, in momentum-space
\begin{eqnarray}
\label{MomentumGaugeConditions}
k^\mu {B}_\mu (k) =0
\end{eqnarray}

The contribution of the 4th term in ${\vec {\cal J}}_\mu$ in (\ref{EwSMIsospinCurrentPrime}) to the LHS of the Master Equation is shown to vanish in Sub-subsection \ref{GaugeFixingLagrangian}.

\subsubsection{Total $SU(2)_L$ sub-current contribution to surface terms on LHS of Master equation}
\label{TotalLagrangianPrime}
The total $SU(2)_L$ sub-current contribution to surface terms on 
the LHS of the Master equation is
\bea
\label{EwSMMasterLHSPrime}
&&\lim_{k_\lambda \to 0} \int d^4z e^{ikz} \partial_\mu  \big< 0\vert 
	T\Big[J_{L;2\otimes 1}^{\mu;t}(z) h(x_1)\cdots h(x_N) \nonumber \\
&&\quad\times \pi^{t_1}(y_1)\cdots \pi^{t_M}(y_M)\Big]
		\vert 0 \big>_{Connected}  \nonumber\\
&&=\lim_{k_\lambda \to 0} \int d^4z e^{ikz} \partial_\mu  \big< 0\vert 
	T\Big[J_{Schwinger}^{\mu;t}(z) h(x_1)\cdots h(x_N) \nonumber \\
&&\quad\times \pi^{t_1}(y_1)\cdots \pi^{t_M}(y_M)\Big]
		\vert 0 \big>_{Connected} 
\eea

\subsection{${\vec  J}_{L+R;Schwinger}^\mu$ \& ${\vec  J}_{L-R;Schwinger}^\mu$ for even \& odd $M$}
\label{EwSMSchwingerMaster}

Amplitudes connecting the $\phi$-sector with the total $SU(2)_L$ isospin sub-current 
(\ref{EwSMIsospinCurrentPrime}) are constructed to conserve $CP$. Therefore, 
on and off-shell connected amputated T-matrix elements and Green's functions of an odd number of $(CP=-1)$ $\vec \pi$s and their derivatives, are zero.

In analogy with the proof in Subsection \ref{SchwingerMaster}, this allows us to write 2 sets of $SU(2)_L$ Ward-Takahashi identities governing the $\phi$-sector of $SU(2)_L\times U(1)_Y$: one for the vector current ${\vec  J}_{L+R;Schwinger}^\mu$ based on $M$ even, and one for the axial-vector current ${\vec  J}_{L-R;Schwinger}^\mu$ based on $M$ odd. In this paper, we focus on those strong constraints placed on the scalar-sector of the $SU(2)_L\times U(1)_Y$, by the axial-vector WTI.

\subsection{$SU(2)_L\times U(1)_Y$ axial-vector Master Equation in ${\vec  J}_{L-R;Schwinger}^\mu$}
\label{EwSMAxialVectorSchwingerMaster}
We now assemble the axial-vector Master Equation, from which we derive our ``1-soft-$\pi$" $SU(2)_{L-R}$ WTI.
The LHS is from (\ref{EwSMMasterLHSPrime}), the RHS from (\ref{EwSMMasterRHS1},\ref{EwSMMasterRHS2}):
\begin{eqnarray}
\label{EwSMAxialVectorMasterEquation}
&&\lim_{k_\lambda \to 0} \int d^4z e^{ikz} \int d^4z e^{ikz}\partial_{\mu} 
\Big< 0\vert T\Big[ \Big( {\vec J}^{\mu}_{L-R;Schwinger}(z) \Big)  \nonumber \\
&&\quad \quad \times h(x_1)...h(x_N) \pi^{t_1}(y_1)...\pi^{t_M}(y_M)\Big]\vert 0\Big>_{Connected} \nonumber\\
&& =\lim_{k_\lambda \to 0}\int d^4z e^{ikz}\sum_{n=1}^{n=N} \big< 0\vert 
	T\Big[ h(x_1)\cdots h(x_{n-1}) \\
&& \quad\times \delta(z^0-x_n^0)\Big[ J_{L-R;Schwinger}^{0;t}(z),h(x_n)\Big] \nonumber \\
&&\quad\times h(x_{n+1})\cdots h(x_N) \pi^{t_1}(y_1)\cdots \pi^{t_M}(y_M)\Big]
		\vert 0 \big>_{Connected}  \nonumber\\
&& + \lim_{k_\lambda \to 0}\int d^4z e^{ikz}\sum_{m=1}^{m=M} \big< 0\vert 
	T\Big[ h(x_1)\cdots h(x_N)\nonumber \\
&&\quad\times \pi^{t_1}(y_1)\cdots \pi^{t_{m-1}}(y_{m-1})\nonumber \\
&& \quad\times \delta(z^0-y_m^0)\Big[ J_{L-R;Schwinger}^{0;t}(z),\pi^{t_m}(y_m)\Big] \nonumber \\
&&\quad\times \pi^{t_{m+1}}(y_{m+1})\cdots \pi^{t_M}(y_M)\Big]
		\vert 0 \big>_{Connected}\nonumber
\end{eqnarray}
Eq. (\ref{EwSMAxialVectorMasterEquation}) is true for any M, odd or even. 
It is derived for $M$ odd in analogy with 
(\ref{SMMasterEquation},\ref{MasterRHS1},\ref{SMSurfacePionPoleDominance},\ref{MasterLHSPrime}).
It is also satisfied trivially, in analogy with (\ref{AxialCurrentMEven}), for $M$ even.

This paper is based on the de facto conservation, 
in the 1-soft-$\pi$ limit, of ${\vec J}^{\mu}_{L-R;Schwinger}$, 
for $CP$-conserving connected amputated  Green's functions 
and on-shell T-matrix elements, in the $\phi$-sector of $SU(2)_L\times U(1)_Y$.

\subsection{$SU(2)_L\times U(1)_Y$ Ward-Takahashi identities}
\label{EwSMMaster} 
The Master equation (\ref{EwSMAxialVectorMasterEquation}) 
is mathematically identical to that in (\ref{AxialVectorMasterEquation}).
This proves that, for each $SU(2)_{L-R}$ WTI that is true in $SU(2)_L$, 
an analogous $SU(2)_{L-R}$ WTI is true in $SU(2)_L\times U(1)_L$.
Appendix \ref{DerivationWTIAHM} proved $SU(2)_L$ WTI relations 
among 1-$\phi$-R $\phi$-sector  T-Matrix elements $T_{N,M}$,
 as well as $SU(2)_L$ WTI relations among 1-$\phi$-I 
$\phi$-sector Green's functions $\Gamma_{N,M}$, in spontaneously broken $SU(2)_L$.
Analogous $SU(2)_{L-R}$ WTI relations among 1-$\phi$-R $\phi$-sector T-Matrix elements $T^{2\otimes 1}_{N,M}$, 
as well as analogous $SU(2)_{L-R}$ WTI relations among 1-$\phi$-I 
$\phi$-sector  Green's functions $\Gamma_{N,M}^{2\otimes 1}$, 
are therefore true for spontaneously broken $SU(2)_L\times U(1)_Y$.

But there is one huge difference! 
The renormalization of  our $SU(2)_{L-R}$ WTI, 
governing $\phi$-sector $T_{N,M}^{2\otimes 1}$ and $\Gamma_{N,M}^{2\otimes 1}$, 
now includes the all-loop-orders contributions of 
virtual isospin and hypercharge gauge bosons, $\phi$-scalars, anti-ghosts, and ghosts, 
i.e.  ${\vec W}^\mu$ , $B^\mu$ $h$, ${\vec \pi}$, ${\bar {\vec \eta}}$,
 ${\bar \eta}_B$,${\vec \omega}$, and ${ \omega}_B$ respectively. 

The $SU(2)_L\times U(1)_Y$ Master equation 
relates connected time-ordered products, in analogy with (\ref{MasterEquation}):
\begin{eqnarray}
\label{EwSMMasterEquation}
&&\lim_{k_\lambda\to 0}\int d^4z e^{ikz}\Big\{ -\HVEV\partial_{\mu}^z \big< 0\vert T\Big[  \big(\partial^\mu\pi^{t}(z)\big)  \nonumber \\
&&\quad \quad \quad \quad \times h(x_1)...h(x_N) \pi^{t_1}(y_1)...\pi^{t_M}(y_M)\Big]\vert 0\big>_{\rm connected} \nonumber \\
&&\quad - \sum_{m=1}^M \quad  i\delta^{t t_m}\delta^4(z-y_m) \big< 0\vert T\Big[ h(z) h(x_1)...h(x_N) \nonumber \\
&&\quad \quad \quad \quad \times  \pi^{t_1}(y_1)...{\widehat {\pi^{t_m} (y_m)}}...\pi^{t_M}(y_M)\Big]\vert 0\big>_{\rm connected} \nonumber \\
&&\quad + \sum_{n=1}^N \quad  i\delta^4(z-x_n) \big< 0\vert T\Big[ h(x_1)...{\widehat {h(x_n)}}...h(x_N) \nonumber \\
&&\quad \quad \quad \quad \times  \pi^{t}(z){\pi^{t_1}}(y_1)...\pi^{t_M}(y_M)\big]\vert 0\big>_{\rm connected} \Big\} \nonumber \\
&&\quad =0
\end{eqnarray}
where the ``hatted" fields ${\widehat {h(x_n)}}$ and ${\widehat {\pi^{t_m} (y_m)}}$ are to be removed.

Isospin indices will become increasingly cumbersome; 
we therefore  again adopt 
B.W. Lee's \cite{Lee1970} convention of supressing isospin indices, 
allowing momenta to implicitly carry them.

The Adler self-consistency relations, 
but now for the $SU(2)_L\times U(1)_Y$ gauge theory 
(rather than global $SU(2)_L \times SU(2)_R$ \cite{Adler1965,AdlerDashen1968}), 
are obtained by putting the external $\phi$ particles on mass-shell:
\begin{eqnarray}
\label{EwSMAdlerSelfConsistency} 
&&\HVEV T^{2\otimes 1}_{N,M+1}(p_1...p_N;0q_1...q_M)\nonumber \\
&& \quad \quad \times (2\pi)^4\delta^4 \Big(\sum_{n=1}^N p_n +\sum_{m=1}^M q_m \Big) \Big\vert^{p_1^2 =p_2^2...=p_N^2=m_{BEH}^2}_{q_1^2 =q_2^2...=q_M^2=0}  \nonumber \\
&& \quad \quad =0 
\end{eqnarray}

With some exceptions, 
the $\phi$-sector connected amputated transition matrix $T^{2\otimes 1}_{N,M}$ 
can be split in two by cutting an internal $h$ or $\vec \pi$ line, 
and are designated 1-$\phi$-R. 
In contrast, 
the $\phi$-sector connected amputated Green's functions $\Gamma^{2\otimes 1}_{N,M}$ 
are defined to be 1-$\phi$-I, 
i.e. they cannot be split by cutting an internal $h$ or $\vec \pi$ line.
\begin{eqnarray}
\label{1SPReducibility}
T^{2\otimes 1}_{N,M} = \Gamma^{2\otimes 1}_{N,M} + (1-\phi-R)\,.
\end{eqnarray}

Both $T^{2\otimes 1}_{N,M}$ and $\Gamma^{2\otimes 1}_{N,M}$ 
are 1-$({\vec W}_\mu , B_\mu)$-Reducible (1-${\vec W}_\mu ,B_\mu$-R), 
i.e. they can be split 
by cutting an internal transverse ${\vec W}_\mu$ or ${B}_\mu$ line. 

The special 2-point functions 
$T^{2\otimes 1}_{0,2}(;q,-q)$ and $T^{2\otimes 1}_{2,0}(p,-p;)$, 
and the 3-point vertex  $T^{2\otimes 1}_{1,2}(q;0,-q)$, 
are 1-$\phi$-I, not 1-$\phi$-R.
They are therefore equal to the corresponding 1-$\phi$-I 
connected amputated Green's functions. 
The 2-point functions
\begin{eqnarray}
\label{EwSMTMatrix2PointA}
T^{2\otimes 1}_{2,0}(p,-p;)&=&\Gamma^{2\otimes 1}_{2,0}(p,-p;)=\big[\Delta_{BEH}(p^2)\big]^{-1} \nonumber \\
T^{2\otimes 1}_{0,2}(;q,-q)&=&\Gamma^{2\otimes 1}_{0,2}(;q,-q)=\big[\Delta_{\pi}(q^2)\big]^{-1}  
\end{eqnarray}
are related to the $(1h,2\pi)$ 3-point $h{\vec \pi}^2$ vertex 
\begin{eqnarray}
\label{EwSM3PointVertex}
T^{2\otimes 1}_{1,2}(p;q,-p-q) = \Gamma^{2\otimes 1}_{1,2}(p;q,-p-q) 
\end{eqnarray}
by a 1-soft-pion theorem analogous with (\ref{SoftPionTMatrixID})
\begin{eqnarray}
\label{EwSMTMatrix2and3Point}
&&\HVEV T^{2\otimes 1}_{1,2}(q;0,-q) -T^{2\otimes 1}_{2,0}(q,-q;)+T^{2\otimes 1}_{0,2}(;q,-q) \nonumber \\
&&\quad \quad =\HVEV \Gamma^{2\otimes 1}_{1,2}(q;0,-q) -\big[\Delta_{BEH}(q^2)\big]^{-1} +\big[\Delta_{\pi}(q^2)\big]^{-1} \nonumber \\
&&\quad \quad =0 \,.
\end{eqnarray}

The Lee-Stora-Symanzik (LSS) theorem, 
in spontaneously broken $SU(2)_L\times U(1)_Y$ 
in $R^{2\otimes 1}_\xi (\xi=0)$ Landau gauge, 
is the $N=0,M=1$ case of that SSB gauge theory's Adler self-consistency relations 
(\ref{EwSMAdlerSelfConsistency})
\begin{eqnarray}
\label{EwSMAppendixGoldstoneTheorem} 
\HVEV T^{2\otimes 1}_{0,2}(;00)&=&0 \nonumber \\
\HVEV \Gamma^{2\otimes 1}_{0,2}(;00)&=&0 \nonumber \\
\HVEV \big[ \Delta_\pi (0) \big]^{-1} &=&-\HVEV\mpisq=0\,,
\end{eqnarray}
proving that $\vec \pi$ is massless. 
That all-loop-orders renormalized masslessness 
is protected/guaranteed by the $CP$-conserving global $SU(2)_{L-R}$ symmetry 
of the  physical states of the gauge theory 
after spontaneous symmetry breaking.

In analogy with (\ref{DefineInternalTMatrixA}), 
separate
\begin{eqnarray}
\label{EwSMDefineInternalTMatrixA}
&&T^{2\otimes 1}_{N,M+1}(p_1...p_N;0q_1...q_M) \nonumber \\
&&\quad \quad =T^{2\otimes 1}_{External;N,M+1}(p_1...p_N;0q_1...q_M) \nonumber \\
&&\quad \quad +T^{2\otimes 1}_{Internal;N,M+1}(p_1...p_N;0q_1...q_M)  
\end{eqnarray}
so that 
\begin{eqnarray}
\label{EwSMInternalTMatrix}
&&\HVEV T^{2\otimes 1}_{Internal;N,M+1}(p_1...p_N;0q_1...q_M) \nonumber \\
&&\quad \quad =\sum_{m=1}^M T^{2\otimes 1}_{N+1,M-1}(q_mp_1...p_N;q_1....{\widehat{q_m}}...q_M)  \nonumber \\
&&\quad \quad -\sum_{n=1}^N T^{2\otimes 1}_{N-1,M+1}(p_1...{\widehat{p_n}}...p_N;p_nq_1...q_M) \quad \quad 
\end{eqnarray}
in analogy with (\ref{InternalTMatrix}).

In analogy with (\ref{GreensFWTI}), 
removing the 1-$\phi$-R  graphs from both sides of (\ref{EwSMInternalTMatrix}) 
yields the recursive identity
\begin{eqnarray}
\label{EwSMGreensFWTI}
&&\HVEV \Gamma^{2\otimes 1}_{N,M+1}(p_1...p_N;0q_1...q_M) \nonumber \\
&&\quad \quad =\sum_{m=1}^M \Gamma^{2\otimes 1}_{N+1,M-1}(q_mp_1...p_N;q_1....{\widehat{q_m}}...q_M)  \nonumber \\
&&\quad \quad -\sum_{n=1}^N \Gamma^{2\otimes 1}_{N-1,M+1}(p_1...{\widehat{p_n}}...p_N;p_nq_1...q_M) \quad \quad
\end{eqnarray}

We observe that the LSS theorem makes tadpoles vanish.
\begin{eqnarray}
\label{EwSMTadpoles}
&&\big<0\vert h(x=0)\vert0\big>_{Connected} \\
&& \qquad \qquad = i \Big[i\Delta_{BEH}(0)\Big]\Gamma^{2\otimes 1}_{1,0}(0;) \nonumber
\end{eqnarray}
but the $N=0,M=1$ case of (\ref{EwSMGreensFWTI}) reads
\begin{eqnarray}
\label{EwSMGoldstoneTadpoles}
\Gamma^{2\otimes 1}_{1,0}(0;)&=&  \HVEV \Gamma^{2\otimes 1}_{0,2}(;00) \nonumber \\
&=&0 \,,
\end{eqnarray}
 so that tadpoles all vanish automatically, 
 and separate tadpole renormalization is un-necessary.
Since we can choose the origin of coordinates anywhere we like
\begin{eqnarray}
\label{EwSMGoldstoneTadpolesVanishExtended}
\big<0\vert h(x)\vert0\big>_{Connected} &=& 0
\end{eqnarray}

The Renormalized $\HVEV$ obeys
\begin{eqnarray}
\label{EwSMHVEV}
\big<0\vert H(x)\vert0\big>_{Connected} &=&\big<0\vert h(x)\vert0\big>_{Connected} +\HVEV \nonumber \\
&=& \HVEV \nonumber \\
\partial_\mu \HVEV &=&0
\end{eqnarray}

 \end{document}